\newcommand{\refs}{\par\noindent\hangindent=1pc\hangafter=1}

\documentclass[preprint2]{pp7}

\usepackage{graphicx}
\usepackage[hidelinks]{hyperref}
\usepackage{xcolor}
\hypersetup{
    colorlinks=true,
    linkcolor=[RGB]{32,120,182},
    filecolor=[RGB]{32,120,182},      
    urlcolor=[RGB]{32,120,182},
    citecolor=[RGB]{32,120,182},
    allcolors=[RGB]{32,120,182},
    }
\usepackage[switch]{lineno}
\usepackage{CJKutf8}
\usepackage[T1]{fontenc}

\input{pp7.h}

\begin{document}

\title{\textbf{\LARGE Geophysical Evolution During Rocky Planet Formation }}

\author {\textbf{\large Tim Lichtenberg}}
\affil{\small Atmospheric, Oceanic and Planetary Physics, Department of Physics, University of Oxford, UK\\tim.lichtenberg@physics.ox.ac.uk}
\author {\textbf{\large Laura K. Schaefer}}
\affil{\small Geological Sciences Department, Stanford University, USA\\lkschaef@stanford.edu}
\author {\textbf{\large Miki Nakajima}}
\affil{\small Department of Earth and Environmental Sciences \& Department of Physics and Astronomy,\\ 
University of Rochester, USA; mnakajima@rochester.edu}
\author {\textbf{\large Rebecca A. Fischer}}
\affil{\small Department of Earth and Planetary Sciences, Harvard University, USA\\rebeccafischer@g.harvard.edu}

\begin{abstract}
\baselineskip = 11pt
\leftskip = 1.5cm 
\rightskip = 1.5cm
\parindent=1pc
{\small Progressive astronomical characterization of planet-forming disks and rocky exoplanets highlight the need for increasing interdisciplinary efforts to understand the birth and life cycle of terrestrial worlds in a unified picture. Here, we review major geophysical and geochemical processes that shape the evolution of rocky planets and their precursor planetesimals during planetary formation and early evolution, and how these map onto the astrophysical timeline and varying accretion environments of planetary growth. The evolution of the coupled core–mantle–atmosphere system of growing protoplanets diverges in thermal, compositional, and structural states to first order, and ultimately shapes key planetary characteristics that can discern planets harboring clement surface conditions from those that do not. Astronomical campaigns seeking to investigate rocky exoplanets will require significant advances in laboratory characterization of planetary materials and time- and spatially-resolved theoretical models of planetary evolution, to extend planetary science beyond the Solar System and constrain the origins and frequency of habitable worlds like our own.
 \\~\\~\\~}
\end{abstract}  

%%%%%%%%%%%%%%%%%%%%%%%%%%%%%%%%%%%%%%%%%%%%%%%%%%%%%%%%%%%%%%%%%%%%%%%%%%%%%%%%%%%%%%%%%%%%%%%%%%%%%%%%%%%%%%%%%%%%%%%%%%%%%%%%%%%%
%%%%%%%%%%%%%%%%%%%%%%%%%%%%%%%%%%%%%%%%%%%%%%%%%%%%%%%%%%%%%%%%%%%%%%%%%%%%%%%%%%%%%%%%%%%%%%%%%%%%%%%%%%%%%%%%%%%%%%%%%%%%%%%%%%%%
%%%%%%%%%%%%%%%%%%%%%%%%%%%%%%%%%%%%%%%%%%%%%%%%%%%%%%%%%%%%%%%%%%%%%%%%%%%%%%%%%%%%%%%%%%%%%%%%%%%%%%%%%%%%%%%%%%%%%%%%%%%%%%%%%%%%
\section{\textbf{INTRODUCTION}} \label{sec:introduction}  \label{sec:1}
%\bigskip
%\noindent

\begin{figure}[tb]
 \epsscale{1.0}
 \plotone{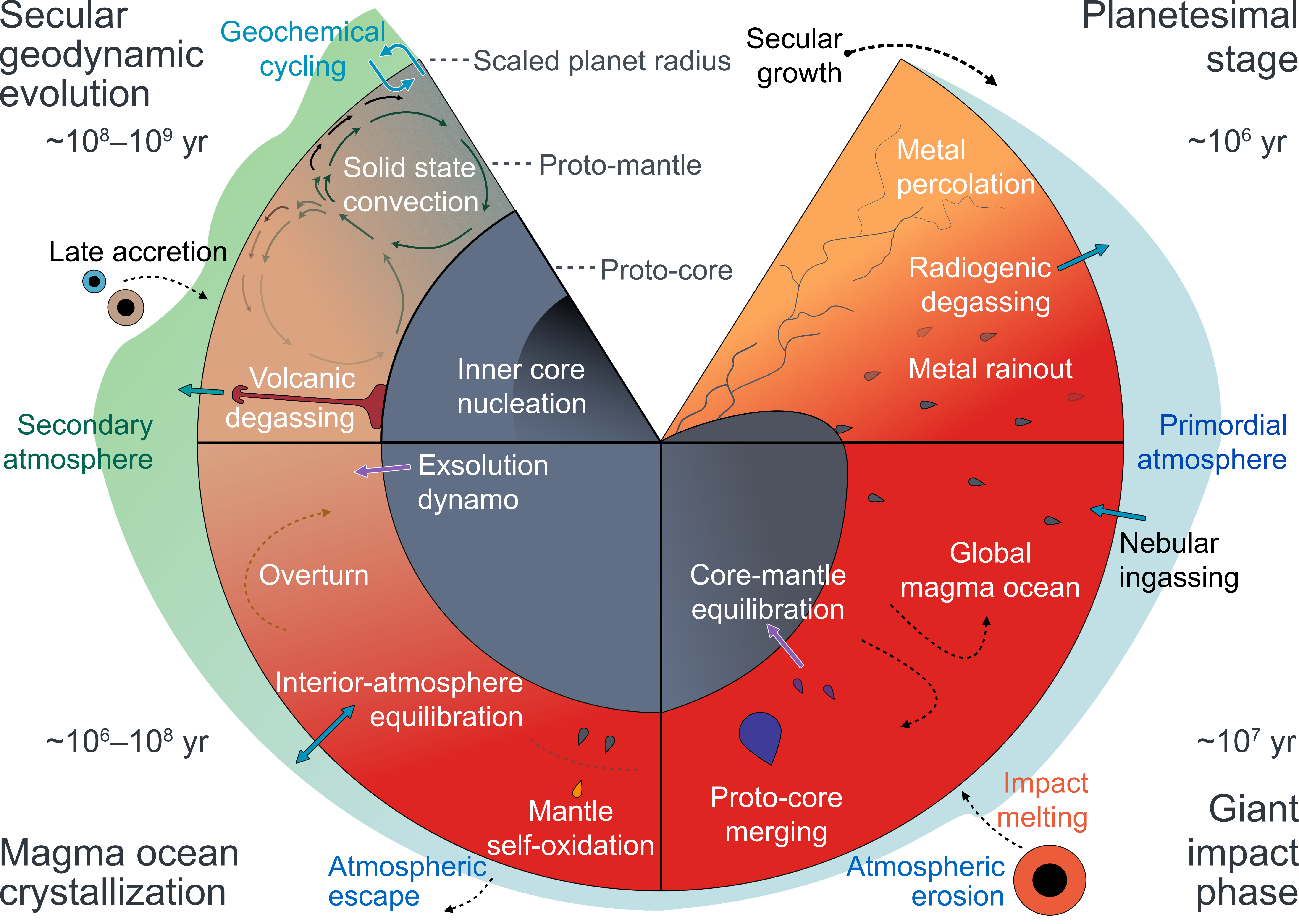}
 \caption{\small Chronology of geophysical and geochemical processes that affect the interior dynamics, structure, and climate of rocky planets. Planet size is normalized to current growth stage, starting from accretion in the disk (planetesimal stage), post-disk (giant impact phase), planet solidification and atmosphere formation (magma ocean crystallization), to the long-term evolution of interior and climate (secular geodynamic evolution). Processes schematically depicted here are introduced in \S\ref{sec:2}, and discussed in the context of accretion in \S\ref{sec:3}. \href{https://osf.io/sxrqz/}{\includegraphics[scale=0.35]{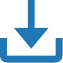}}}
 \label{fig1}
\end{figure}
The past years have seen tremendous advances in our understanding of the formation and evolution of Earth and its planetary siblings, and have expanded our picture from the Solar System to extrasolar planetary systems – both forming and mature ones. Increasing resolution in imaging protoplanetary disks and extensive transit and radial velocity surveys vastly increase the census of known planets. One of the key goals of astronomical surveys is  to better understand how terrestrial planets form and evolve, and ultimately provide answers to the question: \emph{how unique is our own habitable world?} Because data gained by remote sensing is limited in detail, however, observations of rocky planets must be interpreted in the context of data and theories that originate from the study of Earth and its terrestrial siblings. In order to meaningfully evaluate new observations in light of constraints derived from the terrestrial planets and moons of the Solar System, cross-disciplinary efforts are necessary to draw from and extend a common knowledge base by incorporating novel findings from other planetary systems. 

A major hurdle for exoplanet science is that even the most basic requirements for similarity with Earth, such as surface temperature, pressure, and availability of liquid water, are degenerate in terms of key observables like planetary mass, radius, or top-of-atmosphere gas speciation. From a geoscience perspective, different accretion chronologies and stellar environments suggest strongly diverging planetary evolution. The diversity in composition and accretion paths evidenced by disk observations and the scatter in exoplanet compositions hence call into question whether the past evolutionary trajectories of the Solar System terrestrial planets are representative for the exoplanet population. Therefore, in this review, we explore the general evolutionary principles of rocky planets during and following accretion. Specifically, we discuss major geophysical processes that are established to have had first-order consequences for the long-term evolution of rocky planets in the Solar System, and how these may be generalized to understand rocky extrasolar planets. 

In contrast to the system-focused viewing angle that is usually taken in astronomical studies, we here review the physics and chemistry of rocky planets from a planetary-centric point of view (Fig.~\ref{fig1}), with particular emphasis on the processes that drive thermal and compositional change during accretion (\S\ref{sec:geophysical_processes}). These serve as background to illustrate the evolution of rocky planets from their birth in the protoplanetary disk until after mantle solidification and outgassing of a long-lived atmosphere. We then map these  processes onto a general timeline of rocky planet formation (\S\ref{sec:formation_timeline}), dividing the accretion chronology largely into two broad phases: during the presence of the protoplanetary disk (\S\ref{sec:disk_stage}) and planetary growth thereafter (\S\ref{sec:post-disk_phase}). We illuminate how internal processes that operate during accretion shape the structure and thermal history of precursor planetesimals (Fig.~\ref{fig1}, top right) and protoplanets (Fig.~\ref{fig1}, bottom right) and their forming atmospheres. Connecting the evolution during the tail-end of formation to mature planetary systems, we describe how volatile elements are distributed between the interconnected subsystems of core, mantle, and atmosphere of rocky planets (Fig.~\ref{fig1}, bottom left), and the interplay between the style of accretion and the long-term surface environment and climate after planet solidification (\S\ref{sec:transition_to_long-term_evolution}, Fig.~\ref{fig1}, top left) that can potentially be discriminated by astronomical observations. Because the vast majority of geophysical and geochemical constraints are derived from the Solar System planetary objects, many parts of our discussion derive from and focus on our home system. Where possible we include extrasolar systems and extrapolate known trends to parameters relevant for exoplanets. Finally, in  \S\ref{sec:outlook_and_summary} we outline some critical areas that require further community investment, such that astronomical observations can yield their full potential in investigating the diversity of extrasolar planets and constraining the frequency of habitable worlds like our own. 

%%%%%%%%%%%%%%%%%%%%%%%%%%%%%%%%%%%%%%%%%%%%%%%%%%%%%%%%%%%%%%%%%%%%%%%%%%%%%%%%%%%%%%%%%%%%%%%%%%%%%%%%%%%%%%%%%%%%%%%%%%%%%%%%%%%%
%%%%%%%%%%%%%%%%%%%%%%%%%%%%%%%%%%%%%%%%%%%%%%%%%%%%%%%%%%%%%%%%%%%%%%%%%%%%%%%%%%%%%%%%%%%%%%%%%%%%%%%%%%%%%%%%%%%%%%%%%%%%%%%%%%%%
%%%%%%%%%%%%%%%%%%%%%%%%%%%%%%%%%%%%%%%%%%%%%%%%%%%%%%%%%%%%%%%%%%%%%%%%%%%%%%%%%%%%%%%%%%%%%%%%%%%%%%%%%%%%%%%%%%%%%%%%%%%%%%%%%%%%
\section{\textbf{GEOPHYSICAL PROCESSES}}  \label{sec:geophysical_processes} \label{sec:2}
\subsection{\textbf{Thermodynamics}} \label{sec:thermodynamics}  \label{sec:2.1}
\subsubsection{\textbf{Thermal budget}}  \label{sec:thermal_budget}   \label{sec:2.1.1}

\emph{Gravitational Potential Energy.} For planetary-sized objects, release of gravitational potential energy can be a main contributor to the internal heat budget. The maximum potential energy that could be released as heat by accretion of a planet is the planet’s gravitational binding energy $kGM^{2}_{p}/R_{p}$, where $k = \frac{3}{5}$ for a sphere of uniform composition, $G$ is the gravitational constant, $M_{p}$ is the planetary mass, and $R_{p}$ is the planetary radius. This releases energies of $10^{30}$ and $10^{32}$ J for Mars- and Earth-sized bodies, respectively. If temperature is assumed to be homogeneous throughout the object, this would imply temperatures of $\sim$31000 K for Earth and $\sim$6300 K for Mars if they formed instantaneously. However, the temperature structure of a growing body will be dictated by the balance between the rate at which heat is lost by radiation to space and the growth rate of the body, so this maximum temperature will not be achieved. In addition to potential energy released by assembling the materials of a planet together, further energy is released by differentiation, meaning the separation of denser metallic materials to the center of a rocky planet during core formation and contraction upon cooling and densification. Heat generated by core differentiation can be approximated as $U_{H}$ – $U_{D}$, the difference in gravitational potential energy between a homogeneous body and a layered planet (\emph{Breuer \& Moore} 2015). This produces about $10^{29}$ to $10^{31}$ J of heat for Mars- to Earth-sized bodies ($\sim$0.1 of the gravitational binding energy, \emph{Rubie et al.} 2015a; \emph{Solomon} 1979). For an Earth-sized planet, this would produce a temperature increase of $\sim$1300 K if heat is distributed uniformly. 

However, heat generated by core formation can be deposited either within the silicate mantle or transported to the core in the metallic fluid depending on the mechanism and timing of metal separation. As discussed in \S\ref{sec:physics_of_core-mantle_segregation} in more detail, unless the impactor's core directly merges with the target’s core, the core breaks apart into smaller fragments, which sink to the bottom of the magma ocean formed by the impact event. The iron melt would eventually assemble as diapirs that sink through the lower solid mantle to the target’s core. \emph{Samuel et al.} (2010) found that transfer of heat between the mantle and core by viscous heating during such negative diapirism depends on the size of the diapirs, which is influenced by the rheology of the mantle (\emph{Golabek et al.} 2009). Many small diapirs will produce a super-heated core, whereas a few large diapirs will lead to a hot mantle and cold core. \emph{Ke \& Solomatov} (2009) in contrast showed that if metal descends through the mantle in large channels, then most of the heat from differentiation will end up in the core. However, these simulations cannot resolve small-scale physics due to the resolution limitations, which leads to uncertainty in heat distribution. The initial temperature of the Earth’s metallic core remains highly uncertain, with estimates ranging from 4500 to 8000 K (e.g., \emph{Labrosse} 2015; \emph{Nimmo} 2015). The initial core temperature is important for the subsequent thermal evolution of the mantle and the onset of magnetic dynamo generation.

Gas accretion, which is a major source of gravitational potential energy for gas giant planets, is minimal for rocky planets (\emph{Hayashi et al.} 1979). Estimates of gas accretion rates for super-Earths allow envelope-to-core mass fractions of order $10^{-3}$ (\emph{Ginzburg et al.} 2016), which represents a minimal portion of the gravitational accretion energy. 

\emph{Impacts.} Related to the release of gravitational potential energy, accretion produces heat through the conversion of kinetic energy of impactors into heat, $E = ½ mv^{2}$, where $m$ is the impactor mass and $v$ is the impact velocity. Kinetic energy from impactors is partitioned between (i) plastic energy of deformation of solids, (ii) thermal energy of a portion of the target and impactor, and (iii) kinetic energy of the ejecta. A single impact produces a pressurized isobaric core due to the transfer of the incident impactor kinetic energy at the impact site. Shock pressures rapidly decay away from the core. The temperature within the shocked isobaric core often reaches temperatures above melting for larger planets (\emph{Tonks \& Melosh} 1993). A buoyant thermal anomaly is created that can lead to convective-like motions and isostatic adjustment (\emph{Coradini et al.} 1983; \emph{Senshu et al.} 2002), which is vertical and lateral movement of the mantle and crust to regain gravitational equilibrium.

Numerical models have found that the fraction $h$ of the impactor’s kinetic energy that is converted into thermal energy can be highly variable with more efficient conversion at higher impact velocities ($>$10 km/s; \emph{Coradini et al.} 1983; \emph{O’Keefe \& Ahrens} 1977) and less efficient at more grazing impacts (\emph{Nakajima et al.} 2021). Heat is then deposited at different depths depending on the size of the impactor. Models suggest that heat from small impactors remains in the near surface environment where it is quickly radiated back to space in the absence of an atmosphere (\emph{Stevenson} 1989; \emph{Melosh} 1990). Larger impactors bury heat deeper within a planet’s interior (\emph{Safronov} 1978; \emph{Kaula} 1979). More recent work has produced updated scaling laws of heat deposition that also take into account the impact angle (\emph{Nakajima et al.} 2021; \emph{Kegerreis et al.} 2021). These scaling laws can be used to derive the size and shape of impact-generated melt volumes in which metal-silicate equilibration is likely to occur (the time scale ranges from hours to months), prior to isostatic compensation that will cause the melt pool to radially spread out and become a global magma ocean ($10^{2}$ to $10^{5}$ yr, depending on the solid rheology, \emph{Reese \& Solomatov} 2006).

Neglecting radioactive heating and liquid state convection and  considering only gravitational potential and impact energy, the thermal structure of a growing protoplanet can be described by (\emph{Breuer \& Moore} 2015):
\begin{linenomath}\begin{equation}
    T(r)=h \frac{G M(r)}{C_{p} r}\left(1+\frac{r v^{2}}{2 G M(r)}\right)+T_{e}+\Delta T_{a d}(r), \label{eq:1}
\end{equation}\end{linenomath}
where $h$ is the heat retention factor, $M(r)$ is the mass of the planet internal to radius $r$, $C_{p}$ is the specific heat, $v^{2}/2$ is the approach kinetic energy per unit mass, $T_{e}$ is the temperature of the surroundings, and $\Delta T_{ad}(r)$ is the temperature rise due to adiabatic compression as the planet’s radius increases. This model predicts cold internal temperatures and a temperature maximum just below the surface, with temperatures approaching or exceeding melting for planetary-sized bodies. This model can be understood as an end-member example of cold accretion, potentially applicable to late-forming or pebble-accreting planetesimals in the absence of a blanketing atmosphere and internal heat sources. However, early-formed and large bodies melt rapidly, such that Eq. \ref{eq:1} loses its validity. While potential energy contributes strongly to the heating of large bodies, impacts large enough to cause sufficient melting to differentiate an asteroid would likely disrupt these bodies (\emph{Tonks \& Melosh} 1992; \emph{Keil et al.} 1997). Cumulative heating by multiple impacts is also not effective on bodies less than a few hundred kilometers in diameter due to rapid heat loss between subsequent impacts (\emph{Ciesla et al.} 2013). However, impacts on these smaller objects may produce sufficient localized heating for thermal metamorphism and local melting in the walls and floors of craters.

\emph{Short and Long-Lived Radioactive Isotopes.} Short-lived radioactive isotopes (e.g., $^{26}$Al, $^{41}$Ca, $^{60}$Fe, $^{53}$Mn, $^{182}$Hf, $^{129}$I, and $^{244}$Pu) were present in the early Solar System. Excesses of daughter isotopes have been identified in meteorite components, confirming the early presence of these isotopes. Possible sources for these isotopes include AGB stars, supernovae, Wolf-Rayet stars, and spallation reactions (\emph{Lugaro et al.} 2018; \emph{Desch et al.}, this volume), which may have injected material into the nascent solar nebula. Heterogeneities in the initial abundances of some of these isotopes have been identified in early-forming solar nebula components (e.g FUN CAIs, \emph{Holst et al.} 2013; CAIs are Calcium-Aluminum-rich inclusions in meteorites and the oldest dated solids from within the Solar System), which have been attributed either to admixing during the early evolution of the proto-Sun or selective thermal processing of dust grains (\emph{Trinquier et al.} 2009; \emph{Küffmeier et al.} 2016).

Short-lived radioactive isotopes were responsible for substantial heating of small bodies in the Solar System (see \S\ref{sec:3.1}). $^{26}$Al is likely responsible for most of the early heating due to its larger relative abundance ($^{26}$Al/$^{27}$Al $\approx 5.2 \times 10^{-5}$, \emph{Jacobsen et al.} 2008) and stronger heating rate compared to long-lived radioactive isotopes (\emph{Ruedas} 2017). Bodies greater than $\sim$20 km in radius likely melted completely due to $^{26}$Al heating if they accreted within the next 2 Myr after CAI formation, and smaller objects of only a few km may have melted fully if they accreted immediately after CAIs (\emph{Gail et al.} 2014). Because of the short half-life of $^{26}$Al, later formation times mean that insufficient amounts of $^{26}$Al remained to cause melting. For objects that melt, Al can become sequestered into early partial melts and segregate into crustal layers because of its incompatibility (it has either the wrong valence or ionic radius for the cation sites of the solid minerals and will hence preferentially enter the melt phase), potentially leaving internal parts of planetesimals and protoplanets with a depleted radiogenic heat source (\emph{McCoy et al.} 2006). $^{60}$Fe could be another major heat source of extrasolar planetesimals, but its abundance in the Solar System may have been insufficient to contribute substantial heating (\emph{Tang \& Dauphas} 2012; \emph{Cook et al.} 2021). Other short-lived radioactive isotopes ($^{41}$Ca, $^{53}$Mn, $^{182}$Hf, $^{129}$I, $^{244}$Pu) are not thought to have contributed to substantial heating of objects within the Solar System but help provide evidence of the source of the short-lived isotopes and in some instances act as good relative chronometers. The $^{26}$Al abundance in the Solar System is elevated by a factor of $\sim$3–25 relative to the interstellar medium (\emph{Lugaro et al.} 2018). It has been suggested to approximately reflect the galactic mean production (\emph{Jura \& Young} 2014; \emph{Fujimoto et al.} 2018), but enrichment models (\emph{Lichtenberg et al.} 2016b; \emph{Küffmeier et al.} 2016) and observational inferences (\emph{Reiter et al.} 2020; \emph{Forbes et al.} 2021) of $^{26}$Al and $^{60}$Fe distributions in individual star-forming regions suggest widely variable abundance levels and hence planetesimal heating rates across extrasolar planetary systems (\emph{Lugaro et al.} 2018).

Long-lived radioactive isotopes are important heat sources over the evolutionary history of rocky planets. The primary contributors to internal heat generation are $^{235}$U, $^{40}$K, $^{238}$U, and $^{232}$Th. These radioisotopes have half-lives on the order of billions of years. Heating rates for each of these elements are given in \emph{Ruedas} (2017). A number of studies have been conducted to identify what fraction of Earth’s current heat flux is caused by the decay of these long-lived radioactive isotopes (the Urey ratio), but this has not reached a consensus (\emph{Foley et al.} 2020; \emph{KamLAND Collaboration} 2011). Like Al, these elements behave incompatibly during silicate melting, which means that they remain in the melt until lower temperatures than other elements. Therefore, they are likely to become concentrated in the crust of most rocky planets. This transport of heat-producing elements to surface regions has implications for the thermal evolution of rocky planets. Galactic chemical evolution models suggest that continuous nucleosynthetic element production leads to dilution of radiogenic isotopes relative to major rock-forming elements (Si, Mg, Fe, etc.) over time; planets that form earlier in galactic history therefore are more likely to have higher long-term heat production than rocky planets forming later (\emph{Frank et al.} 2014).

Some radioactive elements may become concentrated in metallic phases at high pressure and be segregated into the core. $^{60}$Fe would have been concentrated in core-forming phases if it was still present during core formation, but the initial abundance was likely low enough that it did not strongly influence early core thermal budgets in the Solar System (\emph{Trappitsch et al.} 2018). Uranium and thorium have been shown to partition into metals under highly reducing and sulfur-rich conditions, relevant for Mercury-like objects, or compositions similar to enstatite chondrites (\emph{Wohlers \& Wood} 2015). The U/Th/Pb systematics of the Earth’s crust and mantle do not suggest that significant fractions of these elements partitioned into the Earth’s core. Thus, they are not likely to be a strong contributor to the heat flux of the Earth’s core, but it remains a possibility for other terrestrial planets (\emph{O’Neill et al.} 2020). Minor amounts of potassium in the core are more plausible, with abundances ranging from 26–100 ppm suggested by the observed K-depletion of the Bulk Silicate Earth (BSE) (\emph{Nimmo et al.} 2004; \emph{Blanchard et al.} 2017). These amounts may have been enough to generate an early geodynamo; however, depletions of potassium are consistent with other moderately volatile element depletions, so more evidence is necessary to suggest strong partitioning of K into the core (\emph{O’Neill et al.} 2020).

\emph{Phase Transitions.} Phase transitions within a rocky planet are a minor source (sink) of energy by release (consumption) of latent heat, for instance by crystallization (melting) or condensation (evaporation). Heat from impacts first raises the temperature of mantle materials (sensible heat) to the melting point, then contributes to latent heat of melting, then increases the temperature of the melt, and finally contributes to latent heat of vaporization. Estimates of maximum temperatures that could be reached by giant impacts often neglect latent heat of melting, assuming all energy is converted into specific heat, and therefore overestimate maximum temperatures. Latent heat of crystallization can contribute additional heat to solidifying systems.  Crystallization of the inner core of the Earth provides 15–30\% of the total energy released by inner core formation. Additional effects include gravitational potential energy release and secular cooling (\emph{Nimmo et al.} 2015; \emph{Landeau et al.} 2022). Crystallization within a magma ocean can affect convective patterns and overall cooling timescales (\emph{Solomatov \& Stevenson} 1993). Crystallization of the Earth’s entire silicate reservoir produces $1.7\times10^{27}$ kJ of heat, which translates into an additional 330 K worth of heat that must be removed.

\emph{Other Heat Sources.} Additional heat sources may be important for planetary objects forming under conditions that are exceptional in the Solar System, but may be widespread in exoplanetary systems. Tidal dissipation can be a major source of energy both early in a planet’s history and during its primary evolution phase. For instance, Io, Europa, and Ganymede all currently or early in their history  experienced substantial tidal heating due to their orbital interactions with Europa (for Io), Ganymede (for Europa), or Dione (for Enceladus). Io’s tidal heating rate is several orders of magnitude higher than its heating by long-term radiogenic isotopes (\emph{Foley et al.} 2020). The Earth-Moon system experienced early very strong tidal interactions immediately after Moon-formation during the outward evolution of the Moon’s orbit. Heating by tidal dissipation likely played a role in the lifetimes of the Earth’s and Moon’s magma oceans and the evolution of the lunar orbit (\emph{Zahnle} 2006; \emph{Meyer et al.} 2010; \emph{Zahnle et al.} 2015; \emph{Chen \& Nimmo} 2016). Exoplanets orbiting small stars on close orbits experience strong tidal interactions that circularize their orbits and damp planetary rotation rate until tidal-locking occurs. This also likely generates substantial early heating. Later heating for these planets requires non-zero eccentricity or obliquity, which can be maintained by planet-planet interactions in multi-planet systems (\emph{Bolmont et al.} 2013). Single-planet systems are likely to achieve circularization and tidal-locking, under which circumstances no tidal dissipation will occur (\emph{Jackson et al.} 2008). Tidal heating thus likely matters only for resonant orbits.

Induction heating has been previously suggested as a possible heat source for planetesimals immersed in a stellar wind field (\emph{Sonett et al.} 1970; \emph{Shimazu \& Terasawa} 1995; \emph{Menzel \& Roberge} 2013). \emph{Sonett et al.} (1970) suggested that this mechanism could have been responsible for early differentiation of meteorite parent bodies during the T Tauri phase of the Sun, as an alternative to short-lived radiogenic heating. The calculated heating rate depends strongly on the conductivity of the object, with higher conductivity leading to lower heating (\emph{Shimazu \& Terasawa} 1995). Total heating rates for asteroids can be within factors of $10^{-6}$ to $1$ of heating due to $^{26}$Al, depending on the electrical conductivity (\emph{Menzel \& Roberge} 2013). Because heating rates depend on the induced field strength, they are significantly stronger at close distances to the host star (\emph{Menzel \& Roberge} 2013).  Strong magnetic fields, such as found in some pre-main sequence stars, white dwarfs, or neutron stars will produce stronger heating and may induce orbital decay of close-in asteroids and dwarf planets (\emph{Bromley \& Kenyon} 2019). More recently, magnetic induction has been suggested as a potential heat source for Earth- and super-Earth-sized rocky planets that orbit closely to their stars (\emph{Kislyakova et al.} 2017, 2018; \emph{Kislyakova \& Noack} 2020), and ohmic heating may be an important contributor to hot Jupiter radius inflation (\emph{Thorngren \& Fortney} 2018; \emph{Sarkis et al.} 2021). Io may provide a useful test case for magnetic induction heating in the modern Solar System.

\subsubsection{\textbf{Heat transport}} \label{sec:heat_transport}   \label{sec:2.1.2}

\emph{Conduction.} Conduction is the thermal diffusion of heat across a temperature gradient. Conduction is the dominant internal cooling mechanism of planetesimals when they are below the melting temperature of their materials. Small planetesimals ($<$ a few km) will always cool conductively, but larger planetesimals may cool conductively only until sufficient melting occurs to trigger convection (\emph{Hevey \& Sanders} 2006). Assuming $^{26}$Al is a primary heat source in planetesimals, progressively larger planetesimals may remain purely conductive for later formation times (\emph{Hevey \& Sanders} 2006). Conduction can also occur at boundaries between convecting regions (e.g., between atmosphere and upper mantle, lower mantle and core, etc.). The region where conduction dominates is called a thermal boundary layer; in the Earth this region extends below the lithosphere. The heat flux across a conductive thermal boundary depends directly on the temperature contrast and inversely on the thickness of the boundary layer. Some models suggest that even planetesimals that melt and differentiate maintain a conductive surface boundary layer that remains undifferentiated (\emph{Weiss et al.} 2010; \emph{Elkins-Tanton et al.} 2011).

Conduction timescales depend on the thermal conductivity of the object. Planetesimals may begin as highly porous materials that become progressively more compacted (less porous) as they grow to larger sizes. Porous or particulate materials have much lower thermal conductivities than consolidated rock materials, which can affect the thermal evolution of these objects. Surface regolith can insulate the interior of a consolidated planetesimal and keep cooling rates low (\emph{McSween et al.} 2003). Regolith can be generated either through impacts or remain as a residual layer that does not suffer compaction and sintering due to close contact with space. Cold compaction due to self-gravitation occurs for objects larger than 10 km (\emph{Gail et al.} 2015), which brings the density close to that of the densest random packing of equal-sized spheres. Sintering, in which creep of heated materials causes plastic deformation and filling of remaining voids, occurs rapidly at temperatures of $\sim$700 K at planetesimal pressure conditions (\emph{Gail et al.} 2014). Compacted materials have thermal conductivities 2–3 orders of magnitude larger than highly porous material.

\emph{Convection.} Convection is the transport of heat by physical motion. A variety of materials and convective regimes dominate throughout the growth and evolution of rocky planets. Within planetesimals, heating by $^{26}$Al can trigger melting of different phases that convect throughout the initially porous body. First fluids formed are likely hydrothermal by melting of ices if present. Hydrothermal convection is likely to have been important in planetesimals larger than several tens of kilometers in diameter (\emph{McSween et al.} 2003; \emph{Bland \& Travis} 2017). Significant water volumes have a thermal buffering effect through the large heat of fusion of ice, high heat capacity of water and the ability of circulating water to enhance heat loss, and may keep water-rich planetesimals colder than water-poor planetesimals (\emph{Grimm \& McSween} 1989). Rapid heating may instead drive water outward with enough speed that aqueous alteration is diminished or does not occur.

Further heating of planetesimals leads to sulfide-metal melting, followed by silicate melting. Convection likely becomes more efficient at heat transportation than conduction when enough melt has formed to reduce the viscosity of the system to liquid-like values: $\sim$1 Pa s for liquids compared to $\sim10^{20}$ Pa s for solids, a difference of $\sim$20 orders of magnitude. The critical melt fraction at which this occurs is around 50–60\%, depending on composition (\emph{Abe} 1993; \emph{Solomatov \& Stevenson} 1993; \emph{Costa et al.} 2009). In hot young planets, convection within magma oceans is extremely turbulent due to very low viscosities (\emph{Solomatov} 2015). The convective vigor is characterized by the Rayleigh number, which is the ratio of the energy liberated by buoyant forces to that dissipated by viscous forces (\emph{Turcotte \& Schubert} 2014),
\begin{linenomath}\begin{equation}
Ra=\frac{\rho_{s i l} \alpha g \Delta T z^{3}}{\kappa \eta}, \label{eq:Ra}
\end{equation}\end{linenomath}
where $\rho_{sil}$ is the density of the silicate liquid, $\alpha$ is the coefficient of thermal expansion, $g$ is the gravitational acceleration, $\Delta T$ is the difference between the surface temperature and the potential temperature (in other words: temperature in excess of the adiabatic gradient throughout the layer), $z$ is the depth of the magma ocean, $\kappa$ is the thermal diffusivity, and $\eta$ is the dynamic viscosity. Above a critical value of $Ra_{crit} \sim$ 500–1000 convection occurs. The Rayleigh number of a deep magma ocean on Earth has been estimated to be $10^{27}$--$10^{32}$, indicating vigorous turbulent convection with velocities on the order of meters per second (\emph{Rubie et al.} 2003). At modest $Ra$ convection is controlled by the thermal boundary layer of the upper mantle. For a sufficiently large Rayleigh number (i.e., vigorous convection), the convective regime may transition from soft to hard turbulence, which is characterized by large-scale circulation patterns. In hard turbulence regimes the boundary layer loses control over the heat flow through the mantle, which is controlled by the large-scale eddy circulation (\emph{Solomatov} 2015), so hard turbulence magma ocean periods would be super-luminous, but likely short-lived. Similar to planetary atmospheres, eddy overturn times for magma oceans can thus vary from a few days to a few hundred years, depending on how close the mantle is to the solidus temperature. Liquid-state convection will cease as the magma ocean begins solidifying below the critical melt fraction and the viscosity transitions from liquid- to solid-like values (\S\ref{sec:2.1.3}). The timing and process of the transition from liquid magma ocean convection to solid-state convection remains uncertain. Some models predict that solid state convection is delayed by a density stratification produced by magma ocean crystallization (\emph{Zaranek \& Parmentier} 2004), whereas more recent models predict onset of solid-state convection within the solidified cumulate pile beneath a crystallizing magma ocean (\S\ref{sec:3.3}).

The solid interiors of rocky planets convect as the solid mantle deforms like a viscoelastic fluid on long timescales. Solid state mantle convection is an ongoing phenomenon with overturning timescales for the Earth of $\sim$50–100 Myr for the oceanic crust, $\sim$1–2 Gyr for the upper mantle, and $\sim$6 Gyr for the continental crust and whole mantle, and is likely the dominant heat loss process for rocky planets for most of their lifetimes. Exceptions to this rule would be planets with extremely high silicate viscosities, either due to very different bulk compositions than observed in the Solar System or very low mantle temperatures, which may lose most of their heat through conduction. Convective motions are accommodated by solid-state creep mechanisms that activate at high temperatures and pressures. The tectonics of the lithosphere overlying the convecting mantle will influence the net heat flux out of the mantle. The two most well known tectonic styles are plate tectonics, in which the lithosphere is broken up into distinct plates, and stagnant lid tectonics, in which the entire lithosphere acts as a single plate (\S\ref{sec:3.3.2}). All else being equal, heat fluxes are larger for plate tectonics than stagnant lid planets, implying that plate tectonics planets should typically have lower mantle temperatures and more rapid thermal evolution than stagnant lid planets. It is important to note, however, that rocky planets may evolve between different convection styles throughout their lifetime. See \emph{Foley et al.} (2020) for a review of heat transport in rocky planets.

\emph{Radiative Cooling.} All planets ultimately cool by radiation into space. For bare rocky planets and planetesimals, this will occur as direct emission to space (or the disk gas) from the surface, where the radiative flux is given by $\sigma T_{surf}^{4}$, where $\sigma$ is the Stefan-Boltzmann constant and $T_{surf}$ is the surface temperature. For atmosphere-less objects, the surface temperature is then dependent directly on the incoming stellar flux and the accretion rate. Since the formation location of most planetesimals is unknown, fixed temperatures are often assumed for planetesimal thermal evolution models, with the assumption that surface heating from insolation does not vary significantly over the relevant evolution timescales (\emph{McSween et al.} 2003).

However, planetesimals likely exist for at least a portion of time within the gas of the protoplanetary disk, so that radiative cooling is limited by the gas temperature at their formation location. Accretion of a nebular atmosphere or outgassing of a secondary atmosphere will further blanket the protoplanet and inhibit direct cooling to space/disk. Accreted atmospheres of nebular gas up to a few bars are possible on Mars-sized planets, up to hundreds of bars on Earth-mass planets (\emph{Hayashi et al.} 1979; \emph{Sekiya et al.} 1980, 1981; \emph{Sasaki} 1989; \emph{Sasaki \& Nakazawa} 1990).

Atmospheric heat transport can occur either through radiation or convection. Dense regions of atmospheres will be convective, while optically thin regions will lose heat through radiation. The radiatively-defined temperature structure of the upper atmosphere becomes unstable to convection at the radiative-convective boundary (\emph{Pierrehumbert} 2010). When assuming that gas absorption can be idealized as independent of wavelength (the gray atmosphere approximation), a convective profile will then prevail (\emph{Hayashi et al.} 1979; \emph{Nakazawa et al.} 1985),
\begin{linenomath}\begin{equation}
\frac{\mathrm{d} T}{\mathrm{~d} r}= \begin{cases}-\frac{3 \kappa \rho}{16 \sigma T^{3}} \frac{L}{4 \pi r^{2}} & \text { (radiative) }, \\ \frac{\gamma-1}{\gamma} \frac{T}{P} \frac{\mathrm{d} P}{\mathrm{~d} r} & \text { (convective) },\end{cases}
\end{equation}\end{linenomath}
where $T$ is the temperature, $r$ is the radial distance from a reference point, for instance the planetary surface, $\kappa$ is the Rosseland mean opacity, $\rho$ is the density, $L$ is the luminosity of the surface, $\gamma$ is the heat capacity ratio, and $P$ is the pressure. Ultimate loss of heat to space from an atmosphere will occur at the temperature of the photosphere: $\sigma T_{photo}^{4}$, where the optical depth $\tau = \int_{}^{} \kappa \rho dr = 2/3$. \textit{Hayashi et al.} (1979) found the structure of nebular atmospheres accreted to rocky planets assuming a primary heat source from gravitational potential energy. The photospheric temperature ($T_{photo}$) is marginally less than the radiative temperature of an atmosphere-less planet, due to the dependence of the luminosity-term on $1/r^{2}$, where $r_{photo} > r_{planet}$. In contrast, the surface temperature of an H-He-dominated planet can reach several thousand Kelvin for an Earth-mass planet, depending only slightly on accretion rate. In this regime, surface temperature is a strong function of planetary mass for planets larger than 0.3 $M_{Earth}$, while radiative temperatures are much weaker functions of mass (\emph{Ikoma \& Genda} 2006) but strong functions of gas composition. The atmospheric photosphere temperature is factors of a few lower than the bare planet temperature, resulting in factors of $\mathcal{O}(10)$ longer cooling timescales for H-He atmospheres. Upon loss of the confining pressure of the disk gas as gas is swept from the system, accreted atmospheres of protoplanets less than a few Mars-masses will almost entirely be lost (\emph{Sekiya et al.} 1980, 1981; \emph{Stökl et al.} 2015). The exact loss rates are sensitive functions of temperature, EUV transparency, and rate of late giant impacts.

Early atmospheres may also be formed by shock degassing or internal outgassing of solid-delivered volatiles such as ices, hydrated minerals, and carbonates. Unlike accreted gas, there is no intrinsic limit to the mass of outgassed atmospheres based on the protoplanet mass reached during the disk phase. In this case, major gasses released include H$_{2}$O, CO$_{2}$, N$_{2}$, CH$_{4}$, CO, and other volatile compounds (\S\ref{sec:2.2.3}). H$_{2}$O, CO$_{2}$ and CH$_{4}$, in particular, are strong infrared absorbers that trap heat from the interior within the atmosphere and reduce effective cooling to space (the greenhouse effect). Early models estimated the atmospheric blanketing effect using gray gas models for H$_{2}$O-dominated atmospheres (\emph{Matsui \& Abe} 1986a,b; \emph{Abe \& Matsui} 1986, 1988) that use approximations of the infrared absorption effect to determine the heat flux out of the atmosphere. More sophisticated models take into account line-by-line radiative transfer in the H$_{2}$O atmosphere (e.g., \emph{Hamano et al.} 2015; \emph{Schaefer et al.} 2016; \emph{Marcq et al.} 2017), dry/moist convection and condensation of H$_{2}$O (\emph{Lebrun et al.} 2013) and other greenhouse gasses (\emph{Graham et al.} 2021), non-ideal behavior of atmospheric compounds (\emph{Kasting} 1988; \emph{Dorn \& Lichtenberg} 2021), atmospheric escape (\emph{Abe \& Matsui} 1988; \emph{Zahnle et al.} 1988; \emph{Hamano et al.} 2013), and different atmospheric compositions (\emph{Lupu et al.} 2014; \emph{Lichtenberg et al.} 2021b).

\emph{Advective Cooling.} In some circumstances, a rocky object may physically lose heat through loss of hot impact ejecta or through explosive volcanism. \emph{Wilson \& Keil} (1991) argue that basaltic melt that contains more than a few hundred parts per million of volatiles is lost to space by explosive volcanism for planetesimals $<$100 km in radius, and carry away some heat and plausibly some $^{26}$Al. High vapor pressures associated with ice melting in water-rich planetesimals like carbonaceous chondrites may lead to vapor pressures greater than the confining pressure, which results in fracturing and venting of gasses that carry some heat (\emph{Grimm \& McSween} 1989). Advective cooling has been suggested as a main mechanism of cooling for the early Earth and Io (\emph{Moore \& Webb} 2013; \emph{Moore et al.} 2017; \S\ref{sec:3.3.2}).

Partial melting and differentiation will cause segregation of some radiogenic elements. U, Th, K, and Al are incompatible and likely to end up in crustal layers by buoyant melt segregation. This transport depletes the interior of a differentiating planetesimal or planet of heat-producing elements. $^{60}$Fe, if present, will be transported into core-forming materials. Models of crust formation on Vesta have predicted a reverse thermal gradient because $^{26}$Al is enriched in crustal materials relative to the mantle, causing the crust to attain higher temperatures than the mantle (\emph{McSween et al.} 2003; \emph{Mandler \& Elkins-Tanton} 2013).

\subsubsection{\textbf{Melting and solidification}} \label{sec:melting_and_solidification} \label{sec:2.1.3}

In contrast to gas giant planets, the early evolution of rocky planets is intrinsically governed by the processes of melting and solidification. Melting allows for the large-scale segregation of materials with different properties (e.g., density, volatility, viscosity). This includes the formation of metallic cores at the center of “rocky” planets. Metallic cores are largely (although not exclusively) the regions within a rocky planet that can produce magnetic fields. Melting also allows for massive early outgassing of volatiles otherwise delivered in solid materials such as ices, hydrated silicates, carbonates, or carbides. Large scale melting on rocky planets produces liquids with much lower viscosities than solid materials and permits much more rapid transport of heat and cooling of interiors than would otherwise be possible within objects composed of condensed materials. Silicate mantles can further compositionally differentiate during episodes of partial melt extraction, which produced the crustal materials on the Earth, Mars, Venus, the Moon, and likely Vesta. 

Once segregated by differentiation, materials such as the metallic core, silicate mantle, and felsic crustal materials interact in much more restricted ways during the remainder of the planet’s evolution. Therefore the timing of processes such as volatile delivery, core formation, melting, and solidification of a magma ocean can lead to widely divergent paths for a rocky planet. Planets with identical bulk compositions that had their entire complements of carbon delivered either before or after core segregation might be very different kinds of worlds (\S\ref{sec:2.2}). For instance, reduced carbon is highly siderophile, so if it is accreted before metal segregation it will be locked up in the metallic core, but if accreted after core formation it can remain in the mantle and contribute to the atmosphere.

In a multi-component silicate system, melting begins at the solidus, the temperature at which the first drop of liquid forms, and continues up to the liquidus, the temperature at which the last crystal melts. Crystallization follows the reverse path. There are two end-member scenarios of crystallization: during fractional crystallization minerals with different melting points or partition behavior precipitate and are removed from the melt, which changes the composition of the magma residue. In contrast, during batch crystallization the magma composition remains intact. The crystallization mode of a magma column depends, among other variables, sensitively on composition (refractories, volatiles) and timescale of the solidification process. 

Melt fraction is often assumed to vary linearly with temperature between the solidus and liquidus. Solidus and liquidus temperatures are strongly pressure- and composition-dependent in both silicate and metallic systems. Solidus temperatures of silicates increase by $\sim$1000–1500 K across the Earth’s mantle due to pressure. In silicate systems, lower bulk SiO$_{2}$ abundances produce higher melting temperatures; variations of 30 wt\% SiO$_{2}$ produce differences in melting temperatures of 300–500 K. Liquidus temperatures range from 200–2500 K higher than solidus temperatures in the MgO-SiO$_{2}$ system, depending on the Mg/Si ratio (\emph{Baron et al.} 2017). Silicate melting temperatures are also influenced by iron content. For example, the melting point of MgO at 3 GPa is $\sim$3700 K, while for (Mg$_{0.9}$Fe$_{0.1}$)O, the solidus is $\sim$3250 K and the liquidus is $\sim$3600 K, and for (Mg$_{0.8}$Fe$_{0.2}$)O, the solidus is $\sim$2850 K and the liquidus is $\sim$3500 K (\emph{Zhang \& Fei} 2008). Volatiles such as H$_{2}$O and CO$_{2}$ in silicates and S, P, C in metals lower melting temperatures. Silicates saturated in water or CO$_{2}$ can have solidus temperatures that are reduced by 600–800 K (\emph{Manning} 2018; \emph{Dasgupta \& Hirschmann} 2007). Solid phases are in general less accommodating of volatile impurities than liquids, so as crystallization proceeds within a system, volatiles will typically become increasingly more concentrated in the liquid phase.

Solidification of magma increases silicate viscosity by $\sim$20 orders of magnitude. The viscosity of silicate liquids at the conditions of Earth’s magma ocean is between 0.01 to 1 Pa s (\emph{Karki \& Stixrude} 2010). In comparison, the viscosity of the Earth’s mantle today determined from rates of isostatic rebound data ranges from $10^{20}$ to $10^{22}$ Pa s (\emph{Mitrovica \& Forte} 2004). Formation of crystals during solidification causes an increase in the viscosity of the convecting magma ocean. Above a critical crystal fraction of $\sim$50–60\%, viscosity exhibits an exponential increase to solid-like behavior (\emph{Abe} 1993; \emph{Costa et al.} 2009; \emph{Solomatov} 2015), many orders of magnitude larger. Viscosity exhibits Arrhenius behavior ($\sim exp[(-E+PV)/RT]$), so lower temperatures lead to higher viscosities, whereas higher pressures lead to higher viscosities (\emph{Karato \& Wu} 1993). Therefore colder mantles will convect more sluggishly, while magma oceans below the critical crystal fraction convect very vigorously.

\emph{Thermal Evolution of Metallic Cores.} Metallic cores form through transport of liquid metal and sulfide phases to body centers due to the usual immiscibility of these fluids with silicate materials and their large negative buoyancy. As bodies grow and gravity increases, buoyancy forces also increase. Sulfide phases melt at lower temperatures than other refractory condensed materials, so juvenile cores of planetesimals should be dominated by Fe,Ni-sulfides. Sulfides melt at low enough temperatures that they must migrate through largely solid silicate phases towards central regions. As temperatures increase during accretion, the core-forming fluid will become less sulfur-rich as more metals begin melting. As temperatures increase, silicates also begin melting, so transport of core-forming fluids will occur through partially to eventually fully molten silicate phases. Different transport processes will likely dominate as the phase ensembles change as a result of increasing temperature.

Cores will be liquid when they initially form and must lose heat to the mantle in order to begin solidifying. Core solidification will proceed outwards if the adiabatic gradient is smaller than the liquidus gradient, and inwards (from the core-mantle boundary) if the reverse is true (\emph{Haack \& Scott} 1992). Within the Earth, the adiabatic gradient is smaller than the liquidus, so solids form at the base of the liquid layer to make the inner core. Inner core solids of Fe,Ni alloy have low light-element solubilities, therefore solidification causes enrichment of the liquid core layer in light elements. On small bodies, solidification may begin from the outside of the core layer and move inwards. Solidification may proceed either concentrically or dendritically both inwards and outwards (\emph{Haack \& Scott} 1992). Inward crystallization is likely on bodies smaller than Callisto for relatively low sulfur contents ($<$5 wt\%), but may occur for even larger objects with higher sulfur contents (\emph{Williams} 2009). A transition from bottom-up to top-down crystallization is possible for small bodies like the Moon with moderate sulfur contents; bottom-up fractional crystallization enriches the liquid layer in sulfur until the liquidus slope of the residual liquid becomes shallower than the adiabat and top-down crystallization begins (\emph{Laneuville et al.} 2014; \emph{Scheinberg et al.} 2015). Crystallization at intermediate depths may be possible, with iron crystals ‘snowing’ downwards to form an inner core (\emph{Hauck et al.} 2006; \emph{Stewart et al.} 2007; \emph{Rückriemen et al.} 2018).

\emph{Magma Ocean Evolution and Solidification.} In the simplest case, the slope of the magma ocean adiabat is smaller than that of the solidus and liquidus throughout the mantle (\emph{Elkins-Tanton} 2012; \emph{Stixrude} 2014). For a whole-mantle magma ocean, this leads to the formation of first solids near the base of the mantle. Once crystals form they may either be entrained in the convecting magma or settle towards their neutral buoyancy point. For the Earth, this likely led to equilibrium crystallization for most of the lower mantle, where crystals remain in equilibrium with melt. For the upper mantle, fractional crystallization may dominate, where crystals sink out of the convecting magma and no longer equilibrate with the melt. Fractional crystallization can lead to greater compositional differentiation between melt and solids as incompatible elements become more concentrated in the melt (\emph{Solomatov} 2015). Magma oceans that are dominated by fractional crystallization may solidify to unstable density contrasts and overturn upon crystallization (\emph{Elkins-Tanton et al.} 2003, 2005): Fe-bearing minerals behave incompatibly than other cations in silicates, which means they partition into the liquid phase in a solid-liquid aggregate. When the magma ocean crystallizes from bottom to top, the last remaining melts thus become strongly enriched in dense minerals. Upon solidification this would lead to whole-mantle overturn, during which the dense top layers gravitationally segregate downward, producing large-scale mantle melting and a stable density stratification. It is likely, however, that the solidifying magma ocean would undergo solid state convection during crystallization as opposed to overturn at the end of crystallization. In this scenario, a significant part of the compositional heterogeneity generated by fractional crystallization may be erased (\emph{Maurice et al.} 2017; \emph{Ballmer et al.} 2017; \emph{Boukaré et al.} 2018). See \S\ref{sec:3.3.2} for the consequences on long-term mantle convection.

Molecular dynamics simulations of MgSiO$_{3}$ melt at high pressures suggest that the Grüneisen parameter, which describes how pressure in a material increases with thermal energy on an isentrope,  increases on compression unlike crystalline phases; this leads to greater temperature increases along liquid adiabats than solid adiabats (\emph{Stixrude et al.} 2009; \emph{Stixrude} 2014). The predicted adiabatic gradient in Earth’s lower mantle may exceed the slope of the solidus/liquidus. This would lead to intersection of the adiabat and liquidus at lower pressures, near mid-mantle depths ($\sim$70 GPa) on Earth (\emph{Stixrude et al.} 2009). Melts are also predicted to be denser than crystals at lower mantle pressures, implying that the melts are negatively buoyant. These combined effects suggest that crystallization may begin from mid-mantle depths and proceed both towards the surface and towards the core-mantle boundary with a basal magma ocean possibly persisting for a significant fraction of a planetary lifetime (\emph{Labrosse et al.} 2007). The consequences of this crystallization scenario on the compositional evolution of rocky planets may be significant because basal magma oceans can potentially lock up a fraction of the planet’s volatiles in the interior. Super-Earths may be especially prone to solidify this way (\emph{Stixrude} 2014). The potential connection between basal magma oceans and dynamo generation are discussed in \S\ref{sec:3.2.3}.

\subsection{\textbf{Compositional differentiation}} \label{sec:compositional_differentiation}  \label{sec:2.2}

The separation of a terrestrial planet into distinct chemical layers, such as  metallic core, silicate mantle, and overlying atmosphere, is a complex, multi-stage process. It involves both physical mechanisms, such as melting, disaggregation of droplets, percolation, and diapirism, and chemical reactions, such as equilibration between metallic and silicate melts at high pressures and temperatures, and evolution of redox state. The physics and chemistry of core formation and mantle melting are inherently intertwined and dependent on one another. They have important consequences for other aspects of planetary composition, including the delivery, partitioning, and outgassing of volatiles that govern long-term atmospheric and surface composition of rocky planets and thus link the internal geophysics and -chemistry to astronomical observables.

\subsubsection{\textbf{Physics of core–mantle segregation}} \label{sec:physics_of_core-mantle_segregation}  \label{sec:2.2.1}

\begin{figure}[htb!]
 \plotone{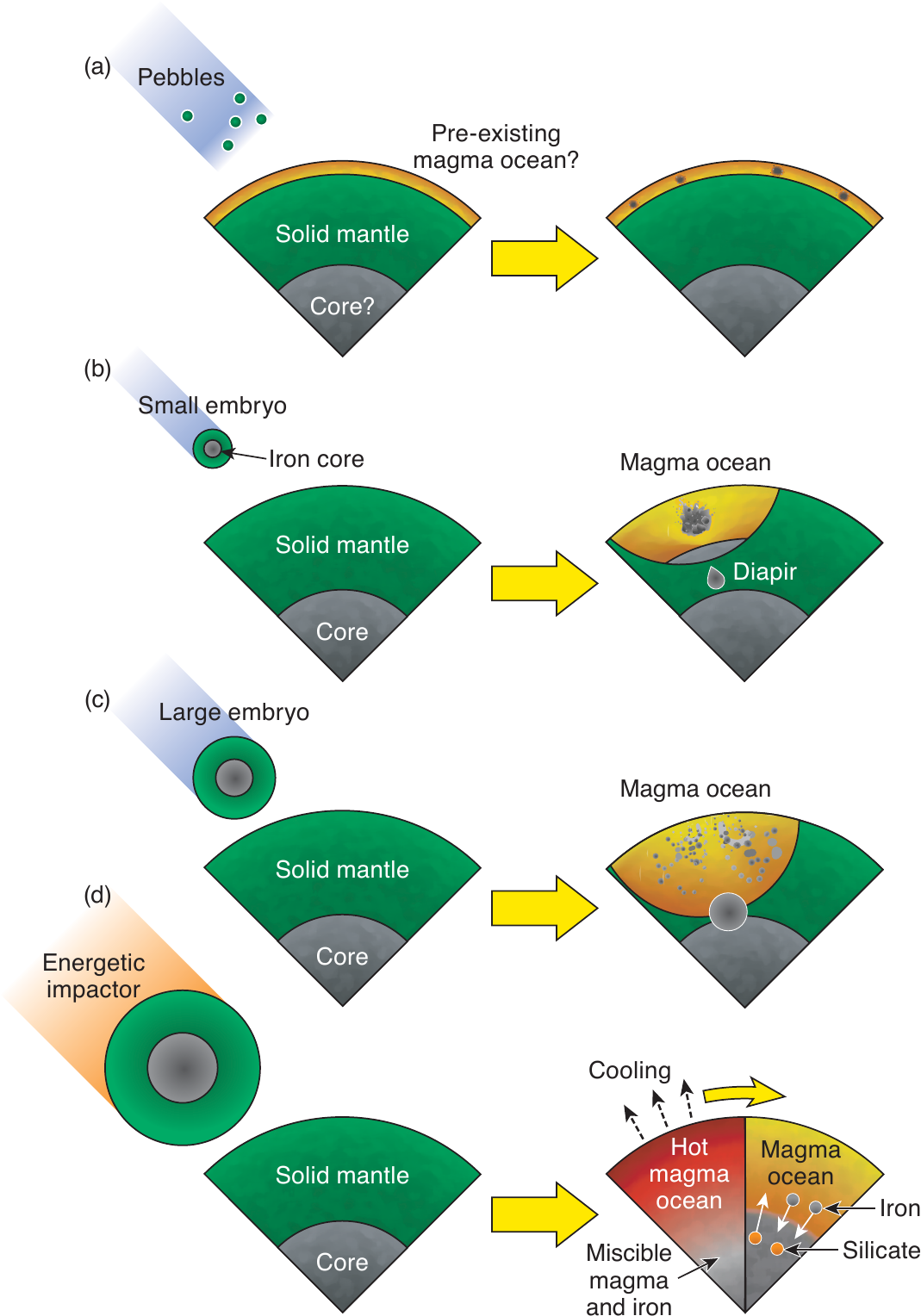}
 \caption{\small Physical mechanisms of metal–silicate differentiation. \textbf{(a)} Accretion of coagulated, $\sim$mm-sized dust aggregates (pebbles) may not generate a magma ocean on $\sim$Mars-sized protoplanets, but they would emulsify if melt is already present. It is unclear if metal cores would form for small planets in the absence of substantial silicate melt. \textbf{(b)} A small embryo can be energetic enough to generate a localized magma pond, where the iron core of the embryo can emulsify. The iron will eventually settle at the bottom of the magma ocean and the metal may sink through the solid lower mantle via diapirism, fractures, and/or percolation. \textbf{(c)} A large impact generates large-scale melting, and exhibits little emulsification or mixing of the impactor core. \textbf{(d)} An extremely energetic impact can raise the core-mantle temperature high enough, so that silicate and iron become miscible. Once the hot magma ocean cools, iron and silicate would separate. \href{https://osf.io/vjrsq/}{\includegraphics[scale=0.35]{icon_download.pdf}}}
 \label{fig2}
\end{figure}
A variety of energy sources are available to cause large-scale melting of rocky planets during their formation (\S\ref{sec:2.1.1}), enabling the gravitational separation of metals from silicates (Fig.~\ref{fig2}). The presence of melts is important to the evolving planet’s composition (\S\ref{sec:2.2.2}), since reaction kinetics are far faster in melts than in solids (\textit{Stevenson} 1981) and liquid-gas interactions thus facilitate rapid chemical exchange between planetary sub-reservoirs. The extent of mantle melting likely varied both spatially and temporally (\textit{Elkins-Tanton} 2008; \textit{Tonks \& Melosh} 1993), with evidence from fractionation of Xe, Ru, and other isotopes and the abundance of highly-siderophile elements in the Earth’s crust, as well as numerical modeling of giant impacts, suggesting that Earth’s mantle was largely but not fully molten during its formation (\textit{Fischer et al.} 2017; \textit{Li \& Agee} 1996; \textit{Mukhopadhyay} 2012; \textit{Murthy} 1991; \textit{Nakajima \& Stevenson} 2015; \textit{Nakajima et al.} 2021; \textit{Rubie et al.} 2011, 2015; \textit{Solomatov} 2015; \textit{Williams et al.} 2021), forming a magma ocean or pond on the surface (Fig.~\ref{fig2}). The physics of core formation in a magma ocean depends on the viscosity of the molten silicate (\textit{Karki \& Stixrude} 2010; \textit{Liebske et al.} 2005), which bears considerable uncertainties at very high pressures.

Impactor cores may break up into smaller droplets as they sink (Fig.~\ref{fig2}b,c), depending on the stable droplet size ($d$) and settling velocity ($v_{s}$). The stable droplet size can be calculated based on the dimensionless Weber number ($We$), 
\begin{linenomath}\begin{equation}
W e=\left(\rho_{m e t}-\rho_{s i l}\right) d \, v_{s}^{2} / \sigma_{s},
\end{equation}\end{linenomath}
which indicates a balance between coalescence (lower $We$) and disaggregation (higher $We$) for values of $\sim$10, where $\rho_{met}$ and $\rho_{sil}$ are the density of the metallic liquid and silicate liquid, respectively, and $\sigma_s$ is the surface energy of the metal–silicate interface (\textit{Rubie et al.} 2015a). In a turbulent flow, the settling velocity is
\begin{linenomath}\begin{equation}
v_{s}=\sqrt{\frac{4}{3 \, C_{D}}\left(\frac{\rho_{m e t}-\rho_{s i l}}{\rho_{s i l}}\right) g \, d},
\end{equation}\end{linenomath}
where $C_{D}$ is the drag coefficient, a function of the friction coefficient (\textit{Rubie et al.} 2003). In a laminar flow, the settling velocity can be calculated from Stokes’ Law,
\begin{linenomath}\begin{equation}
v_{s}=\frac{\left(\rho_{met}-\rho_{s i l}\right) g \, d^{2}}{18 \eta}.
\end{equation}\end{linenomath}
At a value of $\sigma_s$ = 1 N/m and turbulent flow velocities estimated from scaling theory (\textit{Solomatov} 2015),
\begin{linenomath}\begin{equation}
v_{\rm MO}=0.6\left(\alpha \, g \, z \, F_{\rm MO} / \rho_{s i l} \, c_{p}\right)^{1 / 3},
\end{equation}\end{linenomath}
with heat capacity $\alpha$, and soft turbulence heat flux
\begin{linenomath}\begin{equation}
F_{\rm MO}=0.089 \, k \, \Delta T \, R \, a^{1 / 3} / D, \label{eq:MO_heat_flux}
\end{equation}\end{linenomath}
typical estimates are $d \approx$ 1 cm and $v_{s} \approx$ 0.5 m/s, much slower than convection velocities (\textit{Rubie et al.} 2003, 2015a). See \textit{Solomatov} (2015), \textit{Rubie et al.} (2003), and \textit{Deguen et al.} (2014) for typical parameters. The value of $\sigma_s$ may be different, depending on the light element (e.g., Si, O, S) content of the metal (\textit{Rubie \& Jacobson} 2016; \textit{Terasaki et al.} 2012). In giant impacts, the target acts as a fluid regardless of material strength, resulting in very high turbulence (\textit{Nimmo \& Kleine} 2015) and the impactor core merging with the target core within hours (\textit{Canup} 2004; and Fig.~\ref{fig2}c). 

On this type of theoretical basis, emulsification has been argued to occur after an impactor’s core falls a distance equal to a few times its original diameter, so all accreting material except the largest impactor cores likely emulsified significantly (\textit{Rubie et al.} 2003, 2015a; \textit{Samuel} 2012; c.f. \textit{Dahl \& Stevenson} 2010). Laboratory fluid dynamics experiments on two analog fluids representing metal and silicate have been used to investigate this question experimentally (\textit{Deguen et al.} 2011, 2014; \textit{Landeau et al.} 2021), finding qualitatively similar results. These suggest high degrees of metal-silicate equilibration when the impactor’s core penetrates deep into the magma ocean (\textit{Deguen et al.} 2014; \textit{Zube et al.} 2019). \textit{Landeau et al.} (2021) performed experiments with an initial velocity for the metal analog fluid to simulate an impact, finding that impactors with diameters of $<$100 km will fully equilibrate, while larger impactors will only partially equilibrate. The effects of impact angle remain poorly constrained experimentally, but are likely to be important. Moreover, an energetic impact can raise the temperature high enough for metal and silicate to be miscible (e.g., $>$4000 K near the surface and $\gtrsim$7000 K at 130 GPa, \textit{Wahl \& Militzer} 2015). 

These data present a basic paradox for the composition and make-up of the Earth: while the close similarity in isotope fractionation of Earth and the Moon seemingly required a hot and miscible state being present during accretion or Moon-formation, geochemical evidence for mantle heterogeneities suggest the opposite (\textit{Canup et al.} 2021). For the Earth, a very energetic impact has thus been suggested to resolve this conundrum through compositional stratification of a vaporized disk that formed the Moon (\textit{Lock et al.} 2018, 2020). For super-Earths, gravitational potential energy alone makes highly energetic accretionary states likely. As a result, rocky planets can have a homogeneous structure (no clear core-mantle boundary) at first, then core and mantle separate as the planet cools (Fig.~\ref{fig2}d). As to our knowledge, no previous work has experimentally evaluated the equilibration of pebbles (coagulated dust aggregates) in magma oceans or rocky planetary mantles (Fig.~\ref{fig2}a). Judging from the trend of increasing degree of equilibration with decreasing impactor sizes, however, pebbles can be expected to equilibrate fully in largely molten layers. An outstanding test for the degree of pebble-dominated growth in the inner Solar System is whether small, $\sim$Mars-sized planets that would form mainly by pebble accretion would melt, and thus be able to form a metal core (\textit{Melosh} 1990). 

In the solid mantle beneath a magma ocean or in partially molten planetesimals, metal–silicate segregation would have proceeded via different mechanisms (Fig.~\ref{fig2}; \textit{Nimmo \& Kleine} 2015; \textit{Rubie \& Jacobson} 2016; \textit{Rubie et al.} 2015a). Fe or Fe alloy liquid can percolate downward along the grain boundaries of solid silicate minerals, at least in the presence of an interconnected melt network (\textit{Yoshino et al.} 2003; \textit{Ghanbarzadeh et al.} 2017), though the efficiency of this mechanism depends on the poorly-constrained dihedral angles of mantle minerals (\textit{Shi et al.} 2013; \textit{Takafuji et al.} 2004; \textit{Terasaki et al.} 2007; \textit{Cerantola et al.} 2015). Alternatively, if the solid silicate is hot enough to deform, large amounts of metal may undergo gravitational instabilities and sink rapidly through the lower mantle in the form of diapirs (Fig.~\ref{fig2}b; \textit{Karato \& Murthy} 1997; \textit{Ricard et al.} 2009; \textit{Samuel et al.} 2010). Liquid Fe may also descend through a solid lower mantle along large fractures or dikes (\textit{Stevenson} 1981).

\subsubsection{\textbf{Chemistry and redox of core formation}} \label{sec:chemistry_and_redox_of_core_formation}  \label{sec:2.2.2}

To first order, reactions between metallic and silicate liquids at high $P$ and $T$ during core formation set the initial compositions of a planet’s core and mantle, which ultimately control the speciation and pressure of the planetary atmosphere. The extent of these reactions are directly affected by the conditions of core formation.

The metal–silicate partitioning behavior of an element $M$ with valence $n$, which defines its combining capacity with other atoms to form chemical compounds, is described by a partition coefficient, 
\begin{linenomath}\begin{equation}
D_{M}=X_{M}^{m e t} / X_{M O_{n / 2}}^{sil}, \label{eq:partition_coefficient}
\end{equation}\end{linenomath}
where $X_{M}^{met}$ and $X_{MO_{n/2}}^{sil}$ are the mole fractions of element $M$ in the metal and $M$ oxide in the silicate, respectively. $D_{M}$ depends on the $P$ and $T$ of the metal–silicate partitioning reaction, which in turn depend on the geophysical setting, such as the extent of melting (e.g., accretion history, presence of an atmosphere), melt geometry (magma “pond” versus a global magma ocean; Fig.~\ref{fig2}), kinetics of metal–silicate equilibration as compared to sinking velocities, or possible super-liquidus heating. For some elements, $D_{M}$ may also depend on the compositions of the equilibrating metallic and silicate liquids, especially the abundances of light elements in the metal such as C, H, N, S, O, or Si. $D_{M}$ is also dependent on the redox state of the reaction environment. 

The redox state of planetary mantles is a measure of the global and local availability of valence electrons, which govern the type of chemical compounds that are present. In general, oxidation (increase in oxidation state) is the loss of electrons from an atom, ion, or certain atoms in a molecule. By contrast, reduction describes a gain of electrons and decrease in oxidation state. On the global scale of rocky planets, the main driver of the evolution of redox potential is gravity, which segregates the most redox active and cosmochemically abundant elements into the core (Fe) and to space (H) (\textit{Wordsworth et al.} 2018). In terrestrial rock compositions, the availability of oxygen is hence the dominant driver of the redox state. Oxygen fugacity ($f\mathrm{O}_{2}$), which describes the chemical potential or availability of oxygen in the system, is thus used as a convenient scale to  evaluate the redox state of rocky planetary materials, the degree to which they are oxidized or reduced, or their potential to occur with a higher or lower charge. If oxygen were an ideal gas, its chemical potential ($\mu_{\rm O_2}$) would be related to its partial pressure ($P_{P}$):
\begin{linenomath}\begin{equation}
\mu_{\mathrm{O}_2}=\mu_{\mathrm{O}_2}^{\circ}+R T \ln \left(P_{P} / P_{0}\right),
\end{equation}\end{linenomath}
where $\mu^{\circ}_{\rm O_2}$ is the standard state chemical potential, $R$ is the ideal gas constant, and $P_{0}$ is the standard state pressure. When dealing with real gasses, rocks, and other substances, an ideal gas is not always a good approximation, so partial pressure is replaced with fugacity ($f$) to correct for non-ideality:
\begin{linenomath}\begin{align}
\mu_{\mathrm{O}_2}=\mu_{\mathrm{O}_2}^{\circ} & + R T \ln \left(f \mathrm{O}_{2} / f^{\circ} \mathrm{O}_{2}\right) \mu_{\mathrm{O}_2}^{\circ} \nonumber \\
& + R T \ln \left(f \mathrm{O}_{2}\right),
\end{align}\end{linenomath}
where the fugacity of pure oxygen at 1 bar is $f^{\circ}$O$_{2}$ = 1 (\textit{Cottrell et al.} 2022). Oxygen fugacity depends on variables including $P$, $T$, and composition and is often defined relative to a buffering reaction in which a metal–oxide pair (or other assemblage of two or more minerals) coexist stably; for example, oxygen fugacity during core formation is often defined in log units relative to the iron–wüstite (Fe–FeO, or IW) buffer as
\begin{linenomath}\begin{align}
\log _{10}\left(f\mathrm{O}_{2}\right)=\Delta \mathrm{IW} & = 2 \log _{10}\left(a_{F e O}^{s i l}/a_{F e}^{m e t}\right) \nonumber \\
& \approx 2 \log _{10}\left(x_{FeO}^{s i l}/x_{F e}^{met}\right),
\end{align}\end{linenomath}
where $a_{FeO}^{sil}$ and $a_{Fe}^{met}$ are the activities of FeO in the silicate and Fe in the metal, respectively. More positive (negative) values relative to IW indicate more oxidized (reduced) conditions. Most common rock-forming elements, aside from Fe, only have one oxidation state. Therefore, the relative abundances of Fe valence states (Fe$^0$, Fe$^{2+}$, or Fe$^{3+}$) are diagnostic of the overall oxidation state of the system. The change in Gibbs free energy of the IW buffer reaction ($\Delta G^{\circ}_{reaction}$) can be expressed as:
\begin{linenomath}\begin{align}
\Delta G_{reaction}^{\circ} & = -R T \ln \left(K_{e q}\right) \nonumber \\
& =-R T \ln \left(\frac{a_{F e O}^{sil_{2}}}{f\mathrm{O}_{2} \cdot a_{F e}^{m e t_{2}}}\right),
\end{align}\end{linenomath}
where $K_{eq}$ is the equilibrium constant (\textit{Cottrell et al.} 2022).

The oxygen fugacity during and after core formation depends in part on the initial oxidation state of accreted material, which is often thought to be a reflection of provenance, with more oxidized material originating farther from the Sun and more reduced material originating closer to the Sun (\textit{Ciesla \& Cuzzi} 2006; \textit{Gradie \& Tedesco} 1982). The provenance of accreted material may evolve with time as a planet grows, depending on the dynamical conditions in the disk; but may also be strongly influenced by migration of the water snow line with time, the orbits of giant planets, and stochastic effects (\textit{Brasser et al.} 2018; \textit{Fischer et al.} 2018; \textit{O’Brien et al.} 2006; \textit{Raymond et al.} 2007). As a planet grows, its oxidation state relative to IW will also evolve as a consequence of water delivery and core formation, usually increasing with time because higher $P$–$T$ tends to promote the reaction 2Fe + SiO$_{2}$ $\longrightarrow$ Si + 2FeO (\textit{Javoy} 1995; \textit{Ringwood} 1959; \textit{Rubie et al.} 2011). A planet’s oxidation state can continue to evolve after the main phase of core formation via other processes, such as iron disproportionation (\textit{Frost et al.} 2004), subduction (\textit{Cottrell et al.} 2022), or hydrogen escape (\textit{Catling et al.} 2001).

The dependence of metal–silicate partitioning on $P$, $T$, $f\mathrm{O}_{2}$, and composition has been experimentally determined for a variety of elements using the piston-cylinder apparatus, multi-anvil apparatus, or laser-heated diamond anvil cell to recreate the high $P$–$T$ of core formation. Some of the best-studied elements include Si, O, Ni, Co, V, and Cr (e.g., \textit{Bouhifd \& Jephcoat} 2011; \textit{Chabot et al.} 2005; \textit{Cottrell et al.} 2009; \textit{Fischer et al.} 2015; \textit{Geßmann \& Rubie} 1998; \textit{Kegler et al.} 2008; \textit{Li \& Agee} 1996; \textit{Ricolleau et al.} 2011; \textit{Righter et al.} 1997; \textit{Siebert et al.} 2012, 2013; \textit{Tsuno et al.} 2013; \textit{Wade \& Wood} 2005), but many other elements have also been investigated (e.g., \emph{Badro et al.} 2016; \emph{Blanchard et al.} 2017; \emph{Chidester et al.} 2017; \emph{Corgne et al.} 2008; \emph{Hillgren et al.} 1996; \emph{Mahan et al.} 2018a, 2018b; \emph{Mann et al.} 2009; \emph{Righter et al.} 2016; \emph{Siebert et al.} 2011, 2018; \emph{Wade et al.} 2012). These studies allow for determination of the chemical behaviors of elements under extreme conditions, though they are subject to uncertainties and often extrapolations in $P$, $T$, and/or $f\mathrm{O}_{2}$. In general, most elements become more lithophile or less siderophile at higher $f\mathrm{O}_{2}$, and many (but not all) lithophile elements become less lithophile and many (but not all) siderophile elements become less siderophile at more extreme $P$–$T$.

In addition to the dependence of partitioning on $P$, $T$, and $f\mathrm{O}_{2}$, the evolving compositions of the silicate and metallic liquids will also depend on the extent to which the silicate and metal can equilibrate, i.e., the fraction of metal that can react with the ambient silicates. The extent of metal equilibration (often denoted $k$ or $k_{\rm core}$) would depend on factors such as the accretion geometry, metal droplet size, sinking velocity, and degree of entrainment (\emph{Dahl \& Stevenson} 2010; \emph{Deguen et al.} 2011, 2014), with, for example, larger, differentiated impactors likely exhibiting lower $k_{\rm core}$ (Fig.~\ref{fig2}b; \S\ref{sec:2.2.1}). The extent of silicate equilibration depends on factors such as the extent of mantle melting and degree of mixing (\emph{Deguen et al.} 2011; \emph{Rubie et al.} 2015a). While these parameters likely vary for varying accretion events, depending on, e.g., impactor size, velocity, angle, and timing (\emph{Landeau et al.} 2021), they are often treated as constants throughout the core formation process in numerical models (\emph{Badro et al.} 2015; \emph{Fischer et al.} 2017; \emph{Rubie et al.} 2011; \emph{Rudge et al.} 2010). Strong tradeoffs are seen in the compositional effects of the extent of metal and silicate equilibration (\emph{Fischer et al.} 2017). The Hf–W isotopic chronometer (\S\ref{sec:3.2}) indicates an average $k_{\rm core} = 0.4$ for the Earth in the case of whole-mantle equilibration (\emph{Fischer \& Nimmo} 2018; \emph{Nimmo et al.} 2010; \emph{Rudge et al.} 2010), with lower degrees of silicate equilibration or faster accretion in different dynamical regimes requiring higher values of $k_{\rm core}$ (\emph{Fischer \& Nimmo} 2018; \emph{Zube et al.} 2019), and a value of close to unity is required by the mantle trace element composition of Mars (\emph{Brennan et al.} 2020). The degree of silicate equilibration is less well-constrained, with only the lowest values being ruled out for both Earth and Mars (\emph{Morishima et al.} 2013; \emph{Fischer et al.} 2017; \emph{Brennan et al.} 2020). The degree of equilibration (silicate + metal) depends on the amount of metal emulsification and mixing with the silicate, with \textit{Deguen et al.} (2014) reporting a diluted partition coefficient $\delta_M = 1 + D_M/\Delta$ (Eq. \ref{eq:partition_coefficient}) and total equilibration factor $k_{\rm core} = k_{\rm core}/\delta_M$.

Numerical models of core formation range in complexity from single-stage, in which the core and mantle are equilibrated at one $P$–$T$–$f\mathrm{O}_{2}$ (\emph{Chabot et al.} 2005; \emph{Li \& Agee} 1996; \emph{Righter} 2011); to multi-stage models based on a prescribed mass evolution (\emph{Badro et al.} 2015; \emph{Rubie et al.} 2011; \emph{Wade \& Wood} 2005); to multi-stage models based on $N$-body simulations of planetary accretion (\emph{Morishima et al.} 2013; \emph{Rubie et al.} 2015b; \emph{Fischer et al.} 2017). In these models, $f\mathrm{O}_{2}$ may be fixed at a constant value (\emph{Chabot et al.} 2005; \emph{Wade \& Wood} 2005), or evolved along a prescribed path (\emph{Badro et al.} 2015; \emph{Wade \& Wood} 2005), or evolved self-consistently based on metal–silicate partitioning (\emph{Fischer et al.} 2017; \emph{Rubie et al.} 2011, 2015b). In general, these calculations can be used to forward-model the composition of a planet’s core based on a requirement to reproduce its mantle composition, or can be used to back out information about the conditions and mechanisms of core formation, or both. For example, the trace element composition of Earth’s mantle implies metal–silicate equilibration at intermediate mantle depths (\emph{Fischer et al.} 2017; \emph{Li \& Agee} 1996; \emph{Rubie et al.} 2011, 2015; \emph{Wade \& Wood} 2005), and metal–silicate equilibration at such depths would result in an O- and Si-rich core (\emph{Badro et al.} 2015; \emph{Fischer et al.} 2017; \emph{Rubie et al.} 2011). These models may also be coupled with models of isotopic evolution, for example in the Hf–W system (\emph{Fischer \& Nimmo} 2018; \emph{Rubie et al.} 2015b). Studies with self-consistent $f\mathrm{O}_{2}$ evolution during core formation have concluded that the Earth self-oxidized by $\sim$1.5 log units (\emph{Fischer et al.} 2017) or more (\emph{Rubie et al.} 2011), with less oxidation expected to occur in smaller bodies.

%%%%%%%%%%%%%%%%%%%%%%%%%%%%%%%%%%%%%%%%%%%%%%%%%%%%%%%%%%%%%%%%%%%%%%%%%%%%%%%% 
\subsubsection{\textbf{Volatile delivery, partitioning, and outgassing}} \label{sec:volatile_delivery_partitioning_and_outgassing}  \label{sec:2.2.3}

\begin{figure*}[h!]
 \centering
 \includegraphics[width=0.99\textwidth]{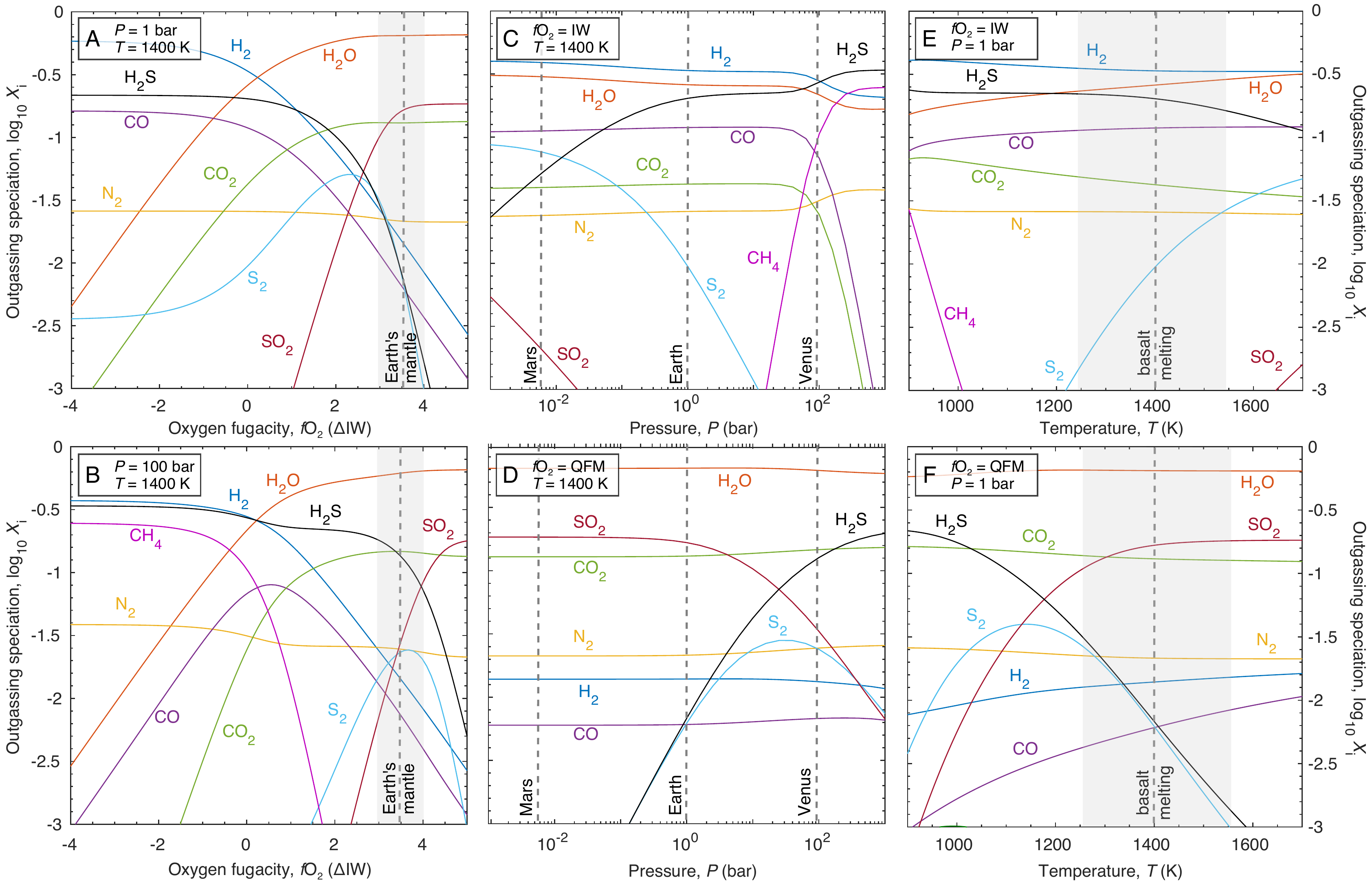}
 \caption{\small Effects of varying redox state \textbf{(A/B)}, pressure \textbf{(C/D)}, and temperature \textbf{(E/F)} on outgassing speciation. Compositions of outgassed atmospheres for fixed elemental abundances in the gas phase (H$_2$O = 1000 ppm, CO$_2$ = 500 ppm, S = 500 ppm, N$_2$ = 50 ppm). Solubility effects are not included. Vertical dashed lines highlight Earth’s upper mantle \textit{f}O$_2$ ($\sim$QFM, \textbf{A/B}), surface atmospheric pressures \textbf{(C/D)}, and basalt melting temperatures \textbf{(E/F)}. \href{https://osf.io/5a84q/}{\includegraphics[scale=0.35]{icon_download.pdf}}}
 \label{fig3}
\end{figure*}
Volatile elements and compounds relevant for the composition of planetary atmospheres (atmophiles) and necessary for life as we know it, such as C, N, and H, were present in solid ice form in the cooler outer regions of the disk (beyond their respective ice lines) and in refractory forms throughout the inner and outer Solar System (\emph{Marty} 2020; \emph{Öberg \& Bergin} 2021). The timing of their delivery to a terrestrial planet in the inner disk may depend on factors including the planet’s semimajor axis, growth mode, evolution of its precursor planetesimals, or the orbits and migration of giant planets, and was likely stochastic (e.g., \emph{Bergin et al.} 2015; \emph{Bond et al.} 2010; \emph{Fischer et al.} 2018; \emph{Morbidelli et al.} 2000; \emph{O’Brien et al.} 2014, 2018; \emph{Raymond \& Izidoro} 2017; \emph{Lichtenberg et al.} 2019a; \emph{Sánchez et al.} 2018; \emph{Krijt et al.}, this volume). There is some controversy about whether most or all of Earth’s volatiles were delivered preferentially later in its growth history, or whether they were delivered throughout Earth’s accretion, based on the apparent discrepancy between isotopic disparity and chemical affinity of the silicate Earth with carbonaceous chondrites (\emph{Braukmüller et al.} 2019). For example, He and Ne seem to imply that at least some early-accreted volatiles were retained in the deep mantle (\emph{Broadley et al.} 2020b; \emph{Tucker \& Mukhopadhyay} 2014). On the basis of C/H, C/N, and C/S ratios, \textit{Hirschmann} (2016) argued that a significant amount of Earth’s volatiles pre-date late accretion, but that they were replenished late. C-S (\emph{Hirschmann et al.} 2021) and Cr (\emph{Bonnand \& Halliday} 2018) systematics of magmatic iron meteorites indicate that the precursor bodies of present-day inner Solar System materials formed volatile-rich and were subsequently depleted in volatiles. \textit{Lichtenberg et al.} (2021a) explain this trend by two waves of volatile-rich  planetesimal formation, first in the inner, then in the outer Solar System. In this scenario of early Solar System accretion, the first planetesimal formation burst was strongly heated by $^{26}$Al decay, devolatilized, and nucleated the growth of the terrestrial planets.

In contrast, \textit{Mahan et al.} (2018a, 2018b) suggested that more volatiles were delivered late based on the metal–silicate partitioning of S, Cu, and Zn, and \textit{Kubik et al.} (2021) reached a similar conclusion based on metal–silicate partitioning of Cd, Bi, Sb, and Ti. Studies of Solar System formation dynamics have suggest that the Earth likely accreted volatile-rich material throughout its growth history (\emph{Izidoro et al.} 2013; \emph{Morbidelli et al.} 2000), though with more material from the outer Solar System (presumably more volatile-rich) accreting later in its history (\emph{O’Brien et al.} 2006, 2014). It has been argued on the basis of isotopic ratios and metal–silicate partitioning that most of the Earth’s budget of atmophile elements came from the Moon-forming impact (\emph{Schönbächler et al.} 2010; \emph{Grewal et al.} 2019; \emph{Budde et al.} 2019). Using isotopes of volatile elements, \textit{Albarède} (2009) suggested that the Earth accreted from very volatile-depleted material, and its volatiles were subsequently added later. \textit{Wang \& Becker} (2013) used the Earth’s S/Se and Se/Te ratios to support a volatile-rich late veneer, delivering 20–100\% of the BSE’s H and C budgets, while, based on updated Se data, \textit{Varas-Reus et al.} (2019) concluded on a lower-mass, concentrated late addition of volatiles. The behavior of volatiles during core formation has been studied experimentally, but often to a lesser degree than other elements due to the unique experimental and analytical challenges they present. The best-studied volatiles include S (\emph{Boujibar et al.} 2014; \emph{Suer et al.} 2017) and C (\emph{Dasgupta et al.} 2013; \emph{Fischer et al.} 2020; \emph{Malavergne et al.} 2019), with N (\emph{Dalou et al.} 2017; \emph{Roskosz et al.} 2013) and H (\emph{Clesi et al.} 2018; \emph{Okuchi} 1997) data being more controversial and limited to lower $P$–$T$. Both C and S become significantly less siderophile (“iron-loving”) at the higher $P$–$T$ of Earth’s core formation (\emph{Fischer et al.} 2020; \emph{Suer et al.} 2017).

Terrestrial planets may inherit parts of their earliest atmospheres and mantle volatiles from nebular ingassing during planet formation (\emph{Ikoma \& Genda} 2006), provided the planet accreted enough mass to trap an atmosphere before the nebula dissipated (\emph{Lammer et al.} 2020a). This argument is  based on He and Ne isotope fractionation patterns in the Earth (\emph{Harper \& Jacobsen} 1996; \emph{Mizuno et al.} 1980; \emph{Yokochi \& Marty} 2004). In addition to light noble gasses, it has been argued that nebular ingassing was also an important source of water and hydrogen (\emph{Olson \& Sharp} 2018, 2019; \emph{Saito \& Kuramoto} 2020; \emph{Sharp} 2017), though \textit{Wu et al.} (2018) found that this process contributed only a small fraction of Earth’s total water budget. \textit{Péron et al.} (2017) argue for solar wind implantation as a potential alternative explanation for the origins of Earth’s He, Ne, and Ar. Nebular ingassing, however, is likely to be important on rocky exoplanets (\emph{Kimura \& Ikoma} 2020), specifically for planets larger than Earth that grow substantially during the lifetime of the protoplanetary disk.

In addition to direct accretion of atmophile elements and compounds, the planetary mantle has an immediate effect on the composition and speciation of subsequently outgassed atmospheres (\emph{Schaefer \& Fegley} 2017; \emph{Gaillard et al.} 2021, 2022), depending on its redox state, relative abundances of volatiles, temperature, and pressure (Fig.~\ref{fig3}). Fig.~\ref{fig3}A,B and Fig.~\ref{fig3}E,F show that at low $f\mathrm{O}_{2}$, an outgassed atmosphere will be dominated by H$_{2}$, with some CO and H$_{2}$O, and very little CO$_{2}$ (plus some CH$_{4}$ at lower temperatures); at higher $f\mathrm{O}_{2}$ (above IW+2), the atmosphere would be comprised mainly of H$_{2}$O, with some CO$_{2}$, and very little H$_{2}$, CH$_{4}$, or CO. \textit{Grewal et al.} (2020) found that very reducing conditions (below IW–3) results in most C and N remaining within the  mantle in accessory phases, with a thin NH$_{3}$- and CH$_{4}$-dominated atmosphere, while moderately reducing conditions (between IW–3 and IW–1.5) produce an atmosphere of NH$_{3}$, H$_{2}$O, and CO, and more oxidized conditions (above IW–1.5) result in outgassing of N$_{2}$, CO$_{2}$, and H$_{2}$O, plus lower amounts of NH$_{3}$ and HCN. \textit{Sossi et al.} (2020) explored the dependence on temperature, finding, for example, less CO$_{2}$ and more H$_{2}$ at higher temperatures depending on volatile solubility in the melt. Fig.~\ref{fig3}C,D show the effect of outgassing pressure on volatile speciation for fixed gas phase abundances, where we do not include gas solubilities in the melt. \textit{Gaillard \& Scaillet} (2014) suggest that pressure from an existing atmosphere (or ocean) on the outgassing vent can limit outgassing of C and H at low pressures due to the effects of melt solubilities (\emph{Holloway et al.} 1992; \emph{Moore et al.} 1998; \emph{Iacono-Marziano et al.} 2012); they find low pressure environments dominated by sulfur-bearing gasses with increasing amounts of H and then C outgassed at progressively higher pressures. Fig.~\ref{fig3}E,F illustrate the effect of temperature, low $T$ favors outgassing of CH$_{4}$ at IW and H$_{2}$S and S$_{2}$ at QFM, while at high $T$ outgassing of S$_{2}$ at IW and SO$_{2}$ at QFM is enhanced.

%%%%%%%%%%%%%%%%%%%%%%%%%%%%%%%%%%%%%%%%%%%%%%%%%%%%%%%%%%%%%%%%%%%%%%%%%%%%%%%%%%%%%%%%%%%%%%%%%%%%%%%%%%%%%%%%%%%%%%%%%%%%%%%%%%%%
%%%%%%%%%%%%%%%%%%%%%%%%%%%%%%%%%%%%%%%%%%%%%%%%%%%%%%%%%%%%%%%%%%%%%%%%%%%%%%%%%%%%%%%%%%%%%%%%%%%%%%%%%%%%%%%%%%%%%%%%%%%%%%%%%%%%
%%%%%%%%%%%%%%%%%%%%%%%%%%%%%%%%%%%%%%%%%%%%%%%%%%%%%%%%%%%%%%%%%%%%%%%%%%%%%%%%%%%%%%%%%%%%%%%%%%%%%%%%%%%%%%%%%%%%%%%%%%%%%%%%%%%%
\section{\textbf{FORMATION TIMELINE}} \label{sec:formation_timeline} \label{sec:3}
\subsection{\textbf{Disk stage}} \label{sec:disk_stage} \label{sec:3.1}

Owing to significant advances in geochemical analyses of extraterrestrial materials and observations of extrasolar planetary systems in recent years, there has been a shift in our understanding of the major accretion phase of rocky planets. The oldest meteorites from the Solar System (Schersté\emph{n et al.} 2006; \emph{Kruijer et al.} 2014), order-of-magnitude dust depletion on $\sim$Myr timescales in protoplanetary disks (\emph{Andrews} 2020), and the prevalence of super-Earths among exoplanets (\emph{Jontof-Hutter} 2019) indicate that the growth of rocky planets in inner planetary systems initiates within a few hundred thousand years after planetary system birth. Total mass addition and individual growth paths during the disk lifetime, however, may vary substantially from system to system and planet to planet, with implications for the diversity of composition, physical structure, and long-term climate of rocky planets within and outside the Solar System.

\subsubsection{\textbf{Evidence}} \label{sec:evidence} \label{sec:3.1.1}

\begin{figure}[htb!]
 \plotone{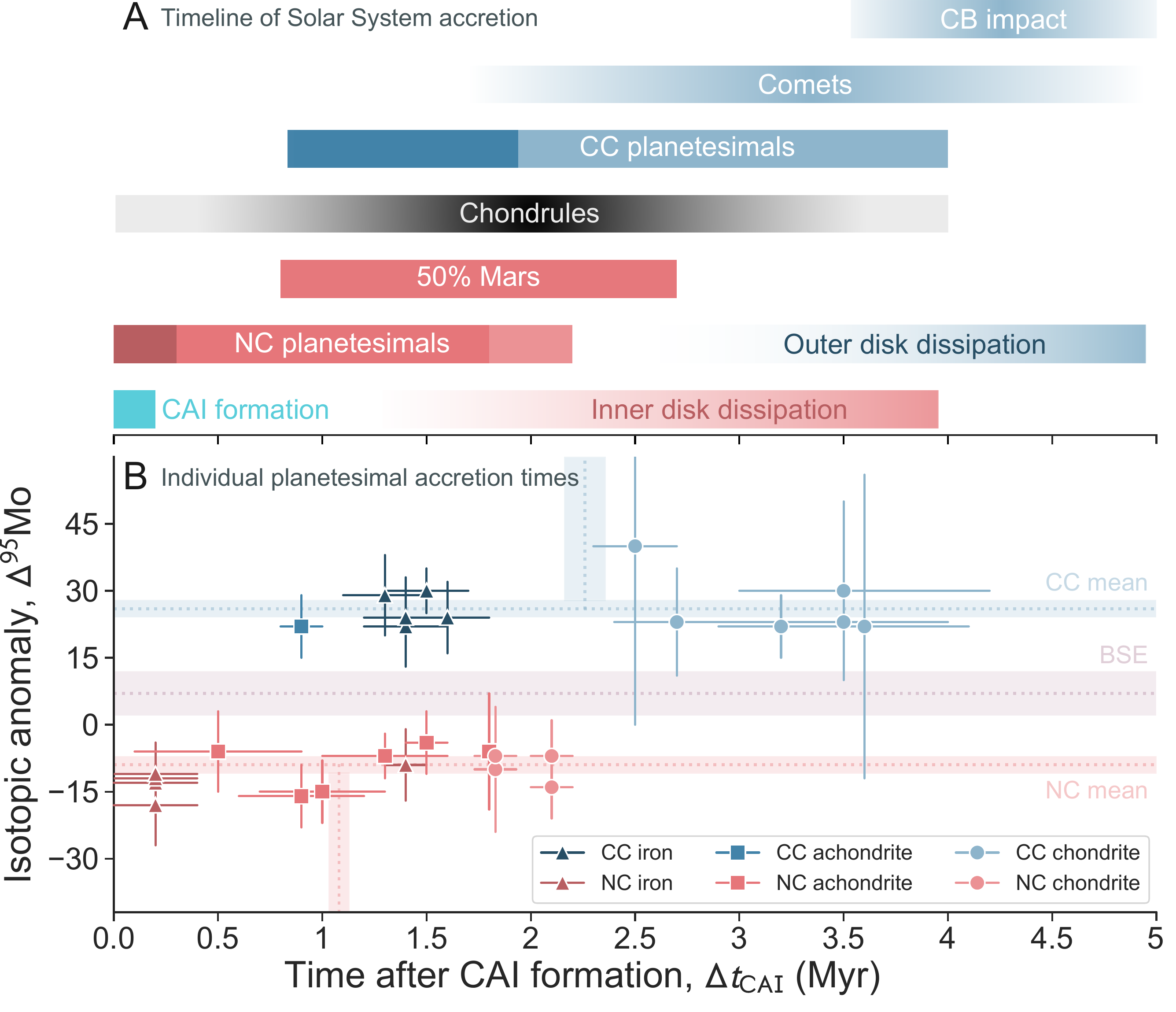}
 \caption{\small Timeline of Solar System formation as indicated by dated extraterrestrial materials. Inner and outer Solar System display an offset of $\sim$1 Myr in mean planetesimal accretion and disk dissipation timescale. \textbf{(A)} Approximate accretion timelines of planetesimals in the inner (NC, red) and outer (CC, blue) Solar System, Calcium-Aluminum-rich inclusions (CAIs, turquoise), chondrules in both NC and CC chondrites combined (gray), comets, chondrule ages from the breakup of the CB parent body that generated extraordinarily metal-rich chondrules, and approximate disk dissipation timescales from paleomagnetism. \textbf{(B)} Modelled planetesimal accretion times and $\Delta^{95}$Mo isotope signature for CC (blue) and NC (red) iron meteorites (triangles), achondrites (squares), and chondrites (circles). Dashed lines indicate the mean isotope signatures and ages of CC (blue), NC (red) and Bulk Silicate Earth (BSE, purple). See text for a discussion. \textsc{References}: CAIs \& chondrules: \textit{Connelly et al.} (2012), \textit{Villeneuve et al.} (2009); disk dissipation: \textit{Weiss et al.} (2021); Mars accretion: \textit{Dauphas \& Pourmand} (2011); comets: \textit{Matzel et al.} (2010); CB impact: \textit{Krot et al.} (2005); $\Delta^{95}$Mo signatures: \textit{Budde et al.} (2019); individual planetesimal accretion times: \textit{Hunt et al.} (2018), \textit{Desch et al.} (2018), \textit{Kleine et al.} (2020), \textit{Kruijer et al.} (2014), \textit{Golabek et al.} (2014), \textit{Neumann et al.} (2018), \textit{Hunt et al.} (2017), \textit{Sugiura \& Fujiya} (2014), \textit{Blackburn et al.} (2017), \textit{Bryson \& Brennecka} (2021), \textit{Doyle et al.} (2015), \textit{Ma et al.} (2021). \href{https://osf.io/4yg63/}{\includegraphics[scale=0.35]{icon_download.pdf}}}
 \label{fig4}
\end{figure}
\textit{Solar System.}  Our understanding of the chronology of terrestrial planet accretion in the inner Solar System (Fig.~\ref{fig4}) is largely based on studies of extraterrestrial matter: samples from other planetary bodies include meteorites that fall to Earth from wider heliocentric orbits or that are collected in-situ on asteroids or comets by spacecraft. Most of the ‘parent bodies’ of meteorites are unknown, and their properties, such as size or formation time, are derived by matching radiometric ages and petrologic properties of materials inside the meteorite specimen with geophysical models of planetesimal evolution (\emph{Gail et al.} 2014). 

From a compositional perspective, meteorites largely classify as differentiated and undifferentiated groups (\emph{Krot et al.} 2014; \emph{Alexander et al.} 2018). Undifferentiated meteorites (chondrites) display approximately solar-like elemental abundances (but importantly differ in atmophile elements like H, C, and N, among others) and are composed of varying mixtures of both high and low temperature mineral phases, including: (a) chondrules, $\lesssim$mm-sized roundish droplets that crystallized from liquid silicates, (b) refractory inclusions including Calcium-Aluminum-rich inclusions (CAIs), the oldest dated solids that originate from within the Solar System, and amoeboid olivine aggregates (AOAs), and (c) matrix, which is fine-grained, volatile-rich and unmelted dust accreted directly from the protoplanetary disk. Differentiated materials originate from melted planetary objects and largely sub-divide into silicate- (achondrite) and metal-dominated (iron) classes. All planetary materials in the Solar System additionally can be classified according to their distinct nucleosynthetic isotope signature (\emph{Dauphas \& Schauble} 2016): the carbonaceous chondrite (CC) and non-carbonaceous (NC) isotope families form two super-groups in stable isotope space (e.g., Ti, Cr, Mo) that each host both differentiated and undifferentiated meteorites and are interpreted as being associated with outer and inner Solar System material, respectively (\emph{Trinquier et al.} 2007; \emph{Warren} 2011). Earth’s isotopic fingerprint is a mix of these two reservoirs (Fig.~\ref{fig4}), but chemically highly depleted in atmophile elements relative to the CC undifferentiated meteorites (\emph{Peslier et al.} 2017; \emph{Zahnle \& Carlson} 2020). No known meteorite class represents the bulk of Earth’s chemical and isotopic composition, and geochemical mixing models suggest that the terrestrial planets accreted significantly from material that is not represented in the present-day meteorite collection (\emph{Mezger et al.} 2020). The strong chemical depletion of the Earth relative to chondritic meteorites in highly volatile (\emph{Bergin et al.} 2015), moderately volatile (\emph{Halliday \& Porcelli} 2001; \emph{Norris \& Wood} 2017; \emph{Sossi et al.} 2019; \emph{Collinet \& Grove} 2020a,b), and refractory elements (\emph{Hin et al.} 2017) provide evidence for open system degassing from planetesimals (\emph{Lichtenberg et al.} 2021a; \emph{Grewal et al.} 2021; \emph{Hirschmann et al.} 2021) and vaporization of the growing protoplanets (\emph{Hin et al.} 2017; \emph{Young et al.} 2019; \emph{Benedikt et al.} 2020) during accretion. Ureilite meteorites have been suggested to come from such a devolatilized (\emph{Sanders et al.} 2017), potentially Mars-sized (\emph{Nabiei et al.} 2018) parent body, which is actively debated (\emph{Zhu et al.} 2020; \emph{Broadley et al.} 2020a; \emph{Collinet \& Grove} 2020c).

The joint analysis of isotopic anomalies, compositional trends, and modeling of accretion times of the parent bodies of meteorites in the last few years have revealed a disparate accretion chronology between the inner and outer Solar System (\emph{Mezger et al.} 2020; \emph{Kleine et al.} 2020; \emph{Bermingham et al.} 2020): the earliest-known planetesimals in the inner Solar System were formed within $\lesssim$0.3 Myr after CAIs, while evidence for the first planetary objects in the outer Solar System appears $\sim$1 Myr later (Fig.~\ref{fig4}). The mean accretion time for the planetesimals with known isotopic signature (\textit{Budde et  al.} 2019) is shifted by $\approx$1.2 Myr in the inner (mean 1.08$\pm$0.05 Myr, interval $\sim$0.0–2.1 Myr) versus the outer (mean 2.27$\pm$0.10 Myr, interval $\sim$1.1–3.6 Myr) Solar System with an overlap in accretion windows of $\sim$1 Myr. Both CC and NC planetesimals show a trend from differentiated (irons and achondrites) to undifferentiated (chondrites) classes, which can be explained in the context of decreasing internal heating by $^{26}$Al ($t_{1/2} \approx 0.72$ Myr), which melted and thermally processed early-formed planetesimals, but left later-formed bodies in increasingly pristine states (\emph{Elkins-Tanton} 2012, 2017; \emph{Monteux et al.} 2018). Internal processing by $^{26}$Al can explain the spectral features of outer main belt asteroids (\emph{Kurokawa et al.} 2021; \textit{Watanabe et al.}, this volume), linking them to the parent bodies of carbonaceous chondrites (\emph{Lichtenberg et al.} 2021a). Uncertainties arise from the unknown bulk composition upon planetesimal formation (meteorites we see today are the thermally and compositionally processed products of accretion and internal evolution), introducing a degeneracy between formation time and abundance of non-Al-hosting materials, such as volatile ices, in the model-derived accretion times. In this context, achondrites are interpreted to represent the (partially) melted mantles, while (magmatic) iron meteorites are thought to represent the metal cores of differentiated planetesimals, and other iron meteorites to result from partial melting and impact events. However, no isotopically fitting iron + silicate/achondrite meteorite from the same parent body has been identified to date. 

Radiochronometry and noble gas analyses that relate planetary bulk abundances to fractionation effects between mantle, core and atmosphere yield approximate accretion timescales for the terrestrial planets. Because of uncertainties in the equilibration between metal and silicates in magma oceans (\emph{Nimmo et al.} 2018; \S\ref{sec:2.2}), however, the earliest accretion phase is only weakly constrained: using extrapolated solar EUV fluxes as a constraint, proto-Earth may have grown at maximum up to $\lesssim$0.6 $M_{\rm Earth}$ (\emph{Lammer et al.} 2020a, 2021) within the disk lifetime of $\sim$4–5 Myr (\emph{Weiss et al.} 2021) and finished accretion later than $\sim$30 Myr (\emph{Kleine \& Walker} 2017), potentially as late as 142$\pm$25 Myr, with the Moon-forming impact (\emph{Maurice et al.} 2020) (see \S\ref{sec:3.2}). Proto-Earth growth in the disk may be even further restricted due to hydrogen deposition into nominally anhydrous minerals found in enstatite meteorites (\emph{Jin \& Bose} 2019; \emph{Piani et al.} 2020; \emph{Stephant et al.} 2021; \emph{Jin et al.} 2021): accretion to large size (+ the H component from enstatites) would create low D/H water by the reaction of nebula H$_{2}$ with mantle FeO (\emph{Ikoma \& Genda} 2006; \emph{Olson \& Sharp} 2018, 2019), possibly exceeding the maximum amount present in Earth (\emph{Peslier et al.} 2017). Measurements of the present-day mantle $^{182}$W/$^{184}$W ratio must be matched by combined accretion and core formation models, which suggests an extended accretion timescale of the proto-Earth on the order of a few tens of Myr after CAIs (\emph{Nimmo \& Kleine} 2015). Mars has been suggested to grow to 50\% of its present-day mass within $1.8^{+0.9}_{-1.0}$ Myr and to essentially finish accretion before $\sim$10–15 Myr (\emph{Dauphas \& Pourmand} 2011; \emph{Marchi et al.} 2020). Early accretion physics of Mars is constrained by the necessity to form a metal core despite its low mass (\emph{Zhang et al.} 2021), and crystallize the Martian mantle by $\sim$20–25 Myr after CAIs (\emph{Kruijer et al.} 2017; \emph{Bouvier et al.} 2018), suggesting that its core-mantle differentiation was powered by $^{26}$Al decay (\emph{Dauphas \& Pourmand} 2011; \emph{Bhatia \& Sahijpal} 2016). Because no samples from Mercury and Venus have been collected yet, there are no equivalent constraints on their accretion times.

\textit{Extrasolar Planetary Systems.} Since the characterization of the first transiting super-Earth CoRoT-7b (\emph{Léger} 2009), population statistics have provided evidence for accretion pathways distinctly different from the Solar System terrestrial planets: the radius valley (\emph{Fulton et al.} 2017; van \emph{Eylen et al.} 2017) between super-Earths, which show minor contribution of volatiles by mass, and sub-Neptunes, which feature lower densities compared to super-Earths, suggests that massive rocky planets with an initially substantial volatile contribution can form (\emph{Owen et al.} 2020) and lose much of their volatiles, producing planets compositionally distinct from either the terrestrial planets or from Neptune and Uranus in the outer Solar System. Most sub-Neptunes may host either substantial H-He envelopes (\emph{Owen \& Wu} 2017; \emph{Ginzburg et al.} 2018) and/or volatile ices (\emph{Zeng et al.} 2019; \emph{Venturini et al.} 2020; \emph{Mousis et al.} 2020), both directly inherited from accretionary processes operating during the protoplanetary disk phase. This extends to Earth-sized planets around M dwarf stars: the TRAPPIST-1 planets have mean densities that are consistent with either $>$wt\% volatile ice fractions or substantial bulk depletion in Fe (\emph{Agol et al.} 2021). This may suggest a significant contribution from outer disk regions and rapid inward migration (\emph{Unterborn et al.} 2018; \emph{Schoonenberg et al.} 2019) or partial devolatilization of initially volatile ice-rich planetesimals (\emph{Lichtenberg et al.} 2019a). The large masses and high volatile concentration of the sub-Neptune planet population provide evidence for a substantial inward flux of dust grains and radial migration of protoplanets during the disk stage (\emph{Bean et al.} 2021).

While progress in disk observations with the Atacama Large mm/sub-mm Array elucidates the chemical inventory of planet-forming systems (\emph{Öberg \& Bergin} 2021; \textit{Ceccarelli et al.}; \textit{Miotello et al.}; \textit{Manara et al.}, this volume), spatial resolution is still insufficient to probe the $\sim$au scales of rocky and terrestrial planet accretion in all but the closest disks (\emph{Andrews} 2020). However, statistical intercomparison of dust mass depletion across star-forming regions of different ages indicates that the solid component that forms rocky planets vanishes from sight on a timescale of $\sim$ $10^6$ yr (\emph{Ansdell et al.} 2016; \emph{Cieza et al.} 2021): dust coagulation and inward-drift commences during the earliest phase of disk build-up within $\sim$ $10^5$ yr (\emph{Harsono et al.} 2018; \emph{Segura-Cox et al.} 2020) of protostar formation. This suggests that the onset of protoplanet accretion operates rapidly, corroborating evidence from Solar System geochronology for an early onset of rocky planet formation contemporaneous with star formation (\emph{Lichtenberg et al.} 2021a).

\subsubsection{\textbf{Growth}} \label{sec:growth} \label{sec:3.1.2}

The astrophysical context of planet formation is discussed in greater detail in \emph{Dr{\k{a}}{\.z}kowska et al.} and \textit{Krijt et al.} (this volume), here we focus on the geophysical evolution of planetesimals and protoplanets during accretion, their chemical differentiation, phase changes (such as melting), and outgassing processes. We sub-divide this into two qualitative stages: (i) planetesimal evolution after formation by gravitational collapse of dust clouds in the disk while the planetary body is not massive enough to hold onto outgassed volatiles or accrete a substantial protoatmosphere from the disk gas; and (ii) the embryo stage when the growing protoplanets are massive enough to accrete an H-He atmosphere that significantly influences the heat transfer between (molten) interior and ambient disk environment. In the following, we briefly summarize these stages, with a focus on how they influence the geophysical evolution and atmospheric build-up of growing planets.

\emph{Planetesimal Stage.} At low levels of turbulence and locally enhanced dust-to-gas ratio (\textit{Lesur et al.}, this volume), coagulating dust grains embedded in the protoplanetary disk gas (pebbles) can self-organize in dense filaments and rapidly collapse under the self-gravity of the pebble cloud to directly form planetesimals of the order of $\sim$100 km (the streaming instability, \emph{Birnstiel et al.} 2015). The exact shape of the initial planetesimal distribution is still debated, but evidence from the asteroid belt (\emph{Delbo et al.} 2017) and Kuiper belt (\emph{Singer et al.} 2019; \emph{McKinnon et al.} 2020), and numerical simulations (\emph{Li et al.} 2019) suggest that planetesimals of $\sim$100–250 km in radius dominate the total integrated mass, while smaller planetesimals dominate in numbers (\emph{Johansen et al.} 2015; \emph{Simon et al.} 2016). The lower size threshold is influenced by turbulent stresses in the disk, while the upper mass end is truncated by the availability of pebbles in local disk regions (\emph{Klahr \& Schreiber} 2020, 2021). The gravitational potential energy of $\sim$100 km planetesimals is low (\S\ref{sec:2.1.1}), and hence these earliest planetesimals accrete with approximately the ambient disk temperature (\emph{Elkins-Tanton} 2012). Localized overdensities in the disk are necessary to trigger planetesimal collapse; this may preferentially happen at specific locations in the disk, for instance at the location of the water snowline (\emph{Dr{\k{a}}{\.z}kowska \& Alibert} 2017; \emph{Schoonenberg et al.} 2017). After gravitational collapse, growth in the inner disk is dominated by mutual collisions among planetesimals because the gravitational perturbation by such low-mass objects is insufficient to accrete a substantial amount of pebbles from the ambient disk (\emph{Visser \& Ormel} 2016). Growth proceeds via planetesimal accretion (\emph{Liu et al.} 2019) until either the local pebble flux increases substantially or the planetary embryos become massive enough to directly attract pebbles from the disk (\emph{Johansen \& Lambrechts} 2017; \emph{Ormel} 2017). 

\textit{Protoplanet Stage.} In this stage, which begins on the order of $\sim$1000 km in size, the growing protoplanets are massive enough to retain ambient disk gas and outgassed volatiles (\emph{Ikoma et al.} 2018). This changes the dynamics of pebbles drifting past the protoplanet orbit, which can instead accrete onto the planetary body: pebble accretion sensitively scales with the mass of the protoplanet and disk scale height, while planetesimal accretion scales with the geometric cross-section (\emph{Ida \& Lin} 2004; \emph{Johansen \& Lambrechts} 2017). Protoplanets of this mass undergo substantial orbital migration due to asymmetric torques from the ambient disk gas (\emph{Kley} 2019; \textit{Paardekooper et al.}, this volume). For rocky and rock-ice protoplanets, migration is typically inward, but recent models suggest the possibility of outward migration induced by planet heating (\emph{Benítez-Llambay et al.} 2015). Protoplanet migration on $\sim$au scales alters the compositional inventory of both pebbles and planetesimals that are accreted onto the planet (\emph{Bitsch et al.} 2019), and the fractionation of volatile compounds that are accreted directly from the disk. Accretion of primordial H-He protoatmospheres correlates positively with mass of the planet (\emph{Ginzburg et al.} 2016). However, for larger protoplanets the increased heating induced by accretion may limit further gas attraction by recycling disk gas into and out of the gravitational sphere of influence of the planet (\emph{Ormel et al.} 2015a,b; \emph{Cimerman et al.} 2017).

\subsubsection{\textbf{Compositional evolution}} \label{sec:compositional_evolution} \label{sec:3.1.3}

\begin{figure}[htb!]
 \plotone{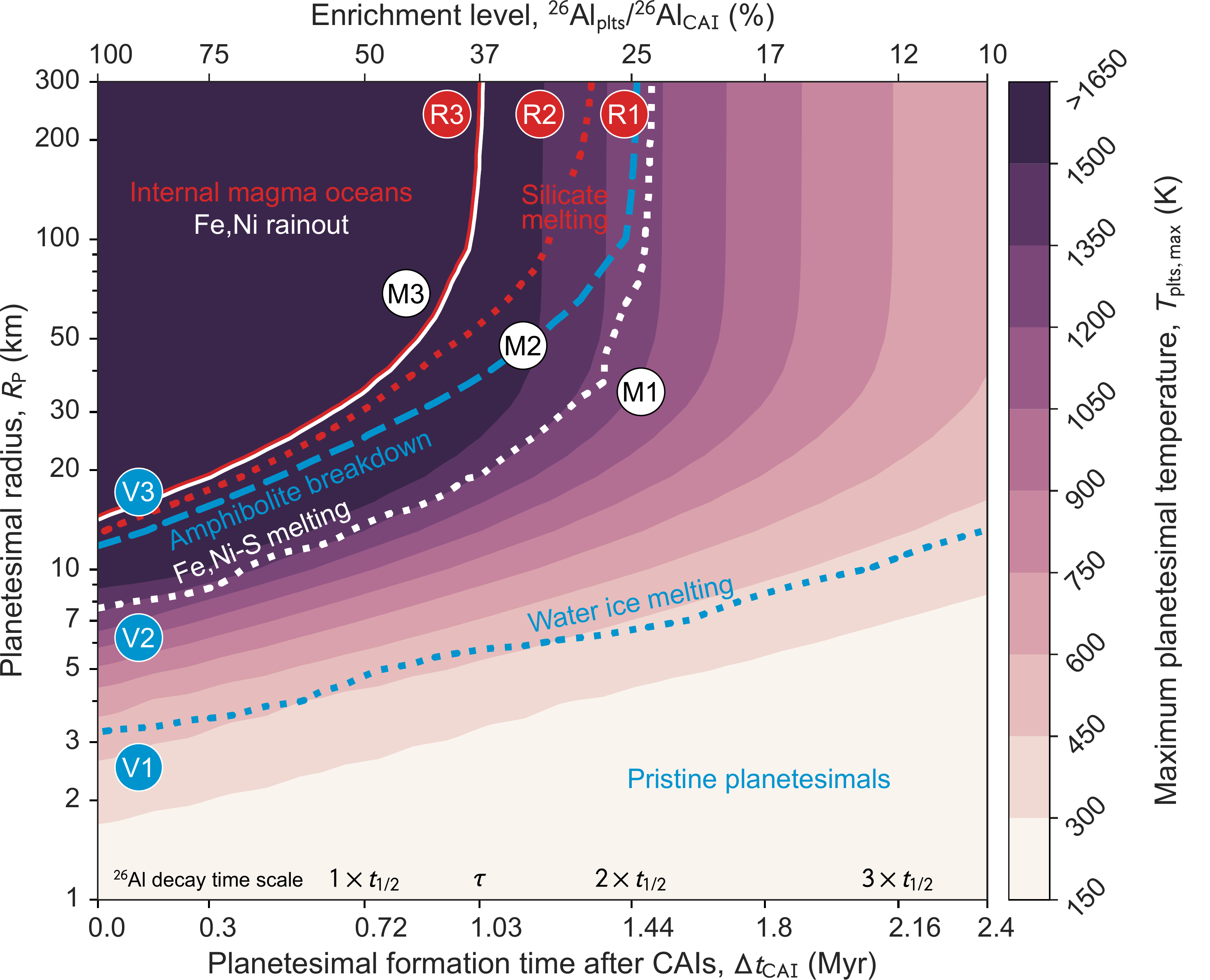}
 \caption{\small Peak temperatures for instantaneously accreting planetesimals up to 300 km in radius with a Solar System-like initial abundance of $^{26}$Al. Blue, white, and red lines and symbols separate qualitative compositional regimes with increasing peak temperature during planetesimal evolution: progressive volatile depletion (blue, V1 to V3), progressive metal core formation  (white, M1 to M3), and progressive rock melting (red, R1 to R3). Thresholds (lines) are chosen such that $>$50 vol\% of the body fall into the respective regimes V/M/R1–3. The dotted red line indicates the first melting of silicates, the solid red-white line indicates melting above the rock disaggregation threshold (internal magma oceans and rainout core formation). The dotted blue line indicates first water ice melting (if present), while the dashed blue indicates complete dehydration of hydrated silicate minerals (amphibolite). The dotted white line indicates the melting point of Fe,Ni-S phases, the potential onset of percolative core formation. Temperatures in the magma ocean regime are buffered by the increased heat flux of vigorous internal convection and do not rise substantially further in the absence of a blanketing protoatmosphere. The upper $x$-axis shows the enrichment level in $^{26}$Al relative to the CAI initial value of the Solar System, which may vary significantly between planetary systems. Indicated on the bottom are $^{26}$Al half-life $t_{1/2}$ and mean lifetime $\tau$. See description in the text and Fig.~\ref{fig6} for a discussion of the compositional evolution and differentiation process. \href{https://osf.io/kz5fc/}{\includegraphics[scale=0.35]{icon_download.pdf}}}
 \label{fig5}
\end{figure}
\begin{figure}[htb!]
 \plotone{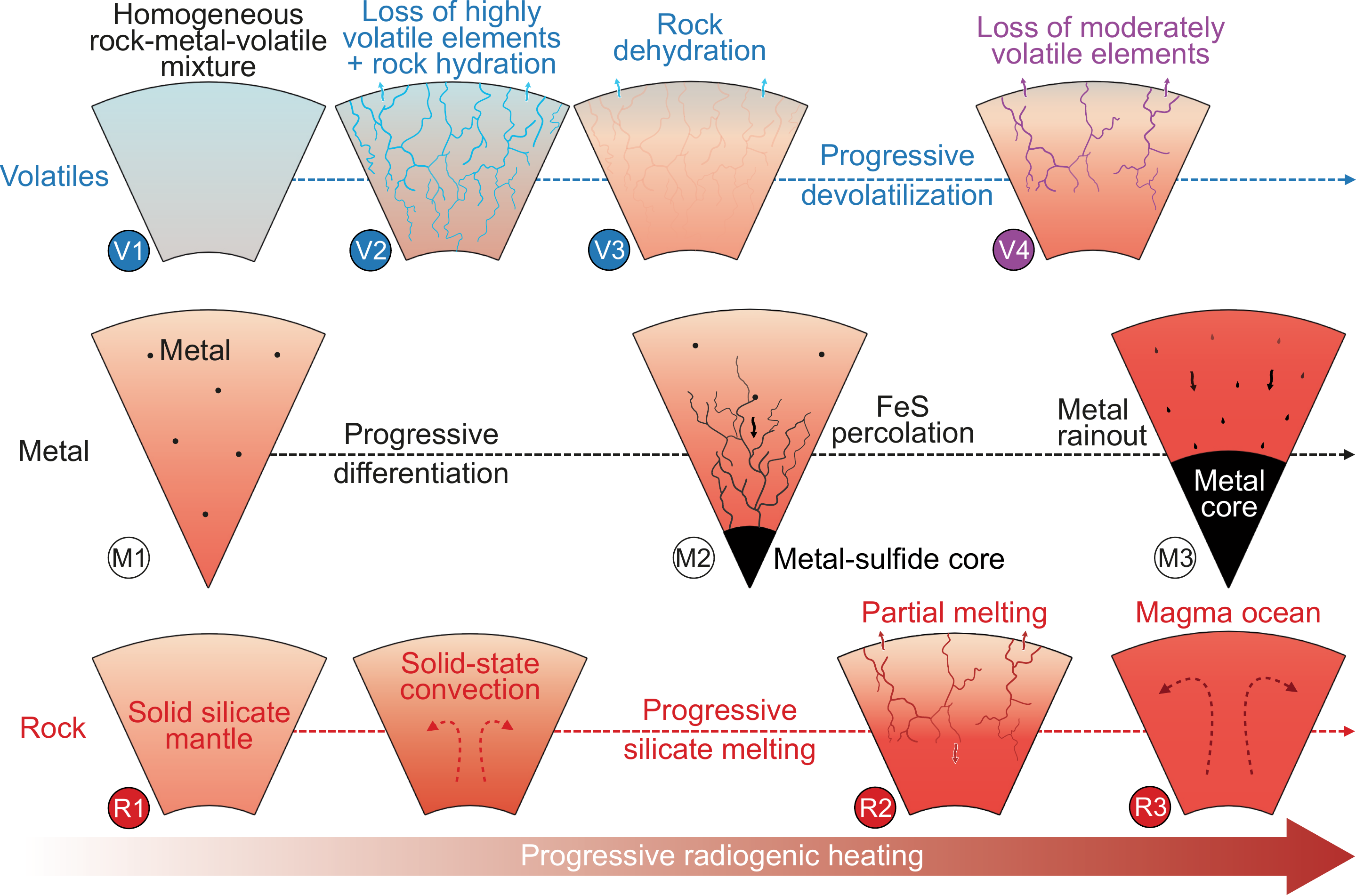}
 \caption{\small Thermal, compositional, and structural evolution of planetesimals during progressive heating. The individual stages indicated by labeled circles show progressive devolatilization of highly and moderately volatile elements (V1--V3, blue, and V4, purple), silicate convection and melting (R1--R3, red), and metal-silicate segregation (M1--M3, black). The three rows are independent of each other, compositional evolution of rock, metal, and volatiles can occur at the same time. The sketches here illustrate the thermal regimes in Fig.~\ref{fig5}. \href{https://osf.io/zxsc4/}{\includegraphics[scale=0.35]{icon_download.pdf}}}
 \label{fig6}
\end{figure}
After the first planetesimals form, their thermal evolution is mainly driven by heating from radioactive decay and subsequent accretion. Until about $\sim$1000 km in size the gravitational potential energy of accretion is of minor global importance for heating (\emph{Srámek et al.} 2012) (\S\ref{sec:2.1.1}). The peak heating temperatures of planetesimals with no atmosphere of a fixed size and composition driven by the radioactive decay of $^{26}$Al with Solar System-like initial abundances (\emph{Nittler \& Ciesla} 2016; \emph{Lugaro et al.} 2018; \emph{Parker} 2020) is shown in Fig.~\ref{fig5}, based on the simulations from \textit{Lichtenberg et al.} (2021a). Early-formed and large planetesimals reach the highest temperatures because their interiors are insulated best against radiative cooling to the ambient disk. Planetesimals above $\sim$50 km in radius display approximately similar thermal peaks and near-isothermal interior temperature profiles (\emph{Castillo-Rogez \& Young} 2017). This is a result of the conduction time scale relative to the $^{26}$Al decay time scale. The exterior disk temperature only influences the uppermost $\sim$3 km of the surface, inner regions are effectively shielded against temperature variations on the outside; variations in disk temperature therefore affect only the uppermost layers of accreting planetesimals. In addition, the pressure gradient in planetesimals is small, and hence the adiabatic slope in the interior is close to an isotherm. In general, the thermal evolution of planetesimals below $\sim$50 km in radius within the first $\sim$2 Myr after CAIs, is dominated by changes in size,  below $\sim$10 km in radius temperatures stay below $\sim$1000 K, irrespective of $^{26}$Al enrichment, and thus silicates do not melt. 

Symbols and lines in Fig.~\ref{fig5} correlate with the sketches in Fig.~\ref{fig6}. V1--V3 (blue) illustrate the devolatilization trend of planetesimals with increasing size and $^{26}$Al heating. Below the dotted blue line (V1) planetesimals retain a homogeneous mixtures of rocks, metals, and ices directly accreted from the protoplanetary disk and no substantial alteration takes place. Between the dotted and dashed blue lines (V2) volatile ices melt (melting temperature is shown for water ice, $T$ $\sim$ 273 K), which can lead to pore-water convection, hydrothermal activity (\emph{Young et al.} 1999; \emph{Wakita \& Sekiya} 2011) or mud convection (\emph{Bland \& Travis} 2017). Loss of highly volatile elements in this regime can be significant (\emph{Fu \& Elkins-Tanton} 2014; \emph{Fu et al.} 2017) and affect the total volatile budget of protoplanets that accrete such devolatilizing planetesimals (\emph{Grimm \& McSween} 1993; \emph{Lichtenberg et al.} 2019a, 2021a). The speciation of outgassed volatiles in this stage sensitively depends on the composition of planetesimals derived from the disk (e.g., the relative abundances of C, O, and S, \emph{Lichtenberg \& Krijt} 2021; \emph{Hirschmann et al.} 2021) and their redox state (\textit{Schaefer \& Fegley Jr.} 2007, 2010, 2017). H, C, and N transport in open system planetesimals is governed by the gas phase and can be rapid (\emph{Hashizume \& Sugiura} 1998). No complete models that treat both thermal evolution and detailed volatile loss via the gas and fluid phase of planetesimals during accretion have been developed to date. 

Volatiles that are not outgassed at this stage react with the ambient rock to form phyllosilicates and other forms of hydrated minerals (in the temperature interval of $T$ $\sim$ 573–673 K, \emph{Nakamura} 2006; \emph{Nakato et al.} 2008). For instance, water that is initially present as H$_{2}$O ice would not typically outgas as a complete compound, but would react with metal (H$_{2}$O + Fe → H$_{2}$ + FeO) and silicates, such that outgassing would be dominated by H$_{2}$ instead. Reaction of water with Fe metal during metamorphism on the ordinary chondrite parent bodies may have produced progressive oxidation signatures identified with petrographic type (\emph{McSween \& Labotka} 1992; \textit{Lewis \& Jones} 2016). Above the dashed blue line in Fig.~\ref{fig5} (V3) no hydrous rock phases are stable anymore at peak heating, which causes even the most temperature resistant hydrous phases, such as amphibolites, to decompose (breakdown temperature $T$ $\sim$ 1250 K). Planetesimals in this stage may be oxidized from their prior volatile-rich composition, but are chemically depleted in highly volatile elements. In terms of hydrogen atoms, such a devolatilized planetesimal would deliver a similar amount of water as the driest known chondrites and achondrites.

Metal-silicate differentiation is indicated in black/white in Figs.~\ref{fig5} and \ref{fig6}. Below the dotted white line, no metal-silicate differentiation takes place (M1, Fig.~\ref{fig5}). Reduced metal phases, such as Fe$^0$ or Fe-Ni-S compounds, remain mixed with their ambient planetesimal assemblage, but can chemically react with other constituents. If metal-sulfide phases remain abundant between the dotted and dashed white lines (M2), they may form an interconnected network of liquid Fe-Ni-S that gravitationally percolates downward to form a core (\emph{Yoshino et al.} 2003; \emph{Ghanbarzadeh} 2017). The efficiency of this mechanism depends on the availability of S, which lowers the melting point of reduced metal phases and allows them to melt, interconnect, and segregate. Percolation is, however, dependent on the spatial connectivity of ambient rock minerals and their redox state (\emph{Nimmo \& Kleine} 2015; \emph{McCoy \& Bullock} 2017). Laboratory experiments find percolation to be of limited effectiveness in forming a metal core (\emph{Bagdassarov et al.} 2009; \emph{Cerantola et al.} 2015). Complete metal-silicate differentiation may therefore be reached only in the magma ocean regime (M3), when the rock matrix is broken down and metallic phases can rain out from the surrounding liquid magma (\emph{Stevenson} 1990). The efficacy of core formation in this regime depends on the gravity and turbulent stresses in the planetesimal, which enable rapid differentiation of planetesimal sub-volumes that reach the rheological transition (\emph{Lichtenberg et al.} 2018). The interplay between silicate melt ascent, percolation, and magma ocean diffusion likely create a complex interplay between different phases of core formation, within single planetesimals (\emph{Neumann et al.} 2018; \emph{Hunt et al.} 2018) and across accreting planetesimal populations (\emph{Ricard et al.} 2017; \emph{Lichtenberg et al.} 2021a).

The red lines and symbols in Fig.~\ref{fig5} indicate silicate melting. Planetesimals right and below the dotted red line (R1) do not melt, but the macroporosity of the initially fluffy dust aggregate decreases by compaction and sintering in the interior (\emph{Henke et al.} 2012; \emph{Gail et al.} 2015; \S\ref{sec:2.1}). Macroporosity of the planetesimal lid is retained in the absence of impact processes. The effects of macroporosity, however, are small on a population level because sintering effects quickly compact the body before $^{26}$Al has released the majority of its heat contribution (\emph{Lichtenberg et al.} 2016a). Before the onset of silicate melting, heat transport operates by conduction and potentially solid-state convection on the largest and most long-lived planetesimals (\emph{Tkalcec et al.} 2013; \emph{Kaminski et al.} 2020). Melting of silicates for chondritic assemblages occurs at $\sim$1400 K, again depending on composition in terms of the ratio of refractory to more volatile components. Upon first melting (in-between the dotted and dashed red lines of Fig.~\ref{fig5}, R2) small melt pockets develop. The subsequent dynamics of the mantle in this partial melting regime is highly sensitive to the density of the magma (\emph{Neumann et al.} 2014) and grain size distribution (which affects the permeability and hence percolation of the first melts that form) of the ambient rock (\emph{Lichtenberg et al.} 2019b), and retention of highly volatile elements (\emph{Wilson \& Keil} 2017). Volatile retention is constrained by the low ambient pressure in the disk, which favors rapid release in the gas phase prior to silicate melting (\emph{Fu et al.} 2017). If few volatiles remain at the time of silicate melting, the main drivers of magma ascent in planetesimals are the FeO content and crystal grain size: planetesimals forming at $\sim$1 Myr after CAIs and with large grain sizes may develop buoyancy-driven volcanic activity, which may either form sub-lid magma chambers or extrude onto the surface (\emph{Moskovitz \& Gaidos} 2011; \emph{Mandler \& Elkins-Tanton} 2013). Extrusive silicate volcanism on planetesimals is therefore sensitive to secondary accretion via the pebble accretion mechanism (\emph{Lichtenberg et al.} 2019b; \emph{Kaminski et al.} 2020). With further heating planetesimals enter the magma ocean regime (above the dashed red line in Fig.~\ref{fig5}, R3): the silicate melt fraction exceeds the rheological transition and the internal heat flow is governed by turbulent diffusion in the internal silicate melt (see \S\ref{sec:2.1.2}). The heat flux in this regime is much higher than in the partially molten regime (R2 in Fig.~\ref{fig5}), which enables some of the largest and most-heated planetesimals to undergo near-complete silicate differentiation (\emph{Hevey \& Sanders} 2006; \emph{Lichtenberg et al.} 2016a). Planetesimals in-between the thresholds R2 and R3 become hot enough to degass moderately volatile elements (V4, purple), which may either be associated with magmatic activity during partial melting (\emph{Collinet \& Grove} 2020a,b), magma ocean–protoatmosphere equilibration (\emph{Lammer et al.} 2020), or impact devolatilization among growing planetesimals (\emph{Sossi et al.} 2019).

During growth from their initial sizes, accreting planetesimals are increasingly shaped by the release of potential energy (\emph{Asphaug} 2010). In specific dynamic settings, such as when the growth of giant planets gravitationally perturbs the accreting planetesimal population, mutual impact velocities are increased in comparison with a self-stirred population. On a population level this can lead to erosion of the outer layers of planetesimals (\emph{Bonsor et al.} 2015) and process a substantial fraction of the whole population in melting and vaporizing impacts (\emph{Carter \& Stewart} 2020; \emph{Davies et al.} 2020). The thermal and compositional effects on both fragments and intact planetesimals results from both the internal evolution due to radiogenic heating and external effects such as impacts: their timescales are comparable and result in complex transitions between melting, differentiation, and retention of primitive materials in accreting planetesimal families. Protoplanets growing from planetesimals inherit their prior compositional history (\emph{Grimm \& McSween} 1993), so that volatile loss due to planetesimal internal heating translates into devolatilized planets. The anticipated variability in short-lived radionuclides between exoplanetary systems (\S\ref{sec:2.1}) may thus lead to inter-system variability in volatile delivery, such as water and carbon compounds, on the $\sim$wt\% level (\emph{Lichtenberg et al.} 2019a; \emph{Lichtenberg \& Krijt} 2021). Difference in accretion of planetesimals of variable redox state (related to initial ice content and dehydration mechanisms) alters the potential for rocky planets and exoplanets to form iron cores (\emph{Elkins-Tanton \& Seager} 2008) and the composition of outgassed atmospheres.

Once planetesimals are large enough to retain a protoatmosphere, heat loss is dependent on the opacity of the gaseous envelope (\emph{Ikoma et al.} 2018; see \S\ref{sec:2.1.2}). Peak temperatures in this regime are not buffered by effective heat loss due to vigorous convection in the magma ocean (Eq. \ref{eq:MO_heat_flux}) anymore, such that temperatures $\gg$2000~K can be reached (\emph{Brouwers et al.} 2018; \emph{Olson \& Sharp} 2018).

\subsection{\textbf{Post-disk phase}} \label{sec:post-disk_phase} \label{sec:3.2}

After gas disk dissipation, orbits of planetary embryos become unstable because of lack of dynamical friction from planetesimals and embryos' eccentricities and inclination damping due to gas drag. This leads to the giant impact stage, where these bodies collide with each other and reach their final masses. Dynamically, the giant impact stage is essential in the collisional accretion model given that terrestrial planets need to grow larger than their isolation masses, when protoplanets clear their own orbits from all smaller bodies, which occurs at approximately Mars masses in the inner Solar System (\emph{Kokubo \& Ida} 1998; \emph{Morbidelli et al.} 2013). The role of giant impacts is less clear in the pebble accretion scenario, because growing planets may reach their final masses with fewer late giant impacts (\emph{Chambers} 2016; \emph{Johansen et al.} 2021). As discussed below, however, several lines of evidence suggest that the terrestrial planets experienced large impacts in the Solar System and this likely is the case in extrasolar systems (\emph{Izidoro et al.} 2017; \emph{Bonomo et al.} 2019), indicating that the giant impact stage is a crucial step for planet formation regardless of the dominant mode of planetary growth during the disk phase. 

\subsubsection{\textbf{Observational \& geochemical constraints}} \label{sec:observational_geochemical_constraints} \label{sec:3.2.1}

\textit{Core Formation – Hf-W System.} The most frequently used isotopic system to determine the timing of core-formation is the Hf-W system. Hf is a lithophile element, while W is a siderophile element. As discussed in \S\ref{sec:2.2.1}, iron droplets from the impactor experience metal-silicate equilibration while they descend to the bottom of the mantle. Lithophile elements, such as Hf, prefer to stay in the mantle while siderophile elements, such as W, prefer to stay in the descending iron, which is eventually delivered to the core of the target. $^{182}$Hf decays to $^{182}$W with a half-life of $\sim$8.9 million years, which is comparable to the accretion timescale, and can be used to track core formation. Its largest uncertainties are related to neutron capture effects, early crustal formation, and the unknown degree of equilibration between metals and silicates during core-merging between colliding protoplanets.

For the Earth, the core formation age and thus cessation of the major accretion phase is estimated to be $\sim$30 Myr or later after Solar System formation (e.g., \emph{Kleine et al.} 2002, 2004; \emph{Rubie et al.} 2015a). The lunar mantle has very similar $^{182}$W values to those of Earth, which indicates that the Moon-forming impact occurred after $^{182}$Hf was extinct ($\gtrsim$60 Myr, \emph{Touboul et al.} 2007). \textit{Touboul et al.} (2015) found that the Moon has slightly elevated $^{182}$W compared to the terrestrial value, which can be explained by more chondritic materials being delivered to Earth than to the Moon by late accretion due to the Earth’s larger cross section and larger gravitational effect. Alternatively, the elevated $^{182}$W could be explained by early formation of the Moon ($\sim$ 50 Myr after CAIs, \emph{Thiemens et al.} 2019). The core formation age of Mars is debated due to variable  $^{182}$W observed in Martian meteorites, uncertainties on  Mars’ composition, and the extent of metal-silicate equilibration during core formation. The estimated ages range from 2–4 Myr (\emph{Dauphas \& Pourmand} 2011) to 10–15 Myr after CAIs (\emph{Marchi et al.} 2020). The core formation age of Vesta is $\sim$4 Myr after CAIs (\emph{Kleine et al.} 2002). No age estimates for Mercury or Venus exist because of lack of sample access.

\textit{Crystallization and Differentiation – U-Pb, Sm-Nd, Lu-Hf, Rb-Sr Systems.} Magma crystallization and differentiation processes are recorded in various isotopic systems. U-Pb dating, which is often conducted on zircons, provides absolute ages of crystallization. $^{235}$U decays to $^{207}$Pb with a half-life of 704 Myr, while $^{238}$U decays to $^{206}$Pb with a half-life of 4.47 Gyr. Cross-calibrating these two independent decay paths provides absolute age determination for a given sample (\textit{Connelly et al.} 2017). The Sm-Nd, Lu-Hf, Rb-Sr, and Hf-W systems are summarized in Tab.~1.
\begin{table}[h]
\begin{tabular}{cccc} 
\footnotesize{System} & \footnotesize{Half-life} & \footnotesize{Compatibility} & \footnotesize{Crust} \\
\hline
\footnotesize $^{182}$Hf $\rightarrow$ $^{182}$W & \footnotesize 8.9 Myr & \footnotesize Hf$>$W & \footnotesize Low $^{182}$W/$^{184}$W \\
\footnotesize $^{146}$Sm $\rightarrow$ $^{146}$Nd & \footnotesize 106 Myr & \footnotesize Sm$>$Nd & \footnotesize Low $^{142}$Nd/$^{144}$Nd \\
\footnotesize $^{176}$Lu $\rightarrow$ $^{176}$Hf & \footnotesize 37.1 Gyr & \footnotesize Lu$>$Hf & \footnotesize Low $^{176}$Hf/$^{177}$Hf \\
\footnotesize $^{87}$Rb $\rightarrow$ $^{87}$Sr & \footnotesize 48.8 Gyr & \footnotesize Rb$<$Sr & \footnotesize High $^{87}$Sr/$^{86}$Sr
\end{tabular}
\caption{\small Isotopic systems that infer silicate differentiation.}
\end{table}

\begin{figure}[bt!]
 \plotone{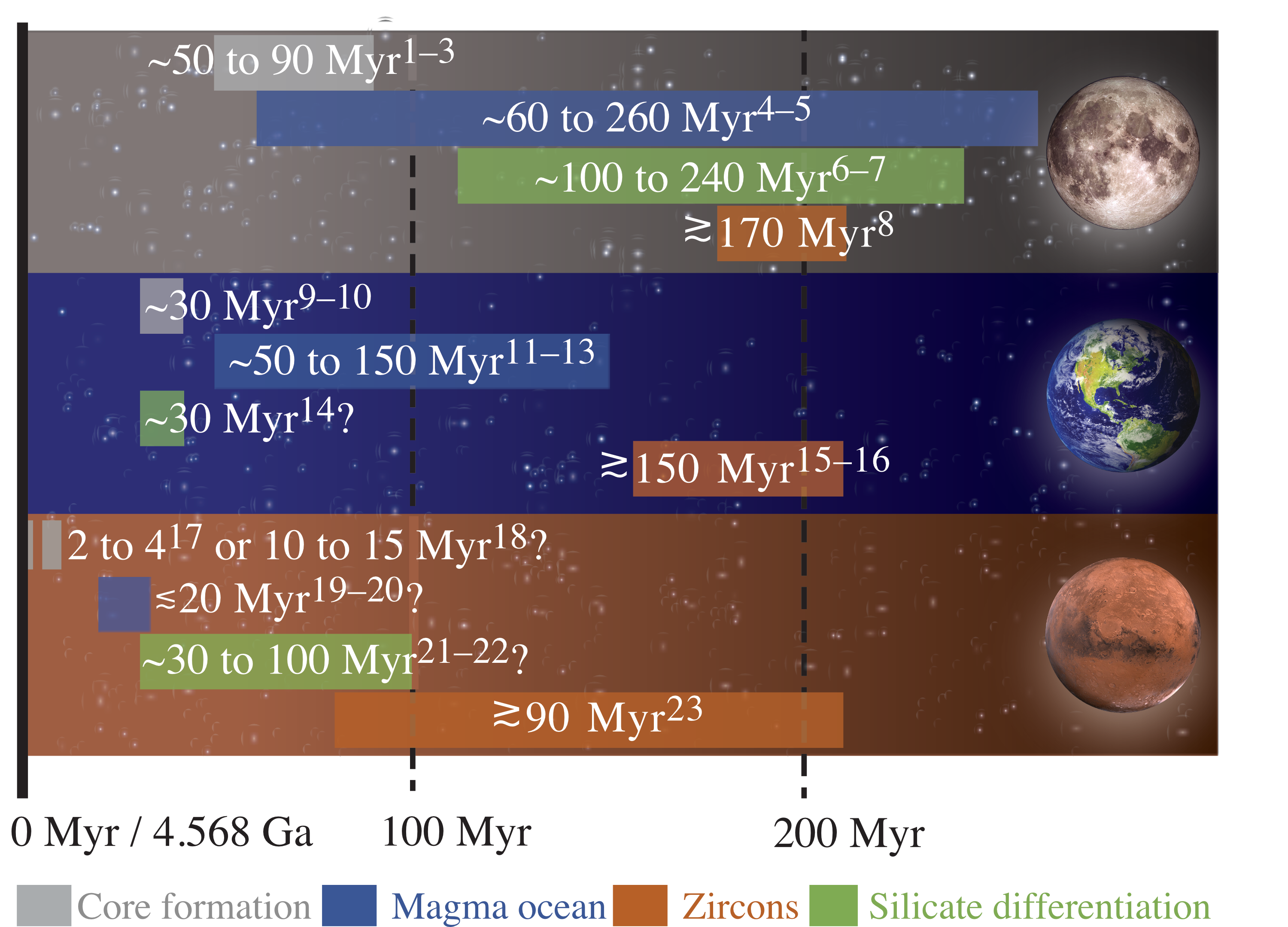}
 \caption{\small Timeline of core formation, mantle crystallization, silicate differentiation, and crust formation inferred from geochemically-dated samples for the Moon (top), Earth (middle), and Mars (bottom). Zircon ages for the Earth indicate the presence of liquid water (oceans) at the Earth’s surface as early as $\gtrsim$150 Myr after CAIs. \textsc{References}: (1) \textit{Touboul et al.} (2007), (2) \textit{Thiemens et al.} (2019), (3) \textit{Kruijer et al.} (2021), (4,5) \textit{Meyer et al.} (2010), \textit{Maurice et al.} (2020), (6,7) \textit{Norman et al.} (2003), \textit{Nyquist et al.} (1995), (8) \textit{Nemchin et al.} (2009), (9,10) \textit{Kleine et al.} (2002, 2004), (11–13) \textit{Abe} (1997), \textit{Lebrun et al.} (2013), \textit{Solomatov} (2000), (14) \textit{Boyet \& Carlson} (2005), (15-16) \textit{Wilde et al.} (2001), \textit{Mojzsis et al.} (2001), (17,18) \textit{Dauphas \& Pourmand} (2011), \textit{Marchi et al.} (2020), (19, 20) \textit{Bouvier et al.} (2018), \textit{Kruijer et al.} (2020), (21,22) \textit{Borg et al.} (2016), \textit{Debaille et al.} (2017), (23) \textit{Costa et al.} (2020). \href{https://osf.io/d5sf2/}{\includegraphics[scale=0.35]{icon_download.pdf}}}
 \label{fig7}
\end{figure}
So called incompatible elements are partitioned into melt, which is typically lighter than the ambient rock, and therefore ascends upward and forms the planetary crust. Thus, comparing incompatible elements with compatible ones, which remain in the solid rock, indicates the timing of silicate differentiation, such as between the mantle and crust. As an example, a list of the core formation age and early crystallization age estimates for the Moon, Earth, and Mars are shown in Fig.~\ref{fig7}.

Earth’s super-chondritic $^{142}$Nd/$^{144}$Nd ratios were proposed to reflect early differentiation ($>$4.53 Ga, \emph{Boyet \& Carlson} 2005) and a hidden reservoir enriched in low $^{142}$Nd/$^{144}$Nd at the base of the mantle, which would hint at substantial collisional reprocessing of precursor planetesimals (\emph{Bonsor et al.} 2015). However, this can be explained by nucleosynthetic effects (\emph{Burkhardt et al.} 2016). The crystallization age estimates in Fig.~\ref{fig7} have a wide range due to a number of uncertainties including (a) planetary bulk composition, (b) compositional heterogeneities within the planet, (c) fractionation effects during later volcanic and partial melting, (d) nucleosynthetic isotope heterogeneities, and (e) the lifetime and frequency of magma ocean episodes. Thus, it is essential to use multiple isotopic systems to determine the formation and crystallization ages of the planet, and to consider the limitations of each method appropriately. I-Pu-Xe is also an important system that tracks planetary formation; $^{129}$I decays to $^{129}$Xe with 15.7 Myr and $^{244}$Pu, whose half life is 80 Myr, produces Xe isotopes including $^{136}$Xe by spontaneous fission (e.g., \emph{Avice \& Marty} 2014). I-Pu-Xe informs about the planet-atmosphere closure time as well as the extent of atmospheric loss.

\textit{Crustal Crystallization – Zircons and Phosphates.} Zircons are extremely resilient minerals that constrain trace amounts of U and Pb, which enable absolute radiogenic age dating of crystallization. On Earth, crustal fragments that formed during the Hadean eon ($\sim$4.0–4.567 Ga, meaning before present) is extremely limited (the oldest known whole rocks are 4.02 Ga in the Acasta Gneiss Complex, \emph{Johnson et al.} 2018), and therefore most of our understanding of the crust originates from detrital zircons at that time (as old as $\sim$4.4 Ga, originating from the Jack Hills in Australia, \emph{Wilde et al.} 2001; \emph{Mojzsis et al.} 2001). These zircons provide localized key information of early Earth, including the timing of Earth’s crust formation, the redox state of the Hadean Earth’s atmosphere (\emph{Trail et al.} 2011), when plate tectonics started (e.g., \emph{Turner et al.} 2020), and when Earth’s magnetic field started (e.g., \emph{Tarduno et al.} 2015; \emph{Borlina et al.} 2020). Microfractures and phase changes (reidite) in zircons inform us (\emph{Timms et al.} 2017) about the pressure-temperature conditions that the zircons experienced. In addition to zircons, phosphates, such as apatite and merrilite that can be dated by the U-Pb system, have also been used to determine their formation ages on the Moon and elsewhere (\emph{Snape et al.} 2016).

\textit{Late Veneer.} When a planet experiences a large impact and forms a deep magma ocean, most of the highly siderophile elements (HSEs) are expected to be stripped from the mantle and delivered to the planetary core. However, the silicate portion of the Earth’s mantle  contains abundant HSEs, much more than estimated based on partition coefficients and the efficiency of core formation (\S\ref{sec:2.2.1}). Hence, the abundance of elements such as iron, gold, or platinum on Earth’s (near-)surface is a conundrum if the Earth had completed all of its accretion in a globally molten state. To account for this discrepancy, the late veneer model posits that after the last giant impact (i.e. the Moon-forming impact for Earth) chondritic materials were added to the planet after the planet was (partly) solidified (e.g., \emph{Chou} 1978; \emph{Kimura et al.} 1974). The additional amounts are (at maximum) 0.5–0.8 wt\% for Earth (\emph{Walker} 2009), 0.02–0.035 wt\% for the Moon (\emph{Day \& Walker} 2015; \emph{Kruijer et al.} 2015), and $\sim$0.7\% for Mars (\emph{Walker} 2009). To put these numbers into context, if the material of the maximum late veneer were to be added to Earth as a single layer, it would be 20 km thick, gathered into a sphere it would be larger than Pluto (\emph{Zahnle et al.} 2020). 

In order to match all of these constraints, especially the relatively small late veneer contribution to the Moon compared to Earth, it has been hypothesized that relatively large impactors brought the majority of HSEs in a few impact events (several 100s–1000s km in diameter, e.g., \emph{Bottke et al.} 2010; \emph{Brasser et al.} 2016; \emph{Genda et al.} 2017), potentially as left-over debris from the Moon-forming giant impact itself, otherwise the Moon’s surface would record higher HSE abundances than observed. However, it may be possible that these impactors fell into the crystallizing lunar magma ocean by the exsolution and segregation of liquid FeS (\emph{Morbidelli et al.} 2018). A fraction of volatiles may have been delivered by the late veneer, but isotopic constraints (\emph{Dauphas} 2017; \emph{Fischer-Gödde \& Kleine} 2017; \emph{Bermingham et al.} 2018; \emph{Zahnle \& Carlson} 2020) and C, H, N, and S elemental abundances in the bulk silicate Earth and Venus’ atmosphere (\emph{Hirschmann et al.} 2016; \emph{Grewal et al.} 2019; \emph{Gillmann et al.} 2020) indicate that the majority of volatile delivery occurred in earlier accretion stages and hence that the late veneer was dry in composition and chemically reduced (\S\ref{sec:2.2.3}).

\textit{Late Heavy Bombardment.} In the Solar System, the Moon has been the basis of our understanding of the impact history and size distribution of the impactors (asteroids) because the craters are well preserved due to minimal tectonic activity and returned lunar rock samples. Early analysis of K-Ar and Ar-Ar dating of lunar samples indicated a peak of impact shock ages at 4.1–3.8 Ga (\emph{Cohen et al.} 2005), which appeared to be consistent with crystallization ages derived from lunar zircons (\emph{Tera et al.} 1974; \emph{Snape et al.} 2016). This led to the hypothesis that the impact flux during this time period was elevated due to orbital instabilities of Jupiter and Saturn in the Solar System (Nice model, \emph{Gomes et al.} 2005), termed the Late Heavy Bombardment (LHB, \emph{Bottke \& Norman} 2017). The term “late” stems from the fact that this occurred $\sim$500 Myr after Solar System accretion essentially ended, but at this time the population of leftover planetesimals should have been small. This necessitates an orbital instability to account for the perceived spike in impact flux, but from a theoretical point of view, the onset of the instability is a free parameter (\emph{Morbidelli et al.} 2018). This elevated impact flux would have caused intense bombardment during this period, which can explain the clustering of shock ages in lunar samples. However, recent studies have indicated potential issues with age determination by the Ar-Ar technique (\emph{Boehnke \& Harrison} 2016). Moreover, it is possible that one large impact (recorded by the Imbrium basin) contaminated the Apollo samples. Thus, it remains controversial if the Solar System experienced a large-scale instability during this time period (\emph{Zellner} 2017). The impact record could be explained by gradual decline of the impact flux over time (\emph{Morbidelli et al.} 2018). Because the proposed giant planet instability nevertheless can explain a number of other dynamical properties of the Solar System, more recent models shift the giant planet instability to earlier times (\emph{Clement et al.} 2018, 2019), disconnected from any potential LHB impact spike.

\subsubsection{\textbf{Evolutionary consequences of impacts}} \label{sec:evolutionary_consequences_of_impacts} \label{sec:3.2.2}

Giant impact events at the end of the planetary accretion phase lead to formation of magma oceans on planetary bodies and determine the initial states and subsequent evolution of planets (\S\ref{sec:2.1}), including their volatile budgets, interior structures, and oxidation states (\S\ref{sec:2.2}). Smaller impacts that occur after the accretion phase can also have a significant influence on the planetary environment, such as atmospheric composition and mass budget, geodynamo and hydrothermal activity, which can strongly affect the climate and potential for surface life.

\textit{Core-Mantle Ratio, Refractory Elements.} Planetary impacts not only heat and melt the mantle (see \S\ref{sec:2.2.1}), but can alter global planetary compositions. First, the crust can be more easily stripped than the mantle as a result of high velocity impacts. The crust is enriched in heat-producing elements and removing it can alter the thermal history of the planet. Moreover, Earth’s super-chondritic Mg/Si ratio can be explained if the protocrust, which would have had lower Mg/Si ratio, continued to be removed (\emph{O’Neill \& Palme} 2008; \emph{Boujibar et al.} 2014). Numerical simulations suggest that up to $\sim$30 wt\% of the planetary crustal mass could be removed by collisional stripping, which could remove $\sim$20\% of heat producing elements. But this loss could be diminished if re-accretion of the stripped materials onto the planet is efficient (\emph{Carter et al.} 2018). If collisional erosion was efficient, the elevated Mg/Si ratio of the Earth could be explained. However, the details depend on the mantle melting (\emph{Lichtenberg et al.} 2019b) and partitioning process (\S\ref{sec:2.2.1}), and impact statistics, such as impactor size, velocity, and angle.

A large and high velocity impact can alter the core-mantle ratio of a planet (\emph{Marcus et al.} 2010). Mercury, which has a fractionally large core, may have experienced an energetic impact that blasted off a large portion of the mantle (\emph{Benz} 2007; \emph{Asphaug \& Reufer} 2014). While some of the stripped mantle would have been re-accreted, the rest could have been lost by the Poynting-Robertson effect (\emph{Benz} 2007) or solar winds (\emph{Spalding \& Adams} 2020). Alternatively, the building blocks of Mercury could have been highly reduced, which removes the need for an impact origin (\emph{Malavergne et al.} 2014). A large density variation in a multiple exoplanetary system also provides evidence for the effects of giant impacts on core-mantle ratios (\emph{Bonomo et al.} 2019), as discussed later in this section. Additionally, a large impact can raise the temperature high enough such that silicates and metals become miscible (\emph{Wahl \& Militzer} 2015), which would diffuse the otherwise clear boundary between the core and mantle (Fig.~\ref{fig2}d). This adds additional heat from gravitational segregation when silicate and iron separate after cooling (\S\ref{sec:2.1.1}), as is the case for gas giants, where He and H separate (\emph{Brygoo et al.} 2021). Moreover, in this scenario, metal-silicate equilibration can occur at the core-mantle boundary when the core and mantle cool enough to be miscible again. 

\textit{Volatile Stripping.} Impacts can remove planetary atmospheres and alter the bulk volatile budget of rocky planets. Typically, it is easier to remove atmospheres when the planetary object is relatively small due to its smaller escape velocity. However, it is still possible to remove parts of the atmosphere after the planet reaches its final mass. A large impact can remove atmospheric mass when the atmospheric velocity surpasses the escape velocity due to shock waves that travel through the ground (\emph{Genda \& Abe} 2003) and ejecta near the impact point (\emph{Schlichting et al.} 2015). The presence of an ocean enhances impact-induced atmospheric loss due to evaporation of the ocean and lower impedance of the ocean (\emph{Genda \& Abe} 2005). More recent work directly calculates the atmospheric loss by impact simulations, which result in typically several wt\% of atmospheric loss, but as high as $\sim$30 wt\% for  Mars-sized impactors (\emph{Lammer et al.} 2020b). Additionally, impact-induced heating of a planet can heat the planetary mantle and surface, which may trigger  hydrodynamic escape of the atmosphere (\emph{Biersteker \& Schlichting} 2019). From a geochemical perspective, the strong depletion of Earth in Ne and Ar relative to Venus and chondrites, despite Venus’ experiencing much greater instellation (cf.~\S\ref{sec:3.2.3}) supports the effectiveness of giant impacts in eroding planetary atmospheres at a late stage.

The Moon experienced significant volatile loss during its formation and accretion phase as evidenced by its depletion  in volatiles with respect to the Earth (see review by \emph{Canup et al.} 2021). Originally, it was thought that lunar volatiles would have been lost from the Earth-Moon system to space during the impact, but this scenario is not likely because the gravity of Earth would have prevented volatiles from escaping to space (\emph{Nakajima \& Stevenson} 2014,  2018). However, it is possible that some lunar volatiles escaped from the protolunar disk (\emph{Canup et al.} 2015; \emph{Lock et al.} 2018; \emph{Mullen \& Gammie} 2020; \emph{Charnoz et al.} 2021) or from the lunar magma ocean  (\emph{Kato et al.} 2015; \emph{Kato \& Moynier} 2017; cf. \emph{Tang \& Young} 2020).  The lunar volatile loss mechanism is an active area of research. Some may argue that Mercury would not have experienced a large impact because of its volatile-rich surface composition (\emph{Peplowski et al.} 2011) compared to the Moon, but it should be noted that the formation environment of the Moon was very unique and it cannot be directly compared with Mercury. Given that the Earth also experienced a large impact, but is not as depleted as the Moon, shows that a volatile-rich surface does not exclude an impact origin for Mercury. 

\textit{Magnetic Fields.} Impacts play a role in shaping the development of planetary magnetic fields by determining the abundances of light elements in the core. As discussed in \S\ref{sec:2.2.2}, metal-silicate equilibration in growing protoplanets determine the budget of light elements in the core. Currently, Earth’s magnetic field is thought to be generated by crystallization of the inner core, by releasing light elements and latent heat at the bottom of the liquid outer core, which facilitates core convection and generates dynamo action. Some of the light elements, such as Si and Mg, could have been added to the core at late stages of accretion and hence under high pressure (\emph{Siebert et al.} 2016; \emph{O’Rourke \& Stevenson} 2016). This could have established a compositional stratification in the core and hence prevented core convection and dynamo generation early in planetary evolution. However, this stratification could be reset and form a homogeneous core structure if the planet experienced a large impact (\emph{Stevenson} 2014; \emph{Jacobson et al.} 2017). 

In addition to impacts at the end of the accretion phase, later basin-forming impacts also affect  the planetary magnetic field. A giant impact or a large basin-forming impact ($\sim$1000s of km) can cause thermal stratification in the core by heating the top of the core, which suppresses the magnetic field (\emph{Roberts \& Arkani-Hamed} 2017; \emph{Arkani-Hamed \& Olson} 2010). Alternatively, if an impact can generate a hot and thick iron layer primarily coming from the impactor at the top of the target’s core, a magnetic field can be generated within as long as the layer is thick enough (10–30 km for Mars, \emph{Reese \& Solomatov} 2010). These studies were originally meant to explain the Martian dynamo that may have ended around $\sim$4.1 Ga (\emph{Lillis et al.} 2013), because large Martian impact basins that formed between $\sim$4.1–3.8 Ga do not show crustal magnetic records (Hellas, Utopia, Argyre, and Isidis).  An alternative explanation for this observation is that the impact excavated deeper parts of the mantle, which feature less magnetic materials, while a Martian dynamo was present throughout the early history (4.5–3.7 Ga, \emph{Mittelholz et al.} 2020). 

\textit{Impact-Induced Hydrothermal Activity.} A large crater-forming impact can provide enough heat to the surface, which can lead to the formation of hydrothermal systems where silicate and water interact under high temperatures. Since hydrothermal systems provide key ingredients for life, including heat, water, chemicals, and nutrients for a prolonged time, they may have the potential to support life in the vicinity on habitable planets (\emph{Osinski et al.} 2013). On Earth, hydrothermal systems have been observed in $\sim$80 craters out of 180 (see review by \emph{Osinski et al.} 2020) with a wide range of crater sizes ($\sim$2–250 km). The lifetime of an impact-induced hydrothermal system depends sensitively on the size of the crater, but it can last from hundreds to a few millions of years for Sudbury-sized impact craters ($\sim$250 km) based on numerical studies (e.g., \emph{Abramov \& Kring} 2004). Many of detected hydrous and hydrated minerals on Mars are associated with impact craters (\emph{Mustard et al.} 2008; \emph{Carter et al.} 2013; \emph{Ehlmann et al.} 2011), which may indicate past impact-induced hydrothermal activities.  The observed bright spots on Ceres, which are likely sodium carbonate, could have formed by impact-induced hydrothermal activity (\emph{Castillo-Rogez et al.} 2019). 

\textit{Giant Impacts in Extrasolar Systems.} In addition to Mercury, a large density variation in the same exoplanetary system may be interpreted as evidence for a giant impact. For instance, Kepler-107b and c have similar planetary radii (1.5–1.6 $R_{\rm Earth}$),  but Kepler-107c (12.6 g cm$^{-3}$) is more than twice as dense as Kepler-107b (5.3 g cm$^{-3}$, \emph{Bonomo et al.} 2019). This large difference cannot be explained by XUV-induced hydrodynamic escape, and hence a large impact may be necessary to explain the density dichotomy. Moreover, additional indirect evidence of giant impacts has been observed. Spikes and short-term variability in infrared radiation and SiO vapor emission indicate debris formation by large impacts and subsequent collisional cascades in debris disks (e.g., \emph{Meng et al.} 2014; \emph{Thompson et al.} 2019; \emph{Chen et al.} 2019). The resulting magma ocean atmospheres may be observable with future astronomical surveys (\emph{Lupu et al.} 2014; \emph{Hamano et al.} 2015; \emph{Bonati et al.} 2019). 

A potential future line of evidence for the frequency of impacts in exoplanetary systems are the presence of exomoons. In general, moons can form by (a) formation in a circumplanetary disk, (b) a large impact, or (c) gravitational capture. Theoretically, it has been proposed that the satellite-to-planet mass ratio  ($M_s/M_p$) may be at most $10^{-4}$ in a circumplanetary disk due to the balance between supply of moon-forming materials and disk mass loss due to gas drag (\emph{Canup \& Ward} 2006). A large impact tends to form fractionally larger moons, such as Earth’s Moon ($M_s/M_p\sim 0.01$) and the Pluto-Charon system ($M_s/M_p\sim 0.1$). There is no known mass limit of the planet-mass ratio for the binary capture scenario, but it is challenging to capture large moons because the orbital kinetic energy needs to be converted into heat in a very short time period (during flybys) or requires a third body (\emph{Agnor \& Hamilton} 2006). Thus, if a fractionally large exomoon is detected, it may indicate that it formed by a giant impact (\emph{Nakajima et al.} 2022). Several exomoon candidates have been found, but no confirmed exomoon exists to date (\emph{Teachey \& Kipping} 2018; \emph{Kreidberg et al.} 2019; \emph{Kipping et al.} 2022). The detected moon-forming disk in PDS 70b (\emph{Benisty et al.} 2021) is not likely formed by an impact because the disk mass is consistent with the circumplanetary disk formation scenario, with $M_s/M_p \sim 10^{-4}$.

\subsubsection{\textbf{Magma ocean crystallization}} \label{sec:magma_ocean_crystallization} \label{sec:3.2.3}

At the end of the planetary accretion phase, rocky planets of the size of Earth and above are covered by global magma oceans (\S\ref{sec:2.1}). The depth of the magma ocean affects the oxidation state of the mantle.  Importantly, in a deep magma ocean, like the one resulting from the Moon-forming impact (\emph{Nakajima \& Stevenson} 2015), the reaction  FeO(melt) + $\frac{1}{4}\mathrm{O}_2$ = FeO$_{1.5}$ (melt) leads to disproportionation of Fe$^{2+}$ to Fe$^{3+}$ (\emph{Frost \& McCammon} 2008). This reaction would not have been efficient for smaller planetary objects, such as the Moon and Mars, due to their smaller pressure ranges. Assuming Fe$^{3+}$/Fe is homogenized in the magma ocean by vigorous convection, it is predicted that the shallower magma ocean is more oxidized compared to the deeper part of the magma ocean (\emph{Hirschmann} 2012). This oxygen fugacity gradient becomes more prominent for larger planets. This could explain why Earth’s mantle is more oxidized than smaller planetary objects, such as the Moon and Mars (\emph{Deng et al.} 2020), but this effect may be less prominent for super-Earths because vigorous convection in magma oceans of super-Earths can suppress  iron rainout  (\emph{Lichtenberg} 2021). An oxidized magma ocean, where iron disproportionation is effective, leads to outgassing of oxidized species, such as H$_{2}$O and CO$_{2}$, while a reduced magma ocean (with either a lot of elemental Fe or H, or a low number density of O), would dominantly outgas reduced species (\S\ref{sec:2.2.3}, Fig.~\ref{fig3}), such as H$_{2}$, CO, NH$_{3}$ or CH$_{4}$. The former is at present regarded to be the standard scenario for Earth (\emph{Zahnle et al.} 2010), which is consistent with bulk silicate D/H ratios (\emph{Pahlevan et al.} 2019), even though trace element partitioning data during metal-silicate equilibration indicate a reverse trend in redox state (\emph{Badro et al.} 2013; \emph{Fischer et al.} 2020). Possible retention of a primordial H$_{2}$ (\emph{Lammer et al.} 2020a) atmosphere may further alter this picture (\emph{Saito \& Kuramoto} 2020).

Directly after the impact, the  silicate-rich atmosphere would have a high photospheric temperature ($>$2000 K), which leads to rapid cooling $\sim \sigma T^4_{photo} \sim 10^6$ W/m$^2$, where $\sigma$ is the Stefan-Boltzmann constant and $T_{photo}$ the photospheric temperature. As the magma ocean cools, and once volatiles, such as H$_{2}$O and CO$_{2}$ for an oxidized environment and H$_{2}$, CO, and CH$_{4}$ for a reduced environment, oversaturate in the magma ocean, they are released to the atmosphere in the form of bubbles.  Once the atmosphere is dominated by these volatiles and is opaque, the cooling slows down significantly due to the lower photospheric temperature, from where most of the radiation escapes to space. For a fixed composition, instellation has a dominant role in controlling planetary energy budget. With water vapor present in a magma ocean atmosphere, the $T$-$P$ structure of the upper atmosphere aligns closely with the water dew point, such that the outgoing radiation is limited to $\sim$280 W/m$^2$ in the surface temperature interval between $\sim$300–2000 K (\emph{Goldblatt et al.} 2013; \emph{Leconte et al.} 2013) for an approximately Earth-like volatile inventory.  Once the surface of the magma ocean becomes mostly crystallized ($\gtrsim$60\% at the rheological transition) or forms a floatation crust (see discussion on the Moon above) cooling slows down even further because heat is transferred by conduction and solid state convection. Full planet solidification for Earth-sized planets with a similar volatile inventory and instellation takes on the order of $10^5$ to $10^8$ yr (\emph{Abe et al.} 1997; \emph{Zahnle et al.} 2010, 2015; \emph{Elkins-Tanton} 2012; \emph{Lebrun et al.} 2013), and is sensitively affected by the speciation of volatiles (\emph{Salvador et al.} 2017; \emph{Wordsworth et al.} 2018; \emph{Lichtenberg et al.} 2021b), instellation (\emph{Hamano et al.} 2013; \emph{Schaefer et al.} 2016), and mode of mantle solidification (\emph{Monteux et al.} 2016; \emph{Bower et al.} 2018; \emph{Miyazaki \& Korenaga} 2022). 

\begin{figure}[tbh!]
 \centering
 \plotone{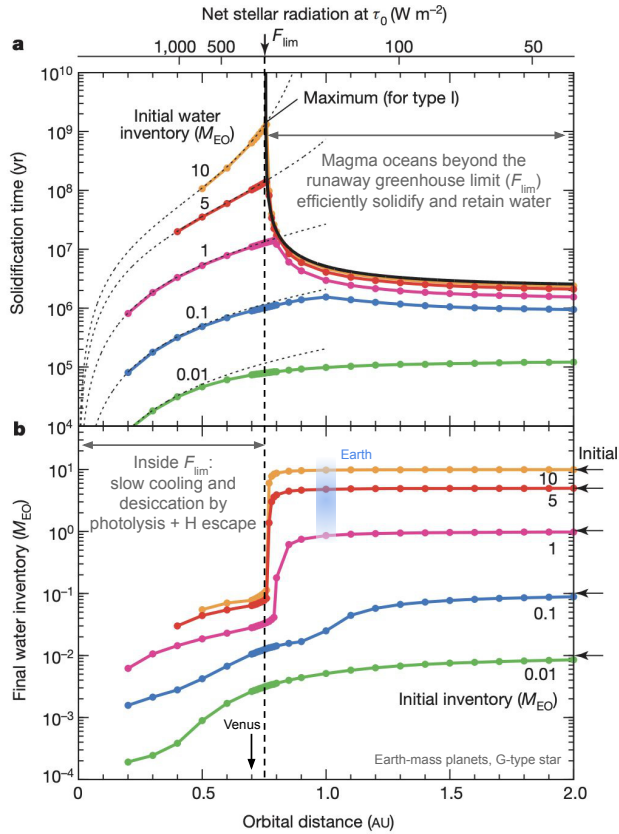}
 \caption{\small Two distinct types of terrestrial planets separated by their evolution during magma ocean solidification. The \textit{x}-axis shows orbital distance (bottom) and initial net stellar radiation (top). The top black arrow in \textbf{(a)} indicates the tropospheric radiation limit ($F_{\rm lim}$), which separates planets that can cool via radiation to space and those that desiccate to near-completion. \textbf{(b)} Total water inventory at the time of complete mantle solidification. Modified and reprinted by permission from Springer Nature, \emph{Nature}, 497, 607–610, \textit{Hamano, K., Abe, Y. \& Genda, H.} (2013).}
 \label{fig8}
\end{figure}

Fig.~\ref{fig8} (\emph{Hamano et al.} 2013) illustrates the sensitive control of initial water abundances and instellation on the magma ocean lifetime of Earth-sized planets dominated by water vapor. Planets that receive irradiation from their central star that is lower than the runaway greenhouse threshold (\emph{Ingersoll} 1969; \emph{Nakajima et al.} 1992; the aforementioned $\sim$280 W/m$^2$) can continuously cool down and solidify. However, planets orbiting closer to their star receive more energy input than they can radiate away (see \S\ref{sec:2.1}). These planets do not cool down until their initial water inventory is substantially reduced by H$_{2}$O photolysis in the upper atmosphere and subsequent H loss by escape. For water inventories of $\sim$10 Earth oceans and higher (Earth’s combined mantle, ocean, and atmosphere water inventory is $\sim$3–11 oceans, \emph{Peslier et al.} 2017) the solidification timescale can become comparable to the main-sequence lifetime of the star (\emph{Hamano et al.} 2013, 2015). Planets that solidify inside the runaway greenhouse threshold become desiccated by this process and, important in an astrophysical context, their initial volatile abundances are decoupled from the volatile abundances accreted during planetary formation. These primary regimes of crystallization shift with different and more complex atmospheric composition (\emph{Salvador et al.} 2017; \emph{Lichtenberg et al.} 2021b; \emph{Graham et al.} 2021) and around other types of stars (\emph{Schaefer et al.} 2016; \emph{Wordsworth et al.} 2018). Cloud feedbacks previously have been suggested to shield solid planets from a runaway greenhouse (\emph{Yang et al.} 2013; \emph{Way et al.} 2016), but high-altitude shortwave absorption on the planetary dayside would induce a net warming effect (\emph{Turbet et al.} 2021). The influence of solidification modes other than bottom-up crystallization and varying redox states on atmospheric loss and retention in the above picture are yet to be explored in sufficient detail (\emph{Lichtenberg et al.} 2021b; \emph{Bower et al.} 2022). 

The radial location of the first crystals solidifying from the magma is determined by where the mantle liquidus and adiabat cross (\emph{Stixrude et al.} 2009), whereas the density difference between melt and solid determines the subsequent evolution process. The density crossover for bridgmanite (lower mantle materials in Earth) occurs at high pressures ($\sim$110–120 GPa, e.g., \emph{Stixrude et al.} 2009; \emph{Caracas et al.} 2019). The location depends on the iron content of the magma ocean, which increases as the magma ocean crystallizes. Given that Earth’s core-mantle boundary (CMB) pressure is $\sim$135 GPa, planets as large as or larger than Earth may experience solidification at the mid-mantle separating a shallow and deep (basal) magma oceans in the case of fractional crystallization (\emph{Stixrude} 2014), but such separation may not occur if the crystallization process is governed by equilibrium fractionation (\emph{Caracas et al.} 2019). 

A basal magma ocean can form by the crystal separation process discussed above or overturn of a dense melt from near the surface to the bottom of the mantle. A basal magma ocean in early Earth might have survived for billions of years due to slow cooling of the overlying mantle (\emph{Labrosse et al.} 2007). The basal magma ocean of Earth is hypothesized to be responsible for producing a geodynamo on early Earth.  Generally speaking, generation of a dynamo is possible when the magnetic Reynolds number, $R_m = \mu_0vL\sigma_{cond}$, is larger than a few 10s.  Here, $\mu_0$ is the magnetic susceptibility, $v$ is the velocity, $L$ is the domain length scale, and $\sigma_{cond}$ is the electrical conductivity. For Earth's basal magma ocean, $\sigma_{cond}$ needs to be larger than $\sim$10,000 S/m (= $\Omega^{-1}$ m$^{-1}$) to generate a dynamo. This is not easily achieved by SiO2- or MgO-enriched liquids at the likely conditions of Earth’s basal magma ocean (\emph{Millot et al.} 2015; \emph{McCoy et al.} 2019), but it may be achievable if the magma ocean is enriched in iron, which is expected due to the relative incompatibility of Fe in silicate melt (\emph{Holmstrom et al.} 2018; \emph{Stixrude et al.} 2020). Basal magma oceans for the Moon and Venus have been hypothesized (\emph{Scheinberg et al.} 2018; \emph{O’Rourke} 2020). Io’s induced magnetic field may indicate the presence of a subsurface magma ocean (\emph{Khurana et al.} 2011). Alternatively, it could be due to “sponge” magma patches (\emph{McEwen et al.} 2019). Generation of a dynamo in magma oceans or basal magma oceans in super-Earths is possible or even more likely because electrical conductivity increases under high pressure, which makes it easier to reach the required  magnetic Reynolds number for dynamo action.

In the absence of an in-situ probe to Io, the Moon provides the most accessible record for a deep magma ocean in the Solar System. The strongest evidence for an ancient deep magma ocean is the anorthositic crust on the Moon’s surface with a thickness of  $\sim$30–40 km (\emph{Wieczorek et al.} 2013). When the lunar magma ocean crystallized, olivine and pyroxene would have sunk because they were denser than the melt, while anorthosite floated at the top of the ocean because of its low density (\emph{Smith et al.} 1970; \emph{Wood et al.} 1970). To explain the anorthositic crustal thickness, the initial magma ocean depth is estimated to be $\sim$1000 km (\emph{Elkins-Tanton et al.} 2011), but this may be challenging to reconcile because the Moon’s accretion would have occurred very quickly (10s–100 years, \emph{Thompson \& Stevenson} 1988; \emph{Salmon \& Canup} 2012; \emph{Lock et al.} 2018) and it appears challenging to avoid a deeper magma ocean. Crystallization progressed at the top and bottom of the mantle and a melt layer existed in between.  The melt layer became enriched in incompatible elements. The layer is considered to be the source of suits of rocks that are enriched in KREEP materials (K for potassium, REE for rare-earth elements, and P for phosphorus). The crystallization ages of lunar rocks are within $\sim$200 Myr after Solar System formation and therefore this is likely the timescale for the lunar magma ocean crystallization. This prolonged crystallization process could have been explained by tides due to Earth in the Moon’s crust (\emph{Meyer et al.} 2010) or magma ocean (\emph{Chen \& Nimmo} 2016) while these processes may extend the lifetime of a magma ocean by up to 10s of millions of years. Alternatively, a low thermal conductivity of the crust could explain the 200 Myr timescale (\emph{Maurice et al.} 2020).

If not during primary accretion, the Earth would have formed a deep magma ocean after the Moon-forming impact (\emph{Solomatov \& Stevenson} 1993; \textit{Nakajima \& Stevenson} 2015), but there is not as clear a geochemical signature for a magma ocean on Earth as there is for the Moon. Earth would not have generated a floating crust because such a crust would likely form only on small and dry planets; on a large planet, plagioclase becomes stable only after the magma ocean solidification is nearly complete and the crystallization front is near the planetary surface; but such a high crystal fraction prevents movement of plagioclase, failing to form a floatation crust. Moreover, water delays plagioclase crystallization, which further supports the idea of no floatation crust on Earth (\emph{Elkins-Tanton} 2012). With an oxidized mantle Earth would have had a thick outgassed CO/CO$_{2}$-H$_{2}$O atmosphere (\emph{Pahlevan et al.} 2019; \emph{Bower et al.} 2022), which would have kept the surface (partially) molten for a prolonged time. Mantle convection in the Earth over time could have erased a large portion of the primordial evidence of a magma ocean. Proposed evidence includes abundances of siderophile elements in Earth’s mantle reflecting the depth of the last magma ocean (\S\ref{sec:2.2.1}),  large low shear velocity provinces (LLSVPs) reflecting solidification and differentiation of a basal magma ocean (\emph{Garnero et al.} 2016), Fe and W isotopic ratios in 3.7-Ga metabasalts that may reflect the magma ocean cumulate (\emph{Williams et al.} 2021), and He/Ne isotopic abundances that reflect mantle ingassing of a nebular H-He component (\emph{Tucker \& Mukhopadhyay} 2014; \emph{Williams \& Mukhopadhyay} 2018).

\subsection{\textbf{Transition to long-term evolution}} \label{sec:transition_to_long-term_evolution} \label{sec:3.3}

The thermal and compositional evolution of the magma ocean stage after planetary formation establishes the initial conditions for the long-term evolution of rocky planets and the build-up of their atmospheres. While the cooling planet is still (mostly) molten, the forming core, liquid mantle, and overlying atmosphere form an interconnected network of subsystems that rapidly equilibrate thermally and compositionally, which is ultimately driven by gravitational and chemical potential energy. With progressing solidification, the timescale for thermal and chemical exchange between these subsystems increases by orders of magnitude. Because the mantle sits in-between the core and atmosphere, its evolution crucially governs both core cooling and atmosphere formation by outgassing. We will now in turn describe the main processes that affect these subsystems, and their combined effect on long-term planetary evolution.

\subsubsection{\textbf{Core}} \label{sec:core} \label{sec:3.3.1}

Seismic observations show that Earth has a solid inner core and liquid outer core, both of which are less dense than pure iron, indicating the presence of light elements, such as H, C, O, Si, and S, but their proportions are unknown (\emph{Hirose et al.} 2013; \emph{Tagawa et al.} 2021). Mercury also has an inner core (\emph{Genova et al.} 2019), but whether Venus (\emph{Margot et al.} 2021) or Mars (\emph{Stähler et al.} 2021) have an inner core remains unclear. \textit{Boujibar et al.} (2020) and \textit{Bonati et al.} (2021) proposed that massive super-Earths are likely to have liquid outer cores and solid inner cores because the large range of CMB temperature leads to the presence of an inner core. The buoyancy forces that facilitate core convection are caused by thermal effects (cooling of the core, radiogenic heating, and latent heat release by core nucleation) as well as compositional effects (light element release by core nucleation). The amount and species of light elements are determined by the recurring metal-silicate equilibration during core formation and thereafter. The presence of light elements in the core also affects the relationship between the core adiabat and melt curve. This determines the inner core-outer core boundary as well as locations of precipitation of iron (iron snow).  Iron snow is proposed for planetary objects such as Mercury, Earth, and Ganymede, and can facilitate core convection and dynamo action (e.g., \emph{Hauck et al.} 2006; \emph{Chen et al.} 2008; \emph{Zhang et al.} 2019; \emph{Breuer et al.} 2015). As discussed in \S\ref{sec:3.2.3}, to generate a dynamo, the magnetic Reynolds number has to exceed a few 10s, meaning that the convective velocity, length scale, and the electrical conductivity need to be large enough. For the convective velocity to be large, the core needs to cool quickly. The cooling rate of the core is often dictated by that of the mantle. Measurements of the thermal conductivity of the Earth’s core indicate that inner core crystallization started between $\sim$0.5–2 Gyr ago (\emph{Williams} 2018). Given the evidence for an ancient magnetic field during the early Archean and Hadean (\emph{Tarduno et al.} 2015), the geodynamo before inner core nucleation must have been driven mainly by secular cooling (\emph{Nimmo} 2015). This requires initial core temperatures $\sim$600–1600 K higher than today, which provides support for a molten lower mantle after magma ocean crystallization (\emph{Labrosse et al.} 2007; \emph{Ulvrova et al.} 2012). Because core cooling and surface heat flux are directly coupled, the tectonic mode of rocky planets (next section) exerts an immediate control on dynamo activity in the core (\emph{Olson} 2016).

\subsubsection{\textbf{Mantle}} \label{sec:mantle} \label{sec:3.3.2}

\textit{Transition From Liquid to Solid State.} As introduced in \S\ref{sec:2}, liquid magma and solid rock differ qualitatively in their viscosity and hence the timescale for fluid motion. Rocks behave as a viscoelastic fluid on geological timescales, displaying characteristics of both a fluid and a solid. The rocky mantle can thus convect and flow on long timescales, but at the same time break due to mechanical forces. Convection is driven by thermal and chemical buoyancy, which is why the compositional evolution during magma ocean crystallization plays a decisive role in what type of long-term convective regime a rocky planet develops. As discussed in \S\ref{sec:2}, fractional magma ocean crystallization may lead to whole-mantle overturn  and a stable density stratification upon mantle solidification. As a standard model for Earth, this creates the problem that the typically produced density stratification ($>$600 kg m$^{-3}$, \emph{Korenaga} 2021) is so strong that it cannot easily be overcome by thermal buoyancy, which would inhibit mantle convection and thus prevent tectonic activity. Recent works have thus turned their attention to the intermediate phase of magma ocean crystallization, when the rheological front traverses through the mantle, and solid state convection initiates while the upper mantle is still largely liquid. \textit{Maurice et al.} (2017), \textit{Ballmer et al.} (2017), and \textit{Boukaré et al.} (2018) suggest that small-scale solid-state convection may efficiently remix the mantle while the crystallization front moves upward, hence preventing global overturn and stable density stratification. In addition, volatiles that are partitioned in the melt in the form of bubbles may be efficiently trapped in solidifying patches, which can affect the amount of, for instance, H$_{2}$O and CO$_{2}$ that is outgassed to form the protoatmosphere (\emph{Hier-Majumder \& Hirschmann} 2017). Phase state and bubble formation in the magma ocean are strong functions of local thermodynamic properties and composition (\emph{Solomatova \& Caracas} 2021).

\textit{Tectonic Mode.} Long-term evolution of the planetary mantle is governed by the vigor of internal convection and hence sensitively affected by the mechanics of its stiff uppermost layer, the lithosphere, which ultimately regulates how heat can escape to the atmosphere and to space. Comparing planetary bodies in the Solar System shows that Earth is unique in this aspect: it is the only planet we know that exhibits an end-member state of mobile-lid convection, the surface expression of which is plate tectonics. All other planetary objects in the Solar System more or less fall into the category of stagnant-lid tectonics (\emph{Stern et al.} 2018), which is defined by (near-)zero surface velocity. These two are the most prominent end-members of tectonic modes, and discussion in the community often depicts these as two distinct regimes that define planets throughout their evolution. However, an immobile surface on stagnant-lid planets does not automatically equal a geologically dead planet – volcanic activity (like on Mars) for instance can transform the surface. Furthermore, transitions between tectonic modes in both time and space seem likely (\emph{Lenardic} 2018; \emph{Stern} 2018; Gerya 2014, 2019a; \emph{Brown et al.} 2020), and increasing efforts of the geodynamic community that go beyond 1-D modeling approaches suggest a flurry of intermediate regimes that are governed by varying timescales of mantle convection and surface expressions (\emph{Noack et al.} 2012; \emph{Foley et al.} 2014).

\begin{figure}[tb!]
 \plotone{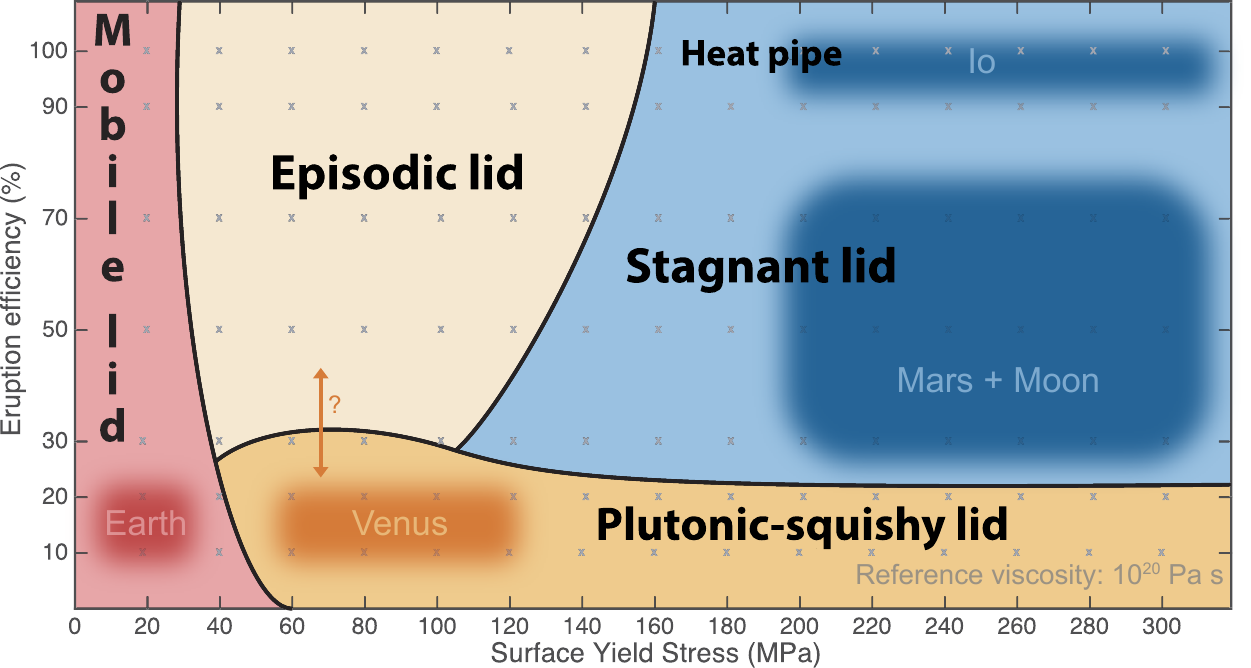}
 \caption{\small Geodynamic regimes of rocky planetary bodies from global mantle convection models. Regime limits are derived for an Earth-like planet with varying eruption efficiency, the fraction of extrusive versus intrusive volcanism, and surface yield stress, a measure of the strength of the lithosphere, the stiff upper lid of the mantle. Modified from \textit{Lourenço et al.} (2020), original figure © 2020. American Geophysical Union. All Rights Reserved. \textsc{References}: Earth: \textit{Crisp} (1984); Venus/Early Earth: \textit{Gerya} (2014), \textit{Byrne et al.} (2021), \textit{Rozel et al.} (2017), \textit{Armann \& Tackley} (2012).}
 \label{fig9}
\end{figure}
Fig.~\ref{fig9} (\emph{Lourenço et al.} 2020) shows a regime diagram of tectonic modes in a 2-D parameter space, varying the fraction of extrusive volcanism (magma erupts on the surface) to intrusive volcanism (magma is emplaced beneath the lithosphere) versus the yield stress, a measure for the resistance of the lithosphere against mechanical failure. Mobile-lid tectonics is characterized by relative surface motion between lithospheric plates that are continuously recycled back into the mantle by subduction.  The thin lithosphere is broken apart in a number of plates, which are defined by a rigid interior and converge or diverge from each other in narrow zones of active deformation, the plate boundaries. This is sometimes called horizontal tectonics because the subducting plates slide on top of each other. In the Solar System, only present-day Earth operates in this tectonic regime. Stagnant-lid tectonics, on the other hand, is characterized by near-zero relative surface motions and a global, thick lid on top of the mantle. Stagnant-lid planets cool slower than mobile-lid ones because their heat loss is limited by conduction through the lithosphere, which is inefficient in comparison with solid-state convection on mobile-lid planets that can efficiently transport warm material close to the surface. If the stiffness of the lithosphere remains high and volcanic activity is mostly extrusive, planets are said to operate in a heat-pipe or vertical tectonics mode. Jupiter’s moon Io is the best-known example of this kind (\emph{Van Hoolst et al.} 2020; \emph{Spencer et al.} 2020). In-between stagnant and mobile-lid planets numerical studies suggest that the lithosphere would thicken over time until stresses in the lithosphere build up to its yield stress (Fig.~\ref{fig9}), leading to mechanical failure and foundering of the uppermost crust in global overturns of the lithosphere and mantle (episodic lid tectonics). These events are geologically rapid ($<10^6$ yr) but can repeatedly drive global magmatic episodes and rejuvenate surface crust. A relatively recent development is a more constant rejuvenation of the surface of early Earth (relevant for the Hadean and Archean eons) in the plume-lid or squishy-lid regime (\emph{Sizova et al.} 2010; \emph{Fischer \& Gerya} 2016; \emph{Rozel et al.} 2017): due to high intrusion efficiency, magma that is emplaced beneath the lithosphere weakens it. This enables constant reworking of the crust due to return flows in the upper mantle and recycling that enables a thin lithosphere (\emph{Lourenço et al.} 2018, 2020). 

\textit{Geodynamic Diversity.} As one of the profound milestones of 20th century Earth Science, the theory of plate tectonics was able to explain a wealth of geologic evidence from Earth’s surface (\emph{Palin et al.} 2020; \emph{Brown et al.} 2020). While sharing some characteristics, no other known planet exhibits the characteristics of modern Earth, likely not even early Earth. The tectonic mode has long been a major discussion in the geodynamic and geological communities. Views on when plate tectonics emerged during our planet’s past diverge widely, ranging from right after magma ocean crystallization (uniformitarianism; \emph{Harrison} 2020; \textit{Korenaga} 2013, 2021) to as late as $\sim$850 Myr ago (\emph{Hamilton} 2011; \emph{Stern} 2018), closely coinciding with the last global glaciation event.  Pre-plate tectonics Earth (and other young rocky planets) may transition from  single, localized subduction events or alternatively follow an evolutionary trajectory from a pre-plate tectonics plume-lid regime to the present-day tectonic mode (\emph{Lenardic} 2018; \textit{Gerya} 2014, 2019a; \emph{Foley et al.} 2014). Regime transitions are actively debated and difficult to precisely define because geological markers in the past are ambiguous. Ancient zircon crystals provide the best evidence for the Hadean eon ($>$4 Gyr in the past, \S\ref{sec:3.2.1}, Fig.~\ref{fig7}), but the oldest analyzed zircons that sample the time $>$4 Ga are dominantly sourced from only one region, the Jack Hills in Western Australia. Global geologic campaigns to sample a wider region of Earth’s surface are necessary to provide further constraints on tectonic regime transitions (\emph{Harrison} 2020).

Additional information may come from other planets, such as Venus. The young surface age of Venus can be explained by an intermediate tectonic mode, either by multiple global overturns in an episodic lid (\emph{Nimmo \& McKenzie} 1998), refreshment of the crust by ongoing plumes of warm material from the core-mantle boundary (\emph{Gülcher et al.} 2020), or lithospheric deformation analogous to a squishy-lid Archean Earth (\emph{Byrne et al.} 2021). Mars, Mercury, and the Moon are examples of stagnant-lid planetary bodies (\emph{Tosi \& Padovan} 2021). While still in its infancy, additional information on possible geodynamic regimes may come from exoplanets. Theoretically it has been debated whether super-Earth exoplanets should be more or less prone to exhibit mobile-lid convection than Earth (\emph{Valencia et al.} 2007; \emph{Valencia \& O’Connell} 2009; \emph{O’Neill \& Lenardic} 2007; \emph{Korenaga} 2010; \emph{Foley et al.} 2012; \emph{Stamenković \& Breuer} 2014;  \emph{Tackley et al.} 2013), but observationally-grounded conclusions remain elusive. Recently, however, detailed phase curve characterizations (\emph{Kreidberg et al.} 2019) have enabled studies of the interior convection of individual, short-period exoplanets (\emph{Meier et al.} 2021), which suggest hemispherically-split geodynamic regimes that are unknown from the Solar System and expand the previously known modes of tectonics.  

\subsubsection{\textbf{Atmosphere}} \label{sec:atmosphere} \label{sec:3.3.3}

\textit{Volatile Sources and Sinks.} Atmosphere formation on rocky planets has long been framed in a rather sharp succession of primary vs. secondary atmospheres, but the expansion of the theoretical context into the realm of exoplanets points toward a smoother transition between initial and long-term volatile envelopes. Protoplanets acquire parts of their atmosphere during the disk phase (\S\ref{sec:3.1}), which then, depending on the stellar XUV environment and melt regime of the planet, is replaced over time by volatiles that are outgassed from the mantle. The relative fraction of these two primary volatile sources over time for rocky protoplanets is unclear, specifically since the results of the Kepler space telescope provide strong evidence for a transition from hydrogen-rich to hydrogen-poor worlds among sub-Neptune-sized planets (\emph{Owen} 2020; \emph{Bean et al.} 2021). If the nebular-sourced (primary), hydrogen-rich protoatmosphere is minor or gets lost during the giant impact phase after the disk stage, the atmosphere is re-established through outgassing from the magma ocean, volcanism, and impact degassing (Fig.~\ref{fig1}, \S\ref{sec:3.2.3}). Build-up and composition of this (often called secondary) atmosphere is governed by the amount of high mean molecular weight volatile compounds (such as CO$_{2}$) that are delivered during planetary growth by various mechanisms, but is sensitively affected by the redox state of the interior and the degassing (surface) pressure (\S\ref{sec:2.2}, Fig.~\ref{fig3}). In the case of an oxidized interior – as expected for Venus, Earth, and similar-sized exoplanets – the dominant volatiles that are outgassed from the interior are CO$_{2}$ and H$_{2}$O (\emph{Gaillard et al.} 2021). Atmosphere buildup in an oxidized environment is then governed by the relative solubilities of these outgassed compounds in the magma: CO$_{2}$ is less soluble than H$_{2}$O, hence the atmosphere is initially carbon-rich (\emph{Sossi et al.} 2020; \emph{Solomatova \& Caracas} 2021). H$_{2}$O outgasses only when most of the mantle is solidified (\emph{Zahnle et al.} 2010), but perhaps most of it remains stuck in the interior (\emph{Hier-Majumder \& Hirschmann} 2017). For more reduced planets the post-magma ocean atmosphere would be rich in hydrogen compounds (\emph{Schaefer \& Fegley} 2017; \emph{Gaillard et al.} 2021, 2022).

For an Earth-like initial volatile content, oxidation state, and instellation, the geothermal heat flux can sustain high surface temperatures only for $10^4$ to $10^5$ yr. After this phase the atmospheric structure and surface heat flux is primarily governed by the equilibrium between incident and outgoing thermal energy. Because the lower atmosphere in this setting is heated from below and by starshine, it is convective, undersaturated in water, and resides on a dry adiabat (\emph{Kasting et al.} 1988; \emph{Pierrehumbert} 2010; \emph{Catling \& Kasting} 2017). At higher altitudes, however, water can condense and rain out toward lower altitudes. Because at such high temperatures water vapor is highly opaque, the heat flux of the protoatmosphere (often called the outgoing longwave radiation, OLR, even though the atmosphere at high surface temperatures also emits significantly in shortwave bands, \emph{Boukrouche et al.} 2021) is fixed to the flux that can escape through the highest regions where water vapor can condense (\emph{Goldblatt et al.} 2013; \emph{Leconte et al.} 2013). Once the surface temperature cools down enough, condensation at the surface enables the build-up of the earliest water oceans on Earth (\emph{Elkins-Tanton} 2011). This usually happens while the interior is still partially molten and high volcanic activity may govern the surface environment. Multiple and cyclic remelting of the surface can be expected, but so far no models treating the complex array of physical and chemical processes of this stage have been developed. Once the surface solidifies, atmospheric CO$_{2}$ is thought to remain in the tens to hundreds of bar (\emph{Zahnle et al.} 2010). Efficient Rayleigh scattering in such a climate would rapidly cool down the atmosphere (\emph{Kasting} 1993) such that surface weathering (see below) may set in, which possibly enabled the earliest hospitable climate on Earth (\emph{Sleep et al.} 2001; \emph{Sleep \& Zahnle} 2001).

\textit{Climate Evolution.} The climate of the early Earth has for long been a puzzle because the bolometric luminosity of G-type stars increases by about 8\% per Gyr. During the first $\sim$Gyr after planetary formation, the insolation of the planet is low – too low for the surface temperature to reside above the freezing point of water with a greenhouse effect comparable to today’s. However, ancient zircons provide evidence for the existence of large amounts of liquid water on the Earth’s surface as early as $\sim$4.4 Ga (\emph{Wilde et al.} 2001; \emph{Mojzsis et al.} 2001), and throughout the Hadean and Archean (\emph{Harrison} 2020). This Faint Young Sun problem (\emph{Pierrehumbert} 2010; \emph{Catling \& Kasting} 2017) is exacerbated by the anticipation that early CO$_{2}$ outgassed during the magma ocean epoch should rapidly react with the early crust and be incorporated into the upper mantle (\emph{Sleep et al.} 2001). Possible solutions to enhance the surface temperature and buffer the low insolation, other than CO$_{2}$ (\emph{Charnay et al.} 2020), include NH$_{3}$ (\emph{Sagan \& Mullen} 1972), CH$_{4}$ (\emph{Pavlov et al.} 2000), H$_{2}$ (\emph{Wordsworth \& Pierrehumbert} 2013a), or variations in cloud (\emph{Rosing et al.} 2010; \emph{Goldblatt et al.} 2021) and organic haze cover (\emph{Wolf \& Toon} 2010).

On Earth, geological and geochemical proxies provide evidence for long-term stability of the surface temperature in the range of $\sim$0–50°C during and after the Archean (\emph{Catling \& Zahnle} 2020), interrupted by a few globally glaciated snowball episodes (\emph{Pierrehumbert et al.} 2011; \emph{Kirschvink et al.} 2017). The most popular theory to explain this evidence in the face of a continuously brightening Sun is the carbonate-silicate cycle (\emph{Walker et al.} 1981; \emph{Sleep \& Zahnle} 2001; \emph{Catling \& Kasting} 2003): weathering of calcium and magnesium silicates in rocks and soils release ions, such as HCO$_3^-$ and CO$_3^{2-}$, that are transported to the seafloor via runoff and precipitation (\emph{Graham \& Pierrehumbert} 2020; \emph{Hakim et al.} 2021), where they are subducted into the mantle and hence (temporarily) removed from the atmosphere. Outgassing from volcanoes at mid-ocean ridges, volcanic arcs, and hotspots (such as Hawai’i) releases CO$_{2}$ back to the atmosphere over geologic time. Because weathering reactions are temperature-dependent (higher temperatures increase weathering and vice versa), this geochemical cycle can provide a negative feedback that regulates the surface temperature via the CO$_{2}$ partial pressure in the atmosphere. The dependence of this mechanism on subduction of sediments highlights its sensitivity on the tectonic mode (as discussed above). The effectiveness of the carbonate-silicate thermostat for stagnant-lid planets (\emph{Tosi et al.} 2017; \emph{Foley \& Smye} 2018; \emph{Dorn et al.} 2018; \emph{Höning et al.} 2019), and varying land-ocean and water-mass fractions (\emph{Abbot et al.} 2012; \emph{Cowan et al.} 2014; \emph{Schaefer \& Sasselov} 2015; \emph{Noack et al.} 2016; \emph{Kite \& Ford} 2018; \emph{Hayworth \& Foley} 2020) for approximately Earth-like planets are actively debated.

An important driver of long-term climate is the global redox evolution of rocky planets (\emph{Wordsworth et al.} 2018; \emph{Gaillard et al.} 2021), which sets the boundary conditions of pathways to organic synthesis and the origin of life on the surface of terrestrial worlds (\emph{Sasselov et al.} 2020; \textit{Krissansen-Totton et al.} 2022). The redox state of the surface is hence an important determinant of whether prebiotic chemistry can operate, and thus potentially distinguish rocky planets that may or may not develop life. The direct connection of the origins of life with the redox state of planetary mantles and their atmospheres therefore links observational signatures from exoplanet surveys with the geophysical and climatic conditions of the prebiotic environment of the earliest Earth.

From a climatic point of view, hydrogen loss by atmospheric escape can be considered the most important effect that oxidizes rocky planets over time (\emph{Catling et al.} 2001; \emph{Catling \& Zahnle} 2020). However, geochemical proxies indicate that the Earth’s crust has been relatively oxidized since the onset of the rock record and possibly earlier (\emph{Delano} 2001; \emph{Trail et al.} 2011). The geochemically favored explanation for this conundrum is the disproportionation of ferrous iron in the solid mantle (\emph{Frost et al.} 2004; \emph{Wade \& Wood} 2005) and possibly magma ocean phase (\emph{Armstrong et al.} 2019), as reviewed in \S\ref{sec:3.2.3}. The pressure dependence of this mechanism suggests that rocky planets of Earth’s size should feature oxidized mantles (irrespective of atmospheric escape), which favor volcanic outgassing of CO$_{2}$ and H$_{2}$O. Mars- or Mercury-sized planetary mantles may thus circumvent this fate and more closely resemble their originally accreted mantle composition. Mars’ climate history has been increasingly constrained from in-situ exploration missions and is reviewed in-depth elsewhere (\emph{Wordsworth} 2016; \emph{Kite} 2019; \emph{McLennan et al.} 2019). Super-Earth interiors may either be more (\emph{Kite \& Schaefer} 2021) or less (\emph{Lichtenberg} 2021) prone to become oxidized by internal redox reactions. Extended magma ocean phases due to greenhouse forcing on short-period exoplanets promote inflated steam atmospheres (\emph{Turbet et al.} 2019; \emph{Mousis et al.} 2020), and may hide a substantial amount of their bulk water budget in the molten mantle (\emph{Dorn \& Lichtenberg} 2021).

\begin{figure}[tb!]
 \plotone{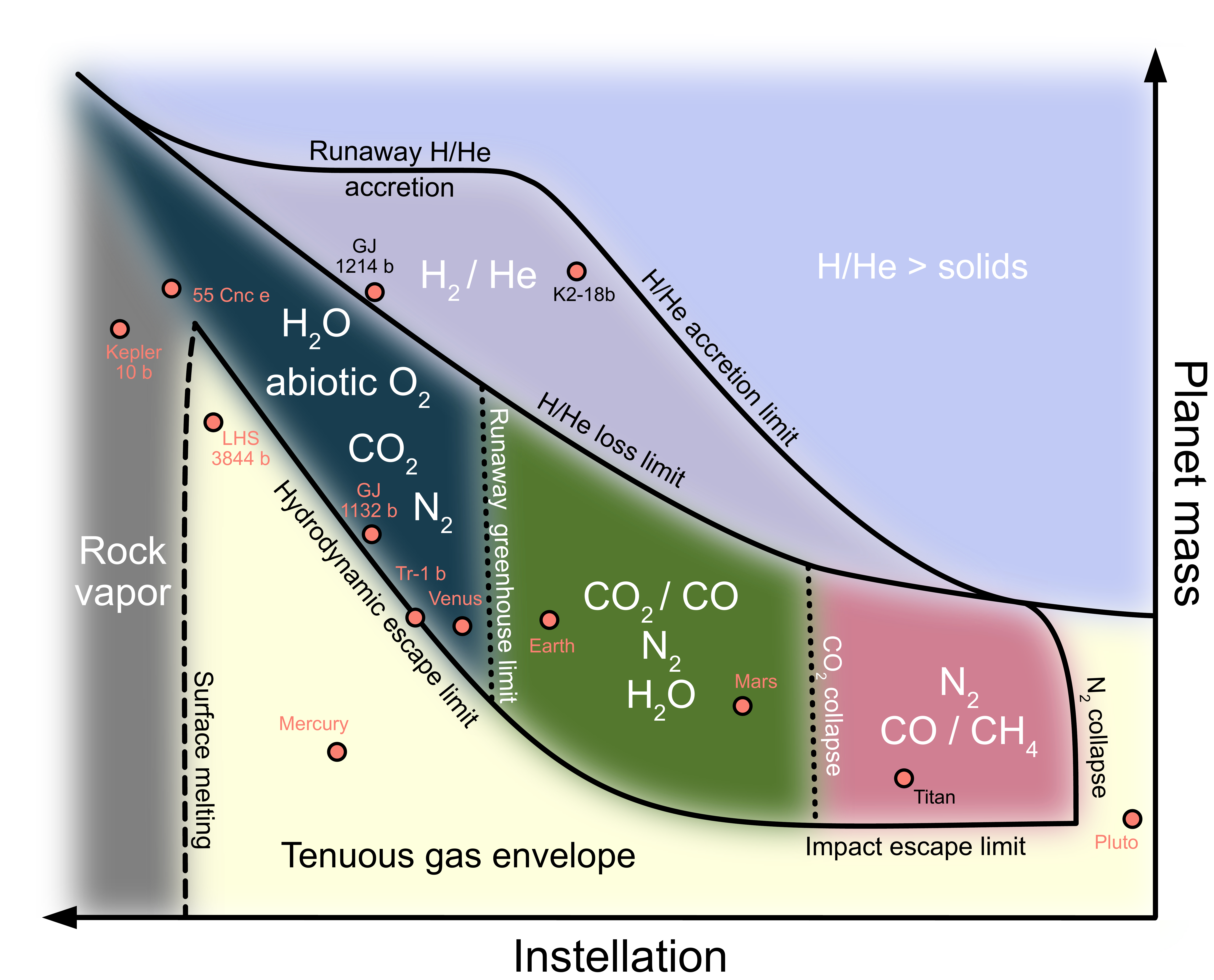}
 \caption{\small Plausible climate regimes of rocky planets as a function of planetary mass and instellation. Only the expected dominant atmospheric species are indicated. Changes in composition and redox state relative to the Solar System would alter the qualitative domains of this plot. Regime boundaries are meant to be illustrative and are influenced by a variety of additional parameters. Some individual planetary objects are shown for reference. Extended from \textit{Forget \& Leconte} (2014). \href{https://osf.io/e326m/}{\includegraphics[scale=0.35]{icon_download.pdf}}}
 \label{fig10}
\end{figure}
Fig.~\ref{fig10} illustrates a qualitative view of possible climate regimes for rocky planets in a 2-D planet mass versus instellation plane. Rocky planets with substantial high mean molecular weight atmospheres (dark blue, green, pink) are bounded by various accretion and loss limits. Above the H/He accretion limit, these gasses make up a significant fraction of the total mass of the planet. Above the H/He loss limit gas accreted during the disk phase can be retained over stellar main-sequence lifetimes (sub-Neptunes). Below the hydrodynamic escape limit (light yellow area) planets are too small to retain any significant gas envelope, which they can also lose from impacts (impact escape limit). Left of the surface melting line the instellation is high enough to melt the planetary surface even in the absence of an atmosphere. If rocky planets end up in a regime where they lose their primordial H/He envelope, but can retain (or outgas) a significant secondary atmosphere, they end up in one of the middle regimes. 

Above a  certain critical flux the radiative feedback effects of water drive the planet into the runaway greenhouse state (or keep it there, \S\ref{sec:3.2.3}). Water in this stage can be photolyzed in the upper atmosphere, which can build up substantial quantities of O$_2$ during hydrogen loss (\emph{Wordsworth \& Pierrehumbert} 2013b; \emph{Luger \& Barnes} 2015, \textit{Schaefer et al.} 2016). Beyond the runaway greenhouse threshold, planets can enter a clement climate, water can condense at the surface, and CO$_{2}$ recycling by weathering and subduction may buffer the surface temperature over long time scales, depending on a number of factors. At lower instellation CO$_{2}$ gas will freeze out on the surface (green area) and CO or CH$_{4}$ become dominant, as does N$_{2}$ for even colder climates (pink), at which point atmospheres become very thin. In the past few years exoplanet observations have revealed a gap in planet occurrence rate for planet sizes between 1.5–2.0 Earth radii (\textit{Lissauer et al.}, \textit{Weiss et al.}, this volume), likely related to loss of their primary envelope (\emph{Bean et al.} 2021; H/He loss limit in Fig.~\ref{fig10}). An important factor in this context is the variable evolution of stellar irradiation with time for different stellar masses: after their initial accretion luminosity burst, Sun-like stars continuously brighten, such that the runaway greenhouse threshold moves outward with time. M stars, on the other hand, feature a bright early phase that dims over time, such that the potential zone for surface liquid water moves inward (\emph{Lissauer} 2007; \emph{Luger \& Barnes} 2015). Progressive characterization of exoplanet diversity and discovery of further such thresholds will refine and significantly alter the potential classes presented in Fig.~\ref{fig10}.

%%%%%%%%%%%%%%%%%%%%%%%%%%%%%%%%%%%%%%%%%%%%%%%%%%%%%%%%%%%%%%%%%%%%%%%%%%%%%%%%%%%%%%%%%%%%%%%%%%%%%%%%%%%%%%%%%%%%%%%%%%%%%%%%%%%%
%%%%%%%%%%%%%%%%%%%%%%%%%%%%%%%%%%%%%%%%%%%%%%%%%%%%%%%%%%%%%%%%%%%%%%%%%%%%%%%%%%%%%%%%%%%%%%%%%%%%%%%%%%%%%%%%%%%%%%%%%%%%%%%%%%%%
%%%%%%%%%%%%%%%%%%%%%%%%%%%%%%%%%%%%%%%%%%%%%%%%%%%%%%%%%%%%%%%%%%%%%%%%%%%%%%%%%%%%%%%%%%%%%%%%%%%%%%%%%%%%%%%%%%%%%%%%%%%%%%%%%%%%
\section{\textbf{OUTLOOK \& SUMMARY}} \label{sec:outlook_and_summary} \label{sec:4}
\subsection{\textbf{Mid- to long-term prospects}} \label{sec:Mid_to_long-term_prospects} \label{sec:4.1}

The previous decade has seen unparalleled advances in detection and progressive characterization of extrasolar planetary systems owing to ground- and space-based observing programs of disks and exoplanets (\textit{Jontof-Hutter} 2019; \emph{Öberg \& Bergin} 2021), and we anticipate an accelerated pace of population statistics and more detailed insights into individual systems and planet properties in the 2020s.

Starting from our cosmic home, ESA’s Juice mission to the Jovian moons will help to clarify distinctions between dominantly rocky versus dominantly icy planetary bodies. This is crucial in the context of extrasolar planets, many of which may be composed of large amounts of volatile ices (\emph{Zeng et al.} 2019; \emph{Venturini et al.} 2020). Dragonfly will investigate the atmospheric conditions and surface of a planetary body dominated by organic haze layers, which may be a good analogue for the Archean Earth (\emph{Arney et al.} 2016; \emph{Krissansen-Totton et al.} 2018) and or super-Earths (\textit{Rimmer et al.} 2021, \textit{Lichtenberg} 2021). Psyche will aim to investigate how iron-enriched asteroids form and together with the Lucy mission reveal new insights into the relationship between meteorites and the physical processes of core formation and chemical segregation in rocky planets (\emph{Elkins-Tanton et al.} 2020). BepiColombo will study the surface and composition of Mercury, revealing crucial insights into the most reduced terrestrial planet of the Solar System (\emph{Rother et al.} 2020; \emph{Genova et al.} 2021). At the time of writing, the InSight mission is starting to reveal the interior structure of Mars using seismic measurements (\emph{Knapmeyer-Endrun et al.} 2021; \emph{Khan et al.} 2021; \emph{Stähler et al.} 2021). Many more observations to other Solar System objects and sample return missions from Mars, the Moon, and asteroids, will help us to refine existing theories on the origin and evolution of the inner Solar System planets, with a level of detail that will be unachievable for extrasolar systems within our lifetime, and put our theoretical understanding of rocky planetary processes on firmer grounding.

Observations of circumstellar disks with ALMA and JWST will produce improved insights into the origins of atmospheric volatiles (\emph{Öberg \& Bergin} 2021; \emph{Wordsworth \& Kreidberg} 2022), and reveal how forming giant planets can shape the redistribution of planetary components during accretion. Detections of circumplanetary disks (\emph{Benisty et al.} 2021) and debris of giant impacts in transition disks (\emph{Thompson et al.} 2019; \emph{Gáspár \& Rieke} 2020) may reveal the difference in formation paths of moons and planets, and constrain the final stages of planetary formation.

At this evolutionary stage, observations of magma ocean atmospheres of young (\emph{Lupu et al.} 2014; \emph{Hamano et al.} 2015; \emph{Bonati et al.} 2019) and old (\emph{Boukrouche et al.} 2021) exoplanets in the runaway greenhouse state would be able to open a crucial window into the abundance and fractionation pattern of atmospheric volatiles and their interaction with the planetary interior over time (\emph{Schaefer et al.} 2016; \emph{Lichtenberg et al.} 2021b), and the operation of the carbonate-silicate cycle on potentially Earth-like worlds (\emph{Ramirez et al.} 2019; \emph{Checlair et al.} 2021; \emph{Bixel \& Apai} 2021). Rocky exoplanet observations on wide, potentially temperate orbits beyond the runaway greenhouse limit, however, will require large-scale direct imaging surveys (\emph{Gaudi et al.} 2019; \emph{Wang et al.} 2019; \emph{Quanz et al.} 2021; \textit{Currie et al.}, this volume). Observations of close-in exoplanets can reveal the presence or absence of atmospheres (\emph{Selsis et al.} 2011; \emph{Koll et al.} 2019; \emph{Mansfield et al.} 2019) and the composition of surfaces on atmosphere-stripped planets (\emph{Hu et al.} 2011; \emph{Kreidberg et al.} 2019), which may reveal the distribution of oxidized (\emph{Kite \& Schaefer} 2021) or reduced (\emph{Lichtenberg} 2021) planetary surfaces and yield clues on the composition and thermal state of sub-Neptune interiors. Detailed characterization of upper-atmosphere composition may probe the presence of shallow, low-pressure surfaces (May \& \emph{Rauscher et al.} 2020; \emph{Yu et al.} 2021) or oceans (\emph{Loftus et al.} 2019; \emph{Hu et al.} 2021; \emph{Tsai et al.} 2021). Population studies on exoplanet densities are required to probe the distribution of volatile-rich planets, internal structure, and phase changes (\emph{Dorn et al.} 2015; \emph{Noack \& Lasbleis} 2020; \emph{Bonati et al.} 2021; \emph{Dorn \& Lichtenberg} 2021). Finally, observed variations in accretion disk chemistry (\emph{McClure et al.} 2020; \textit{Miotello et al.}, \textit{Manara et al.}, this volume) and of exoplanetary debris in evolved planetary systems, such as polluted white dwarfs, promise insights into how planets acquire their differentiated structure (\emph{Bonsor et al.} 2020; \emph{Hollands et al.} 2021) and composition (\emph{Farihi et al.} 2013; \emph{Doyle et al.} 2019, 2021; \emph{Bonsor et al.} 2021). While the Gyr-long evolution of white dwarf systems must be treated with caution, the direct insight into the elemental building blocks of rocky planets opens a novel complementary perspective to present-day exoplanet systems (\emph{Xu \& Bonsor} 2021).

In order to make sense of these observational opportunities, models of planetary physics and chemistry need improved data from laboratory measurements to construct physically-motivated simulations of planetary evolution that are capable of going beyond what we observe in the Solar System, and asking appropriate questions. We see exoplanets at only a snapshot of their evolution, but the population of exoplanets spans ranges of age, instellation, mass, and composition far beyond those that exist in the Solar System: new data will be needed to push models into these parameter spaces. Geophysical models of interior structure require precise measurements of interior phase properties of core and mantle components (\emph{Wicks et al.} 2018; \emph{Miozzi et al.} 2018; \emph{Brugman et al.} 2021; \emph{Militzer et al.} 2021), such as electrical and thermal conductivity (\emph{McWilliams et al.} 2012; \emph{Millot et al.} 2015; \emph{Kislyakova et al.} 2017, 2018), mixing of water and silicates (\emph{Vazan et al.} 2022; \textit{Dorn \& Lichtenberg} 2021), and latent heat of vaporization (\emph{Stewart et al.} 2020; \emph{Davies et al.} 2020). Because the atmosphere is the observational window into the interiors of planets, models of outgassing and cycling of volatiles require better constraints on volatile partitioning and opacities in non-terrestrial environments (\textit{Schaefer \& Fegley Jr.} 2017; \textit{Vazan et al.} 2018; \textit{Fegley Jr. et al.} 2020; \emph{Gaillard et al.} 2021; \emph{Liggins et al.} 2021). In particular relevant to the transition between gas- and solid-dominated planets is the solubility of hydrogen in silicate melts (\emph{Hirschmann et al.} 2012), which is poorly constrained for pressures above a few GPa (\emph{Chachan \& Stevenson} 2018; \emph{Kite et al.} 2019, 2020; \emph{Schlichting \& Young} 2021).

From a modeling perspective, experimental constraints on pressure broadening for multi-species and non-terrestrial atmospheric compositions and pressures will be needed to build generalized climate models that connect the interior to atmospheric observations (\emph{Forget \& Leconte} 2014). Upcoming theoretical models that provide the basis for observational interpretation must move beyond boundary layer theory (effectively 0-D) and 1-D models to incorporate interior phase changes (\emph{Bower et al.} 2019), geochemical reactions (\emph{Kite \& Schaefer} 2021; \emph{Lichtenberg et al.} 2021; \emph{Schlichting \& Young} 2021), and atmospheres that are rich in condensables (\emph{Ding \& Pierrehumbert} 2016; \emph{Graham et al.} 2021; \emph{Loftus \& Wordsworth} 2021). In order to constrain compositional differentiation between metal, silicates, and volatiles, multi-phase fluid dynamical models are required (\emph{Keller \& Suckale} 2019; \textit{Gerya} 2019b). Astrophysical models of planetary growth from a system-perspective will need to be coupled to geophysical and geochemical approaches that resolve their evolution from a planet-centric perspective to understand the structural and compositional consequences of different modes of planetary accretion (\emph{Lichtenberg et al.} 2021a), such as growth dominated by pebbles (\emph{Brouwers et al.} 2018; \emph{Lambrechts et al.} 2019) or planetesimals (\emph{Rubie et al.} 2015b; \emph{Burkhardt et al.} 2021). Ultimately, spatially and temporally-resolved models  of the coupled evolution of core, mantle, and atmosphere of rocky planets are required to make sense of evolutionary changes and to connect the formation of rocky planets to their long-term evolution (\emph{Golabek et al.} 2018; \emph{Kite \& Barnett} 2020; \emph{Lichtenberg et al.} 2021b; \emph{Chao et al.} 2021; \emph{Nakajima et al.} 2021).

\subsection{\textbf{Conclusions}} \label{sec:conclusions} \label{sec:4.2}

Much of our currently perceived knowledge surrounding the formation and evolution of rocky planets is derived from the terrestrial planets of the Solar System, and overwhelmingly biased toward modern Earth. In this review we covered the primary geophysical and some geochemical aspects that drive the evolution of rocky planets from their birth to their early evolution, and which set the boundary conditions for the long-term surface and climatic setting. Rocky planets are not passive receivers of their astrophysical formation environment: their internal and atmospheric evolution in turn shape how and which materials are incorporated into the planet, and how they are chemically segregated into core, mantle, crust, potential ocean, and atmospheric layers.

The chronology of formation is of crucial importance: building a planet first from volatile-rich, then volatile-poor materials results in a different planetary structure and chemical layering than the reverse formation pathway. This ultimately leads to a diverging climate and surface environment that may or may not be conducive to originating life and enabling its long-term survival. We do not yet know if a planet like Earth – or terrestrial planets more generally – are rare or common across planetary systems. Exoplanet observations suggest that large rocky worlds with thick primary envelopes are abundant; these likely differ substantially from any known planet in the Solar System, and we do not currently possess the methodological means to ask the right questions. Detailed characterizations of close-in super-Earths and statistical population studies in the upcoming years have the potential to reveal how common atmosphere-stripped and volatile-rich rocky planets are, and thus reveal clues about the uniqueness of Earth and the frequency of habitable planets in the galaxy.

\bigskip

\noindent\textbf{Acknowledgments.}
We thank Kevin Zahnle, Francis Nimmo, Lena Noack, Robin Wordsworth, Edwin Kite, Bowen Fan, Victoria Meadows, Diogo Lourenço, and Elizabeth Cottrell for comments and suggestions that significantly improved the scope and clarity of the manuscript.  T.L. was supported by the Simons Foundation (SCOL Award No.~611576). M.N. was supported in part by the Alfred P. Sloan Foundation under grant G202114194, by the National Aeronautics and Space Administration (NASA) under Grant No. 80NSSC19K0514 and No. 80NSSC21K1184, and by the Center for Matter at Atomic Pressures (CMAP), a National Science Foundation (NSF) Physics Frontier Center, under Award PHY-2020249. R.A.F. was partially supported by NSF grant EAR-2054912 and NASA grants NNX17AE27G and 80NSSC21K0388. Any opinions, findings, conclusions or recommendations expressed in this material are those of the authors and do not necessarily reflect those of the National Science Foundation. Chapter figures available for download at \href{https://osf.io/rcjt7}{osf.io/rcjt7}.
\bigskip

\noindent\textbf{REFERENCES}
\bigskip
\parskip=0pt
{\small
\baselineskip=11pt
\refs 
Abbot, D.S. et al., 2012, {\it ApJ}, 756, 178
\refs 
Abe, Y. \& Matsui, T., 1985, {\it JGR Suppl.}, 90, C545–C559
\refs 
Abe, Y. \& Matsui, T., 1986, {\it JGR}, 91, E291–E302
\refs 
Abe, Y. \& Matsui, T., 1988, {\it J. Atmos. Sci.}, 45, 3081–3101
\refs 
Abe, Y., 1993, {\it Evolution of the Earth and Planets} (E. Takahashi, R. Jeanloz, D. Rubie, eds., Geophys. Monograph), 74, 41–54
\refs 
Abe, Y., 1997, {\it PEPI}, 100, 27–39
\refs 
Abramov, O. \& Kring, D.A., 2004, {\it JGR}, 109, E10007
\refs 
Agnor, C.B. \& Hamilton, D.P., 2006, {\it Nature}, 441, 192–194
\refs 
Agol, E. et al., 2021, {\it PSJ}, 2, 1
\refs 
Albarède, F., 2009, {\it Nature}, 461, 1227–1233
\refs 
Alexander, C.M.O.D. et al., 2018, {\it SSR}, 214, 36
\refs 
Andrews, S.M., 2020, {\it ARAA}, 58, 483–528
\refs 
Ansdell, M. et al., 2016, {\it ApJ}, 828, 46
\refs 
Arkani-Hamed, J. \& Olson, P., 2010, {\it GRL}, 37, L02201
\refs 
Armann, M. \& Tackley, P. J., 2012, {\it JGR},  117, E12003
\refs 
Armstrong, K. et al., 2019, {\it Science}, 365, 903–906
\refs 
Arney, G. et al., 2016, {\it Astrobiology},  16, 873–899
\refs 
Asphaug, E., 2010, {\it Geochemistry},  70, 199–219
\refs 
Asphaug, E. \& Reufer, A., 2014, {\it Nat. Geosci.}, 7, 564–568
\refs 
Avice, G., \& Marty, B., 2014, {\it Phil. Trans. R. Soc. A.}, 272, 20130260.
\refs 
Badro, J. et al., 2015, {\it PNAS}, 112, 12310–12314
\refs 
Badro, J. et al., 2016, {\it Nature}, 536, 326–328
\refs 
Bagdassarov, N. et al., 2009, {\it EPSL}, 288, 84–95
\refs 
Ballmer, M. D. et al., 2017, {\it GGG}, 18, 2785–2806
\refs 
Baron, M.A. et al., 2017, {\it EPSL}, 472, 186–196
\refs 
Bhatia, G.K., \& Sahijpal, S., 2016, {\it M\&PS}, 51, 138-154
\refs 
Bean, J.L. et al., 2021, {\it JGR:P}, 126, e2020JE006639
\refs 
Benedikt, M.R. et al., 2020, {\it Icarus}, 347, 113772
\refs 
Benisty, M. et al., 2021, {\it ApJL}, 916, L2
\refs 
Benítez-Llambay, P. et al., 2015, {\it Nature}, 520, 63–65
\refs 
Benz, W. et al., 2007, {\it SSR}, 132, 189–202
\refs 
Bergin, E.A. et al., 2015, {\it PNAS}, 112, 8965–8970
\refs 
Bermingham, K. R., 2018, {\it EPSL}, 487, 221–229
\refs 
Bermingham, K.R. et al., 2020, {\it SSR}, 216, 133
\refs 
Biersteker, J.B. \& Schlichting, H.E., 2019, {\it MNRAS}, 485, 4454–4463
\refs 
Birnstiel, T. et al., 2016, {\it SSR}, 205, 41–75
\refs 
Bitsch, B. et al., 2019, {\it A\&A}, 624, A109
\refs 
Bixel, A. \& Apai, D., 2021, {\it AJ},  161, 228
\refs 
Blackburn, T. et al., 2017, {\it GCA}, 200, 201–217
\refs 
Blanchard, I. et al., 2017, {\it Geochem. Perspect. Lett.}, 5, 1–5
\refs 
Bland, P.A. \& Travis, B.J., 2017, {\it Sci. Adv.}, 3, e1602514
\refs 
Boehnke, P. \& Harrison, T.M., 2016, {\it PNAS}, 113, 10802–10806
\refs 
Bolmont, E. et al., 2013, {\it A\&A}, 556, A17
\refs 
Bonati, I. et al., 2019, {\it A\&A}, 621, A125
\refs 
Bonati, I. et al., 2021, {\it JGR:P}, 126, e2020JE006724
\refs 
Bond, J.C. et al., 2010, {\it Icarus}, 205, 321–337
\refs 
Bonnand, P. \& Halliday, A.N., 2018, {\it Nat. Geosci.}, 11, 401–404
\refs 
Bonomo, A.S. et al., 2019, {\it Nat. Astron.} 3, 416–423
\refs 
Bonsor, A. et al., 2015, {\it Icarus}, 247, 291–300
\refs 
Bonsor, A. et al., 2020, {\it MNRAS}, 492, 2683–2697
\refs 
Bonsor, A. et al., 2021, {\it MNRAS}, 503, 1877–1883
\refs 
Borg, L.E. et al., 2016, {\it GCA}, 175, 150–167
\refs 
Bottke, W.F. \& Norman, M.D., 2017, {\it AREPS}, 45, 619–647
\refs 
Bottke, W.F. et al., 2010, {\it Science}, 330, 1527–1530
\refs 
Bouhifd, M.A. \& Jephcoat, A.P., 2011, {\it EPSL}, 307, 341–348
\refs 
Boujibar, A. et al., 2014, {\it EPSL} 391, 42–54
\refs 
Boujibar, A. et al., 2020, {\it JGR:P}, 125, e2019JE006124
\refs 
Boukaré, C. E. et al., 2018, {\it EPSL}, 491, 216–225
\refs 
Boukaré, C.E. et al., 2018, {\it EPSL} 491, 216–225
\refs 
Boukrouche, R. et al., 2021, {\it ApJ}, 919, 130
\refs 
Bouvier, L.C. et al., 2018, {\it Nature}, 558, 586–589
\refs 
Bower, D.J. et al., 2018, {\it PEPI}, 274, 49–62
\refs 
Bower, D.J. et al., 2019, {\it A\&A}, 631, A103
\refs 
Bower, D.J. et al., 2022, {\it PSJ}, {\it arXiv}:2110.08029
\refs 
Boyet, M. \& Carlson, R.W., 2005, {\it Science}, 309, 576–581
\refs 
Brasser, R. et al., 2016, {\it EPSL} 455, 85–93
\refs 
Brasser, R. et al., 2018, {\it GRL}, 45, 5908–5917
\refs 
Braukmüller. N. et al., 2019, {\it Nat. Geosci.}, 12, 564–568
\refs 
Brennan, M.C. et al., 2020, {\it EPSL}, 530, 115923
\refs 
Breton, T. et al., 2015, {\it EPSL}, 425, 193–203
\refs 
Breuer, D. \& Moore, B., 2015, {\it Phys. Terr. Planets Moons},  10, 255–305
\refs 
Breuer, D. et al., 2015, {\it Prog. Earth Planet. Sci.}, 2, 39 
\refs 
Broadley, M.W. et al., 2020a, {\it GCA}, 270, 325–337
\refs 
Broadley, M.W. et al., 2020b, {\it PNAS}, 117, 13997–14004
\refs 
Bromley, B.C. \& Kenyon, S.J., 2019, {\it ApJ}, 876, 17
\refs 
Brouwers, M.G. et al., 2018, {\it A\&A}, 611, A65
\refs 
Brown, M. et al., 2020, {\it AREPS}, 48, 291–320
\refs 
Brugman, K.K. et al., 2021, {\it JGR:P}, e2020JE006731
\refs 
Brygoo, S. et al., 2021, {\it Nature}, 593, 517–521
\refs 
Bryson, J.F.J. \& Brennecka, G.A., {\it ApJ}, 912, 163
\refs 
Budde, G. et al., 2019, {\it Nat. Astron.}, 3, 736–741
\refs 
Burkhardt, C. et al., 2016, {\it Nature}, 537, 394–398
\refs 
Burkhardt, C. et al., 2021, {\it Sci. Adv.}, 7, eabj7601
\refs 
Byrne, P. K. et al., 2021, {\it PNAS}, 118, e2025919118
\refs 
Canup, R.M. \& Ward, W.R., 2006, {\it Nature}, 441, 834–839
\refs 
Canup, R.M. et al., 2015, {\it Nat. Geosci.} 8, 918–921
\refs 
Canup, R.M. et al., 2021, {\it New Views of the Moon II}, in press
\refs 
Canup, R.M., 2004, {\it Icarus}, 168, 433–456
\refs 
Caracas, R. et al., 2019, {\it EPSL} 516, 202–211
\refs 
Carter, J. et al., 2013, {\it JGR:P}, 118, 831–858
\refs 
Carter, P.J. \& Stewart, S.T., 2020, {\it PSJ}, 1, 45
\refs 
Carter, P.J. et al., 2018, {\it EPSL} 484, 276–286
\refs 
Castillo-Rogez, J. \& Young, E.D., 2017, {\it Planetesimals: Early Differentiation and Consequences for Planets} (L.T. Elkins-Tanton and B.P. Weiss, Cambridge University %%
\refs 
Press), 92–114
\refs 
Castillo-Rogez, J.C. et al., 2019, {\it GRL}, 46, 1963–1972
\refs 
Catling, D.C. \& Kasting, J.F., 2003, {\it ARAA}, 41, 429–463
\refs 
Catling, D.C. \& Kasting, J.F., 2017, {\it Atmospheric Evolution on Inhabited and Lifeless Worlds}, Cambridge University Press
\refs 
Catling, D.C. \& Zahnle, K.J., 2020, {\it Sci. Adv.}, 6, eaax1420
\refs 
Catling, D.C. et al., 2001, {\it Science}, 293, 839–843
\refs 
Cerantola, V. et al., 2015, {\it EPSL}, 417, 67–77
\refs 
Chabot, N.L. et al., 2005, {\it GCA}, 69, 2141–2151
\refs 
Chachan, Y. \& Stevenson, D.J., 2018, {\it ApJ}, 854, 21
\refs 
Chambers, J. E., 2016), {\it ApJ}, 825, 63
\refs 
Chao, K.H. et al., 2021, {\it Geochemistry},  125735
\refs 
Charnay, B. et al., 2020, {\it SSR}, 216, 1–29
\refs 
Charnoz, S. et al., 2021, {\it Icarus} 364, 114451
\refs 
Checlair, J. et al., 2021, {\it AJ},  161, 150
\refs 
Chen, B., Li, J., and Hauck, S. A., 2008, {\it GRL}, 35, L07201
\refs 
Chen, E.M.A. \& Nimmo, F., 2016, {\it Icarus} 275, 132–142
\refs 
Chen, L. et al., 2019, {\it ApJL}, 887, L32
\refs 
Chidester, B.A. et al., 2017, {\it GCA}, 199, 1–12
\refs 
Chou, C.-L., 1978, {\it LPSC}, 9th, 219–230
\refs 
Ciesla, F.J. \& Cuzzi, J.N., 2006, {\it Icarus}, 181, 178–204
\refs 
Ciesla, F.J. et al., 2013, {\it M\&PS}, 48, 2559–2576
\refs 
Cieza, L.A. et al., 2021, {\it MNRAS}, 501, 2934–2953
\refs 
Cimerman, N.P. et al., 2017, {\it MNRAS}, 471, 4662–4676
\refs 
Clement, M. S. et al., 2018, {\it Icarus}, 311, 340–356
\refs 
Clement, M. S. et al., 2019, {\it Icarus}, 321, 778–790
\refs 
Clesi, V. et al., 2018, {\it Sci. Adv.}, 4, e1701876
\refs 
Cohen, B.A. et al., 2005, {\it , M\&PS} 40, 755–777
\refs 
Collinet, M. \& Grove, T.L., 2020a, {\it GCA}, 277, 334–357
\refs 
Collinet, M. \& Grove, T.L., 2020b, {\it GCA}, 277, 358–376
\refs 
Collinet, M. \& Grove, T.L., 2020c, {\it M\&PS}, 55, 832–856
\refs 
Connelly, J.N. et al, 2012, {\it Science}, 338, 651-655.
\refs 
Connelly, J.N. et al., 2017, {\it GCA}, 201, 345–363
\refs 
Cook, D. L. et al., 2021, {\it ApJ}, 917, 59
\refs 
Coradini, A. et al., 1983, {\it PEPI}, 31, 145–160
\refs 
Corgne, A. et al., 2008, {\it GCA}, 72, 574–589
\refs 
Costa, A. et al., 2009, {\it GGG}, 10, Q03010
\refs 
Costa, M. M. et al., 2020, {\it PNAS}, 117, 30973-30979
\refs 
Cottrell, E. et al., 2009, {\it EPSL}, 281, 275–287
\refs 
Cottrell, E. et al., 2022, {\it Magma Redox Geochemistry} (R. Moretti, D.R. Neuville, eds., Geophys. Monograph), 33-61
\refs 
Cowan, N. B. \& Abbot, D. S., 2014, {\it ApJ}, 781, 27
\refs 
Crisp, J. A., 1984, {\it J. Volcanol. Geotherm. Res.}, 20, 177–211
\refs 
Dahl, T.W. \& Stevenson, D.J., 2010, {\it EPSL}, 295, 177–186
\refs 
Dalou, C. et al., 2017, {\it EPSL}, 458, 141–151
\refs 
Dasgupta, R. \& Hirschmann, M.M., 2007, {\it Am. Mineral.}, 92, 370–379
\refs 
Dasgupta, R. et al., 2013, {\it GCA}, 102, 191–212
\refs 
Dauphas, N. \& Pourmand, A., 2011, {\it Nature}, 473, 489–492
\refs 
Dauphas, N. \& Schauble, E.A., 2016, {\it AREPS}, 44, 709–783
\refs 
Dauphas, N., 2017, {\it Nature}, 541, 521–524
\refs 
Davies, E.J. et al., 2020, {\it JGR:P}, 125, e2019JE006227
\refs 
Day, J.M.D. \& Walker, R.J., 2015, {\it EPSL} 423, 114–124
\refs 
Deguen, R. et al., 2011, {\it EPSL}, 310, 303–313
\refs 
Deguen, R. et al., 2014, {\it EPSL}, 391, 274–287
\refs 
Delano, J.W., 2001, {\it Orig. Life Evol. Biosph.}, 31, 311–341
\refs 
Delbo’, M. et al., 2017, {\it Science}, 357, 1026–1029
\refs 
Deng, J. et al., 2020, {\it Nat. Commun.}, 11, 2007
\refs 
Desch, S.J. et al., 2018, {\it ApJ}S, 238, 11
\refs 
Ding, F. \& Pierrehumbert, R.T., 2016, {\it ApJ}, 822, 24
\refs 
Dorn, C. \& Lichtenberg, T., 2021, {\it ApJL}, 922, L4
\refs 
Dorn, C. et al., 2015, {\it A\&A}, 577, A83
\refs 
Dorn, C. et al., 2018, {\it A\&A}, 614, A18
\refs 
Doyle, A.E. et al., 2019, {\it Science}, 366, 356–359
\refs 
Doyle, A.E. et al., 2021, {\it ApJL}, 907, 10
\refs 
Doyle, P.M. et al., 2015, {\it Nat. Comm.}, 6, 1–10
\refs 
Dr{\k{a}}{\.z}kowska, J. \& Alibert, Y., 2017, {\it A\&A}, 608, A92
\refs 
Ehlmann, B. L., 2011, {\it Science}, 479, 53-60.
\refs 
Elkins-Tanton, L.T. \& Seager, S., 2008, {\it ApJ}, 688, 628
\refs 
Elkins-Tanton, L.T., 2011, {\it Astrophys. Space Sci.}, 332, 359–364
\refs 
Elkins-Tanton, L.T. et al., 2003, {\it M\&PS}, 38, 1753–1771
\refs 
Elkins-Tanton, L.T. et al., 2005, {\it JGR},  110, E12S01
\refs 
Elkins-Tanton, L.T. et al., 2011, {\it EPSL} 304, 326–336
\refs 
Elkins-Tanton, L.T. et al., 2020, {\it JGR:P}, e2019JE006296
\refs 
Elkins-Tanton, L.T., 2008, {\it EPSL}, 271, 181–191
\refs 
Elkins-Tanton, L.T., 2012, {\it AREPS}, 40, 113–139
\refs 
Elkins-Tanton, L.T., 2017, {\it Planetesimals: Early Differentiation and Consequences for Planets} (L.T. Elkins-Tanton and B.P. Weiss, Cambridge University Press), 365–375
\refs 
Farihi, J. et al., 2013, {\it Science}, 342, 218–220
\refs 
Fegley Jr, B. et al., 2020, {\it Geochemistry},  80, 125594
\refs 
Fischer-Gödde, M. \& Kleine, T., 2017, {\it Nature}, 541, 525–527
\refs 
Fischer, R. \& Gerya, T., 2016, {\it J. Geodyn.}, 100, 198–214
\refs 
Fischer, R.A. \& Nimmo, F., 2018, {\it EPSL}, 499, 257–265
\refs 
Fischer, R.A. et al., 2015, {\it GCA}, 167, 177–194
\refs 
Fischer, R.A. et al., 2017, {\it EPSL}, 458, 252–262
\refs 
Fischer, R.A. et al., 2018, {\it EPSL}, 482, 105–114
\refs 
Fischer, R.A. et al., 2020, {\it PNAS}, 117, 8743–8749
\refs 
Foley, B. et al., 2020, {\it Planetary Diversity} (E.J. Tasker et al., IOP), 4–60
\refs 
Foley, B.J. \& Smye, A. J., 2018, {\it Astrobiology},  18, 873–896
\refs 
Foley, B.J. et al., 2012, {\it EPSL}, 331, 281–290
\refs 
Foley, B.J. et al., 2014, {\it JGR:SE}, 119, 8538-8561
\refs 
Forget, F. \& Leconte, J., {\it Philos. Trans. R. Soc. A}, 372, 20130084
\refs 
Frank, E.A. et al., 2014, {\it Icarus}, 243, 274–286
\refs 
Frost, D.J. et al., 2004, {\it Nature}, 428, 409–412
\refs 
Frost, D.J. \& McCammon, C.A. 2008, {\it AREPS}, 36, 389–420
\refs 
Fu, R.R. \& Elkins-Tanton, L.T., 2014, {\it EPSL}, 390, 128–137
\refs 
Fu, R.R. et al., 2017, {\it Planetesimals: Early Differentiation and Consequences for Planets} (L.T. Elkins-Tanton and B.P. Weiss, Cambridge University Press), 115–135
\refs 
Fujimoto, Y. et al., 2018, {\it MNRAS}, 480, 4025–4039
\refs 
Fulton, B.J. et al., 2017, {\it AJ},  1543, 109
\refs 
Gail, H.P. et al., 2014, {\it Protostars \& Planets VI} (H. Beuther, R.S. Klessen, C.P. Dullemond, and T. Henning, University of Arizona Press), 571–594
\refs 
Gail, H.P. et al., 2015, {\it A\&A}, 576, A60
\refs 
Gaillard, F. \& Scaillet, B., 2014, {\it EPSL}, 403, 307–316
\refs 
Gaillard, F. et al., 2021, {\it SSR}, 217, 1–54
\refs 
Gaillard, F., 2022, {\it EPSL}, 577, 117255.
\refs 
Garnero, E.J. et al., 2016, {\it Nat. Geosci.} 9, 481–489
\refs 
Gáspár, A. \& Rieke, G.H., 2020, {\it PNAS}, 18, 9712–9722 
\refs 
Gaudi, B.S. et al., Proc. SPIE, 11115, 111150M
\refs 
Genda, H. \& Abe, Y., 2003, {\it Icarus}, 164, 149–162
\refs 
Genda, H. \& Abe, Y., 2005, {\it Nature}, 433, 842–844
\refs 
Genda, H. et al., 2017, {\it EPSL}, 480, 25–32
\refs 
Genova, A. et al., 2021, {\it SSR}, 217, 1–62
\refs 
Gerya, T., 2014, {\it Gondwana Res.}, 25, 442–463
\refs 
Gerya, T., 2019, {\it Geology}, 47, 1006–1007
\refs 
Gerya, T., 2019, {\it Introduction to Numerical Geodynamic Modelling}, Cambridge University Press
\refs 
Geßmann, C.K. \& Rubie, D.C., 1998, {\it GCA}, 62, 867–882
\refs 
Ghanbarzadeh, S. et al., 2017, {\it PNAS}, 114, 13406–13411
\refs 
Gillmann, C. et al., 2020, {\it Nat. Geosci.}, 13, 265–269
\refs 
Ginzburg, S. et al., 2016, {\it ApJ}, 825, 29
\refs 
Ginzburg, S. et al., 2018, {\it MNRAS}, 476, 759–765
\refs 
Golabek, G.J.  et al., 2009, {\it GGG}, 10, Q11007
\refs 
Golabek, G.J. et al., 2014, {\it M\&PS}, 49, 1083–1099
\refs 
Golabek, G.J. et al., 2018, {\it Icarus}, 301, 235–246
\refs 
Goldblatt, C. et al., 2013, {\it Nat. Geosci.}, 6, 661–667
\refs 
Goldblatt, C. et al., 2021, {\it Nat. Geosci.}, 14, 143–150
\refs 
Gomes, R. et al., 2005, {\it Nature}, 435, 466–469
\refs 
Gradie, J. \& Tedesco, E., 1982, {\it Science}, 216, 1405–1407
\refs 
Graham, R.J. \& Pierrehumbert, R., 2020, {\it ApJ}, 896, 115
\refs 
Graham, R.J. et al., 2021, {\it PSJ}, 2, 207
\refs 
Grewal, D.S. et al., 2019, {\it Sci. Adv.}, 5, eaau3669
\refs 
Grewal, D.S. et al., 2020, {\it GCA}, 280, 281–301
\refs 
Grewal, D.S. et al., 2021, {\it Nat. Geosci.}, 14, 369–376
\refs 
Grimm, R.E. \& McSween, H.Y., 1989, {\it Icarus}, 82, 244–280
\refs 
Grimm, R.E. \& McSween, H.Y., 1993, {\it Science}, 259, 653–655
\refs 
Gülcher, A. J. et al., 2020, {\it Nat. Geosci.}, 13, 547–554
\refs 
Haack,H. \& Scott, E.R.D., 1992, {\it JGR},  97, 14727–14734
\refs 
Hakim, K. et al., 2021, {\it PSJ}, 2, 49
\refs 
Halliday, A.N., \& Porcelli, D., 2001, {\it EPSL}, 192, 545–559
\refs 
Hamano, K. et al., 2013, {\it Nature}, 497, 607–610
\refs 
Hamano, K. et al., 2015, {\it ApJ}, 806, 216
\refs 
Hamilton, W. B., 2011, {\it Lithos}, 123, 1–20
\refs 
Harper Jr., C.L. \& Jacobsen, S.B., 1996, {\it Science}, 273, 1814–1818
\refs 
Harrison, T. M., 2020, {\it Hadean Earth}, Springer Nature
\refs 
Harsono, D. et al., 2018, {\it Nat. Astron.}, 2, 646–651
\refs 
Hashizume, K., Sugiura, N., 1998, {\it M\&PS}, 33, 1181–1195
\refs 
Hauck, S.A. et al., 2006, {\it JGR},  111, E09008
\refs 
Hayashi, C. et al., 1979, {\it EPSL}, 43, 22–28
\refs 
Hayworth, B. P. C. \& Foley, B. J., 2020, {\it ApJL}, 902, L10
\refs 
Helled, R. et al., 2020, {\it Nat. Rev. Phys.}, 2, 562–574
\refs 
Henke, S. et al., 2012, {\it A\&A}, 537, A45
\refs 
Hevey, P.J. \& Sanders, I.S., 2006, {\it M\&PS}, 41, 95–106
\refs 
Hier-Majumder, S. \& Hirschmann M. M., 2017, {\it GGG}, 18, 3078–3092
\refs 
Hillgren, V.J. et al., 1996, {\it GCA}, 60, 2257–2263
\refs 
Hin, R.C. et al., 2017, {\it Nature}, 549, 511–515
\refs 
Hirose, K. et al., 2013, {\it AREPS}, 41, 657–691
\refs 
Hirschmann, M.M. et al., 2021, {\it PNAS}, 118, e2026779118
\refs 
Hirschmann, M.M., 2012, {\it EPSL}, 341–344, 48–57
\refs 
Hirschmann, M.M., 2016, {\it Am. Mineral.}, 101, 540–553
\refs 
Hirschmann, M.M., 2018, {\it EPSL}, 502, 262–273
\refs 
Hoffman, P.F. et al., 2017, {\it Sci. Adv.}, 3 e1600983
\refs 
Hollands, M.A., 2021, {\it Nat. Astron.}, 5, 451–459
\refs 
Holloway, J.R. et al., 1992, {\it Eur. J. Mineral.}, 4, 105-114
\refs 
Holmström, E. et al., 2018, {\it EPSL} 490, 11–19
\refs 
Höning, D. et al., 2019, {\it A\&A}, 627, A48
\refs 
Hu, R. et al., 2012, {\it ApJ}, 752, 7
\refs 
Hu, R. et al., 2021, {\it ApJ}, {\it arXiv}:2108.04745
\refs 
Hunt, A.C. et al., 2017, {\it GCA}, 199, 13–30
\refs 
Hunt, A.C. et al., 2018, {\it EPSL}, 482, 490–500
\refs 
Iacono-Marziano, G. et al., 2012, {\it GCA}, 97, 1-23
\refs 
Ida, S. \& Lin, D.N.C., 2004, {\it ApJ}, 616, 567
\refs 
Ida, S. \& Guillot, T., 2016, {\it A\&A}, 596, L3
\refs 
Ikoma, M. \& Genda, H., 2006, {\it ApJ}, 648, 696
\refs 
Ikoma, M. et al., 2018, {\it SSR}, 214, 76
\refs 
Ingersoll, A.P., 1969, {\it J. Atmos. Sci.}, 26, 1191–1198
\refs 
Izidoro, A. et al., 2013, {\it ApJ}, 767, 54
\refs 
Izidoro, A. et al., 2017, {\it MNRAS}, 470, 1750–1770
\refs 
Jackson, B. et al., 2008, {\it ApJ}, 681, 1631
\refs 
Jacobson, S.A. et al., 2017, {\it EPSL} 474, 375–386
\refs 
Javoy, M., 1995, {\it GRL}, 22, 2219–2222
\refs 
Jin, Z. \& Bose, M., 2019, {\it Sci. Adv.}, 5, eaav8106
\refs 
Jin, Z. et al., 2021, {\it PSJ}, 2, 244
\refs 
Johansen, A. \& Lambrechts, M., 2017, {\it ARAA}, 45, 359–387
\refs 
Johansen, A. et al., 2015, {\it Sci. Adv.}, 1, e1500109
\refs 
Johansen, A. et al., 2021, {\it Sci. Adv.}, 7, eabc0444
\refs 
Johnson, T.E. et al., 2018, {\it Nat. Geosci.} 11, 795–799
\refs 
Jontof-Hutter, D., 2019, {\it AREPS}, 47, 141–171
\refs 
Jura, M. \& Young, E. D, 2014, {\it AREPS}, 42, 45–67
\refs 
Kaminski, E. et al., 2020, {\it EPSL}, 548, 116469.
\refs 
KamLAND Collaboration, 2011, {\it Nat. Geosci.}, 4, 647
\refs 
Kane, S. R. et al., 2021, {\it JGR:P}, 126, e2020JE006643
\refs 
Karato, S.-i. \& Murthy, V.R., 1997, {\it PEPI}, 100, 61–79
\refs 
Karato, S.-i. \& Wu, P., 1993, {\it Science}, 260, 771–778
\refs 
Karki, B.B. \& Stixrude, L.P., 2010, {\it Science}, 328, 740–742
\refs 
Kasting, J.F., 1988, {\it Icarus}, 74, 472–494
\refs 
Kasting, J. F., 1993, {\it Icarus}, 101, 108–128
\refs 
Kato, C. \& Moynier, F., 2017, {\it Sci. Adv.} 3, e1700571
\refs 
Kato, C. et al., 2015, {\it Nat. Commun.}, 6, 7617
\refs 
Kaula, W.M., 1979, {\it JGR},  84, 999–1008
\refs 
Ke, Y. \& Solomatov, V.S., 2009, {\it JGR},  114, E07004
\refs 
Kegerreis, J.A. et al., 2018, {\it ApJ}, 861, 52
\refs 
Kegler, Ph. et al., 2008, {\it EPSL}, 268, 28–40
\refs 
Keil, K. et al., 1997, {\it Meteoritics}, 32, 349–363
\refs 
Keller, T. \& Suckale, J., 2019, {\it GJI}, 219, 185–222
\refs 
Khan, A. et al., 2021, {\it Science}, 373, 434–438
\refs 
Khurana, K. K. et al., 2011, {\it Science}, 332, 1186–1189
\refs 
Kimura, K. et al., 1974,  38, 683–701
\refs 
Kimura, T. \& Ikoma, M., 2020, {\it MNRAS}, 496, 3755–3766
\refs 
Kipping, D. et al., 2022, {\it Nat. Astron.}, 6, 367–380
\refs 
Kislyakova, K. \& Noack, L., 2020, {\it A\&A}, 636, 10
\refs 
Kislyakova, K.G. et al., 2017, {\it Nat. Astron.}, 1, 878–885
\refs 
Kislyakova, K.G. et al., 2018, {\it ApJ}, 858, 105
\refs 
Kite, E.S. \& Barnett, M.N., 2020, {\it PNAS}, 117, 18264–18271
\refs 
Kite, E.S. \& Ford, E. B., 2018, {\it ApJ}, 864, 75
\refs 
Kite, E.S. \& Schaefer, L., 2021, {\it ApJL}, 909, L22
\refs 
Kite, E.S. et al., 2019, {\it ApJL}, 887, L33
\refs 
Kite, E.S. et al., 2020, {\it ApJ}, 891, 111
\refs 
Kite, E.S., 2019, {\it SSR}, 215, 10
\refs 
Klahr, H. \& Schreiber, A., 2020, {\it ApJ}, 901, 54
\refs 
Klahr, H. \& Schreiber, A., 2021, {\it ApJ}, 911, 9
\refs 
Kleine, T. \& Walker, R.J., 2017, {\it AREPS}, 45, 389–417
\refs 
Kleine, T. et al., 2002, {\it Nature}, 418, 952–955
\refs 
Kleine, T. et al., 2004, {\it EPSL} 228, 109–123
\refs 
Kleine, T. et al., 2020, {\it SSR}, 216, 55
\refs 
Kley, W., 2019, {\it From Protoplanetary Disks to Planet Formation} (M. Audard, M.R. Meyer, Y. Alibert, eds., Springer), 151–260
\refs 
Knapmeyer-Endrun, B. et al., 2021, {\it Science}, 373, 438–443
\refs 
Koll, D.D.B. et al., 2020, {\it ApJ}, 886, 140
\refs 
Korenaga, J., 2010, {\it ApJL}, 725, L43
\refs 
Korenaga, J., 2013, {\it AREPS}, 41, 117–151
\refs 
Korenaga, J., 2021, {\it Precambrian Res.}, 359, 106178
\refs 
Kreidberg, L. et al., 2019, {\it ApJ} 877, L15
\refs 
Kreidberg, L. et al., 2019, {\it Nature}, 573, 87–90
\refs 
Krissansen-Totton, J. et al., 2018, {\it PNAS}, 115, 4105–4110
\refs 
Krissansen-Totton, J. et al., 2022, {\it Nat. Astron.}, 6, 189–198
\refs 
Krot, A.N. et al., 2014, {\it Treatise on Geochemistry} (H.D. Holland and K.K. Turekian, vol. 1, Elsevier), 1–63
\refs 
Kruijer, T.S. et al., 2014, {\it Science}, 344, 1150–1154
\refs 
Kruijer, T.S. et al., 2015, {\it Nature}, 520, 534–537
\refs 
Kruijer, T.S. et al., 2017, {\it EPSL}, 474, 345–354
\refs 
Kruijer, T.S. et al., 2021, {\it Nat. Geosci.}, 14, 714-715
\refs 
Kubik, E. et al., 2021, {\it GCA}, 306, 263–280
\refs 
Küffmeier, M. et al., 2016, {\it ApJ}, 826, 22
\refs 
Kurokawa, H. et al., 2021, {\it AGU Adv.}, e2021AV000568
\refs 
Labrosse, S. et al., 2007, {\it Nature}, 450, 866–869
\refs 
Lambrechts, M. et al., 2019, {\it A\&A}, 627, A83
\refs 
Lammer, H. et al., 2020a, {\it SSR}, 216, 74
\refs 
Lammer, H. et al., 2020b, {\it Icarus}, 339, 113511
\refs 
Lammer, H. et al., 2021, {\it SSR}, 217, 1–35
\refs 
Landeau, M. et al., 2021, {\it EPSL}, 564, 116888
\refs 
Landeau, M. et al., 2022, {\it Nat. Rev. Earth Environ.}, 1–15
\refs 
Laneuville, M. et al., 2014, {\it EPSL}, 401, 251–260
\refs 
Lebrun, T. et al., 2013, {\it JGR},  118, 1155–1176
\refs 
Leconte, J. et al., 2013, {\it Nature}, 504, 268–271
\refs 
Léger, A. et al., 2011, {\it Icarus}, 213, 1–11
\refs 
Lenardic, A., 2018, {\it Philos. Trans. R. Soc. A}, 376, 20170416
\refs 
Lewis,  J.A. \& Jones, R. H., 2016, {\it M\&PS}, 51, 1886–1913
\refs 
Li, J. \& Agee, C.B., 1996, {\it Nature}, 381, 686–689
\refs 
Li, R. et al., 2019, {\it ApJ}, 885, 69.
\refs 
Lichtenberg, T. \& Krijt, S., 2021, {\it ApJL}, 913, L20
\refs 
Lichtenberg, T. et al., 2016a, {\it Icarus}, 274, 350–365
\refs 
Lichtenberg, T. et al., 2016b, {\it MNRAS}, 462, 3979–3992
\refs 
Lichtenberg, T. et al., 2018, {\it Icarus}, 302, 27–43
\refs 
Lichtenberg, T. et al., 2019a, {\it Nat. Astron.}, 3, 307–313
\refs 
Lichtenberg, T. et al., 2019b, {\it EPSL}, 507, 154–165
\refs 
Lichtenberg, T. et al., 2021a, {\it Science}, 371, 365–370
\refs 
Lichtenberg, T. et al., 2021b, {\it JGR:P}, 126, e2020JE006711
\refs 
Lichtenberg, T., 2021, {\it ApJL}, 914, L4
\refs 
Liebske, C. et al., 2005, {\it EPSL}, 240, 589–604
\refs 
Liggins, P. et al., 2021, {\it arXiv}:2111.05161
\refs 
Lillis, R.J. et al., 2013, {\it JGR:P}, 118, 1488–1511
\refs 
Lissauer, J.J. 2007, {\it ApJ}, 660, L149-L152
\refs 
Liu, B. et al., 2019, {\it A\&A}, 624, A114
\refs 
Lock, S.J. et al., 2018, {\it JGR:P}, 123, 910–951
\refs 
Lock, S.J. et al., 2020, {\it SSR}, 216, 1–46
\refs 
Loftus, K. et al., 2019, {\it ApJ}, 887, 231
\refs 
Loftus, K. \& Wordsworth, R.D., 2021, {\it JGR:P}, 126, e2020JE006653
\refs 
Lourenço, D.L. et al., 2018, {\it Nat. Geosci.}, 11, 322–327
\refs 
Lourenço, D.L. et al., 2020, {\it GGG}, e2019GC008756
\refs 
Lugaro, M. et al., 2018, {\it Prog. Part. Nucl. Phys.}, 102, 1–47
\refs 
Luger, R. \& Barnes, R., 2015, {\it Astrobiology},  15, 119–143
\refs 
Lupu, R.E. et al., 2014, {\it ApJ}, 784, 27
\refs 
Ma, N. et al., 2021, {\it Goldschmidt 2021 Abstract}, 3852
\refs 
Mahan, B. et al., 2018a, {\it JGR:SE}, 123, 8349–8363
\refs 
Mahan, B. et al., 2018b, {\it GCA}, 235, 21–40
\refs 
Malavergne, V. et al., 2014, {\it EPSL} 394, 186–197
\refs 
Malavergne, V. et al., 2019, {\it Icarus}, 321, 473–485
\refs 
Mandler, B.E. \& Elkins‐Tanton, L.T., 2013, {\it M\&PS}, 48, 2333–2349
\refs 
Mann, U. et al., 2009, {\it GCA}, 73, 7360–7386
\refs 
Manning, C.E., 2018, {\it Magmas Under Pressure} (Y. Kono, C. Sanloup, Elsevier), 83–113
\refs 
Mansfield, M. et al., 2019, {\it ApJ}, 886, 141
\refs 
Marchi, S. et al., 2020, {\it Sci. Adv.}, 6, eaay2338
\refs 
Marcq, E. et al., 2017, {\it JGR},  122, 1539–1553
\refs 
Marcus, R. A. et al., 2010, {\it ApJ}, 719, 45–49
\refs 
Marty, B., 2020, {\it Geochem. Perspect.}, 9, 1–272
\refs 
Matsui, T. \& Abe, Y., 1986a, {\it Nature}, 319, 303–305
\refs 
Matsui, T. \& Abe, Y., 1986b, {\it Nature}, 322, 526–528
\refs 
Maurice, M. et al., 2017, {\it JGR:P}, 122, 577–598
\refs 
Maurice, M. et al., 2020, {\it Sci. Adv.}, 6, eaba8949
\refs 
May, E.M. \& Rauscher, E., 2020, {\it ApJ}, 893, 161
\refs 
McClure, M. K., 2020, {\it A\&A}, 642, L15
\refs 
McCoy, C.A. et al., 2019, {\it Phys. Rev. B.}, 100, 014106
\refs 
McCoy, T. \& Bullock, E., 2017, {\it Planetesimals: Early Differentiation and Consequences for Planets} (L.T. Elkins-Tanton and B.P. Weiss, Cambridge Planetary Science), 71–91
\refs 
McCoy, T.J. et al., 2006, {\it Meteorites and the Early Solar System} (D.S. Lauretta \& H.Y. McSween, Jr., eds., U. AZ Press), 733–745
\refs 
McEwan, A. et al., 2019, {\it EOS}, 100, doi:10.1029/2019EO135617 
\refs 
McKinnon, W.B. et al., 2020, {\it Science}, 367, eaay6620
\refs 
McLennan, S.M. et al., 2019, {\it AREPS}, 47, 91–118
\refs 
McSween, H.Y., Jr. et al., 2002, {\it Asteroids III} (W.F. Bottke et al., eds., U. AZ Press) 559
\refs 
McSween Jr, H. Y. \& Labotka, T. C., 1992, {\it Meteoritics}, 27
\refs 
McWilliams, R. S., 2015, {\it PNAS}, 112, 7925–7930
\refs 
Meier, T. G. et al., 2021, {\it ApJL}, 908, L48
\refs 
Melosh, H.J., 1990, {\it LPI Conf. Origin of the Earth}, 69–83
\refs 
Meng, H.Y.A. et al., 2014, {\it Science}, 345, 1032–1035
\refs 
Menzel, R.L. \& Roberge, W.G., 2013, {\it ApJ}, 776, 89
\refs 
Meyer, J. et al., 2010, {\it Icarus} 208, 1–10
\refs 
Mezger, K. et al., 2020, {\it SSR}, 216, 27
\refs 
Miyazaki, Y. \& Korenaga, J., 2022, {\it Nature}, 603, 86–90
\refs 
Militzer, B. et al., 2021, {\it Phys. Rev. E} 103, 013203
\refs 
Millot, M. et al., 2015, {\it Science}, 347, 418–420
\refs 
Miozzi, F. et al., 2017, {\it JGR:P}, 123, 2295–2309
\refs 
Mitrovica, J.X. \& Forte, A.M., 2004, {\it EPSL}, 225, 177–189
\refs 
Mittelholz, A. et al., 2020, {\it Sci. Adv.} 6, eaba0513
\refs 
Mizuno, H. et al., 1980, {\it EPSL}, 50, 202–210
\refs 
Mojzsis, S. J. et al., 2001, {\it Nature}, 409, 178–181
\refs 
Monteux, J. et al., 2016, {\it EPSL}, 448, 140–149
\refs 
Monteux, J. et al., 2018, {\it SSR}, 214, 39
\refs 
Moore, G. et al., 1998, {\it Am. Mineral.}, 83, 36-42
\refs 
Moore, W.B. \& Webb, A.A.G., 2013, {\it Nature}, 501, 501–505
\refs 
Moore, W.B. et al., 2017, {\it EPSL}, 474, 13–19
\refs 
Morbidelli, A. \& Wood, B. J., 2015, {\it The Early Earth: Accretion and Differentiation} (J. Badro, M.J. Walter, eds., Geophys. Monograph), 212, 71–82
\refs 
Morbidelli, A. et al., 2000, {\it M\&PS}, 35, 1309–1320
\refs 
Morbidelli, A. et al., 2012, {\it AREPS}, 40, 251–275
\refs 
Morbidelli, A. et al., 2018, {\it Icarus} 305, 262–276
\refs 
Morishima, R. et al., 2013, {\it EPSL}, 366, 6–16
\refs 
Moskovitz, N. \& Gaidos, E., 2011, {\it M\&PS}, 46, 903–918
\refs 
Mousis, O. et al., 2020, {\it ApJL}, 896, L22
\refs 
Mukhopadhyay, S., 2012, {\it Nature}, 486, 101–104
\refs 
Mullen, P.D. \& Gammie, C.F., 2020, {\it ApJL} 903, L15
\refs 
Murthy, V.R.,1991, {\it Science}, 253, 303–306
\refs 
Mustard, J.F. et al., 2008, {\it Nature}, 454, 305–309
\refs 
Nabiei, F. et al., 2018, {\it Nat. Comm.}, 9, 1327
\refs 
Nakajima, M. \& Stevenson, D.J., 2014, {\it Icarus} 233, 259–267
\refs 
Nakajima, M. \& Stevenson, D.J., 2015, {\it EPSL} 427, 286–295
\refs 
Nakajima, M. \& Stevenson, D.J., 2018, {\it EPSL} 487, 117–126
\refs 
Nakajima, M. et al., 2021, {\it EPSL}, 568, 116983
\refs 
Nakajima, M. et al., 2022, {\it Nat. Commun.}, 13, 568
\refs 
Nakajima, S. et al., 1992, {\it J. Atmos. Sci.}, 49, 2256–2266
\refs 
Nakamura, T., 2006, {\it EPSL}, 242, 26–38
\refs 
Nakato, A. et al., 2008, {\it EPS}, 60, 855–864
\refs 
Nakazawa, K. et al., 1985, {\it J. Geomag. Geoelec.}, 37, 781–799
\refs 
Nemchin, A. et al., 2009, {\it Nat. Geosci.}, 2, 133–136
\refs 
Neumann, W. et al., 2014, {\it EPSL}, 395, 267–280
\refs 
Neumann, W. et al., 2018, {\it Icarus}, 311, 146–169
\refs 
Neumann, W. et al., 2018, {\it JGR:P}, 123, 421–444
\refs 
Nimmo, F. \& Kleine, T., 2015, {\it The Early Earth: Accretion and Differentiation} (J. Badro and M. Walter, AGU/Wiley), 83–102
\refs 
Nimmo, F. \& McKenzie, D., 1998, {\it AREPS}, 26, 23–51
\refs 
Nimmo, F. et al., 2004, {\it Geophys. J.}, 156, 363–376
\refs 
Nimmo, F. et al., 2010, {\it EPSL}, 292, 363–370
\refs 
Nimmo, F. et al., 2018, {\it SSR}, 214, 101
\refs 
Nimmo, F., 2015, {\it Treatise on Geophysics}, 9, 201–219
\refs 
Nittler, L.R. \& Ciesla, F., 2016, {\it ARAA}, 54, 53–93
\refs 
Noack, L. \& Lasbleis, M., 2020, {\it A\&A}, 638, A129
\refs 
Noack, L. et al., 2012, {\it Icarus}, 217, 484–498
\refs 
Noack, L. et al., 2016, {\it Icarus}, 277, 215–236
\refs 
Norman, M. D.et al., 2003, {\it M\&PS}, 38, 645–661
\refs 
Norris, C.A. \& Wood, B.J., 2017, {\it Nature}, 549, 507–510
\refs 
Nyquist, L.E. et al., 1995, {\it GCA}, 59, 2817–2837
\refs 
O’Rourke, J.G., \& Stevenson, J. D., 2016, {\it Nature}, 529, 387–389
\refs 
O'Rourke, J.G., 2020, {\it GRL}, 47, e2019GL086126
\refs 
O’Brien, D.P. et al., 2006, {\it Icarus}, 184, 39–58
\refs 
O’Brien, D.P. et al., 2014, {\it Icarus}, 239, 74–84
\refs 
O’Brien, D.P. et al., 2018, {\it SSR}, 214, 47
\refs 
O’Keefe, J.D. \& Ahrens, T.J., 1977, {\it LPSC}, 8, 741
\refs 
O’Neill, C. \& Lenardic, A., 2007, {\it GRL}, 34, 19
\refs 
O’Neill, C. et al., 2020, {\it Icarus}, 352, 114025
\refs 
O’Neill, H.S.C. \& Palme, H., 2008, {\it Philos. Trans. R. Soc. A}, 366, 4205–4238
\refs 
Öberg, K.I. \& Bergin, E.A., 2021, {\it Phys. Rep.}, 893, 1–48
\refs 
Okuchi, T., 1997, {\it Science}, 278, 1781–1784
\refs 
Olson, P., 2016, {\it GGG}, 17, 1935–1956
\refs 
Olson, P. \& Sharp, Z.D., 2018, {\it EPSL}, 498, 418–426
\refs 
Olson, P.L. \& Sharp, Z.D., 2019, {\it PEPI}, 294, 106294
\refs 
Ormel, C.W. et al., 2015a, {\it MNRAS}, 446, 1026–1040
\refs 
Ormel, C.W. et al., 2015b, {\it MNRAS}, 447, 3512–3525
\refs 
Ormel, C.W., 2017, {\it Formation, Evolution, and Dynamics of Young Solar Systems} (M. Pessah and O. Gressal, Springer), 197–228
\refs 
Osinski, G.R. et al., 2020, {\it Astrobiology}, 20, 1121–1149
\refs 
Owen, J.E. \& Wu, Y., 2017, {\it ApJ}, 847, 29
\refs 
Owen, J.E. et al., 2020, {\it SSR}, 216, 129
\refs 
Pahlevan, K. et al., 2019, {\it EPSL}, 526, 115770
\refs 
Palin, R. M. et al., 2020, {\it Earth Sci. Rev.}, 207, 103172
\refs 
Parker, R.J., 2020, {\it R. Soc. Open Sci., 7}, 201271
\refs 
Pavlov, A. A. et al., 2000, {\it JGR:P}, 105, 11981–11990
\refs 
Péron, S. et al., 2017, {\it Geochem. Perspect. Lett.}, 3, 151–159
\refs 
Peslier, A.H. et al., 2017, {\it SSR}, 212, 743–810
\refs 
Piani, L. et al., 2020, {\it Science}, 369, 1110–1113
\refs 
Pierrehumbert, R.T., 2010, {\it Principles of Planetary Climate}, Cambridge University Press
\refs 
Pierrehumbert, R.T. et al., 2011, {\it AREPS}, 39, 417–460.
\refs 
Quanz, S.P. et al., 2021, {\it A\&A}, {\it arXiv}:2101.07500
\refs 
Ramirez, R.M. et al., 2019, {\it BAAS} 51, 31
\refs 
Raymond, S.N. \& Izidoro, A., 2017, {\it Icarus}, 297, 134–148
\refs 
Raymond, S.N. et al., 2007, {\it Astrobiology},  7, 66–84
\refs 
Reese, C.C. \& Solomatov, V.S., 2006, {\it Icarus}, 102–120
\refs 
Reese, C.C. \& Solomatov, V.S., 2010, {\it Icarus} 207, 82–97
\refs 
Reiter, M., 2020, {\it A\&A}, 644, L1
\refs 
Ricard, Y. et al., 2009, {\it EPSL}, 284, 144–150
\refs 
Ricard, Y. et al., 2017, {\it Icarus}, 285, 103–117
\refs 
Ricolleau, A. et al., 2011, {\it EPSL}, 310, 409–421
\refs 
Righter, K. et al., 1997, {\it PEPI}, 100, 115–134
\refs 
Righter, K. et al., 2016, {\it EPSL}, 437, 89–100
\refs 
Righter, K., 2011, {\it EPSL}, 304, 158–167
\refs 
Rimmer, P.B., 2021, {\it ApJL}, 921, L28
\refs 
Ringwood, A.E., 1959, {\it GCA}, 15, 257–283
\refs 
Roberts, J.H. \& Arkani-Hamed, J., 2017, {\it EPSL} 478, 192–202
\refs 
Rosing, M. T. et al., 2010, {\it Nature}, 464, 744–747
\refs 
Roskosz, M. et al., 2013, {\it GCA}, 121, 15–28
\refs 
Rothery, D.A. et al., {\it SSR}, 216, 1–46
\refs 
Rozel, A. B. et al., 2017, {\it Nature}, 545, 332–335
\refs 
Rubie, D.C. \& Jacobson, S.A., 2016, {\it Deep Earth: Physics and Chemistry of the Lower Mantle and Core} (H. Terasaki and R.A. Fischer, AGU/Wiley), 181–190
\refs 
Rubie, D.C. et al., 2003, {\it EPSL}, 205, 239–255
\refs 
Rubie, D.C. et al., 2011, {\it EPSL}, 301, 31–42
\refs 
Rubie, D.C. et al., 2015a, {\it Treatise on Geophysics} (G. Schubert, Elsevier), 43–79
\refs 
Rubie, D.C. et al., 2015b, {\it Icarus}, 248, 89–108
\refs 
Rudge, J.F. et al., 2010, {\it Nat. Geosci.}, 3, 439–443
\refs 
Ruedas, T., 2017, {\it GGG}, 18, 3530–3541
\refs 
Rückriemen, T. et al., 2018, {\it Icarus}, 307, 172–196
\refs 
Safronov, V.S., 1978, {\it Icarus}, 33, 3–12
\refs 
Sagan, C. \& Mullen, G., 1972, {\it Science}, 177, 52–56
\refs 
Saito, H. \& Kuramoto, K., 2020, {\it ApJ}, 889, 40
\refs 
Salmon, J. \& Canup, R.M., 2012, {\it ApJ}. 760, 83–101
\refs 
Salvador, A. et al., 2017, {\it JGR:P}, 122, 1458–1486
\refs 
Samuel, H. et al., 2010, {\it EPSL}, 290, 13–19
\refs 
Samuel, H., 2012, {\it EPSL}, 313–314, 105–114
\refs 
Sánchez, M.B. et al., 2018, {\it MNRAS}, 481, 1281–1289
\refs 
Sanders, I. S. et al., 2017, {\it M\&PS}, 52, 690–708
\refs 
Sarkis, P. et al., 2021, {\it A\&A}, 645, A79
\refs 
Sasaki, S. \& Nakazawa, K., 1990, {\it Icarus}, 85, 21–42
\refs 
Sasaki, S., 1989, {\it A\&A}, 215, 177–180
\refs 
Sasselov, D. D. et al., 2020, {\it Sci. Adv.}, 6, eaax3419
\refs 
Schaefer, L. \& Fegley Jr., B., 2007, {\it Icarus}, 186, 462–483
\refs 
Schaefer, L. \& Fegley Jr., B., 2010, {\it Icarus}, 205, 483–496
\refs 
Schaefer, L. \& Fegley Jr., B., 2017, {\it ApJ}, 843, 120
\refs 
Schaefer, L. \& Sasselov, D., 2015, {\it ApJ}, 801, 40
\refs 
Schaefer, L. et al., 2016, {\it ApJ}, 829, 63
\refs 
Scheinberg, A. et al., 2015, {\it Icarus}, 254, 62–71
\refs 
Scheinberg, A.L. et al., 2018, {\it EPSL} 492, 144–151
\refs 
Scherstén, A. et al., 2006, {\it EPSL}, 241, 530–542
\refs 
Schlichting, H.E. \& Young, E.D., 2021, {\it arXiv}:2107.10405
\refs 
Schlichting, H.E. et al., 2015, {\it Icarus} 247, 81–94
\refs 
Schönbächler, M. et al., 2010, {\it Science}, 328, 884–887
\refs 
Schoonenberg, D. \& Ormel, C.W., 2017, {\it A\&A}, 602, A21
\refs 
Schoonenberg, D. et al., 2019, {\it A\&A}, 627, A149
\refs 
Segura-Cox, D.M. et al., 2020, {\it Nature}, 586, 228–231
\refs 
Sekiya, M. et al., 1980, {\it Prog. Theor. Phys.}, 64, 1968–1985
\refs 
Sekiya, M. et al., 1981, {\it Prog. Theor. Phys.}, 66, 1301–1316
\refs 
Selsis, F. et al., 2011, {\it A\&A}, 532, A1
\refs 
Senshu, H. et al., 2002, {\it JGR},  107, 5118
\refs 
Sharp, Z.D., 2017, {\it Chem. Geol.}, 448, 137–150
\refs 
Shi, C.Y. et al., 2013, {\it Nat. Geosci.}, 6, 971–975
\refs 
Shimazu, H. \& Terasawa, T., 1995, {\it JGR},  100, 16923–16930
\refs 
Siebert, J. et al., 2011, {\it GCA}, 75, 1451–1489
\refs 
Siebert, J. et al., 2012, {\it EPSL}, 321–322, 189–197
\refs 
Siebert, J. et al., 2013, {\it Science}, 339, 1194–1197
\refs 
Siebert, J. et al., 2016, {\it Nature}, 536, 326–328
\refs 
Siebert, J. et al., 2018, {\it EPSL}, 485, 130–139
\refs 
Simon, J.B. et al., 2016, {\it ApJ}, 822, 55
\refs 
Singer, K.N. et al., 2019, {\it Science}, 363, 955–959
\refs 
Sizova, E. et al., 2010, {\it Lithos}, 116, 209–229
\refs 
Sleep, N.H. \& Zahnle, K., 2001, {\it JGR:P}, 106, 1373–1399
\refs 
Sleep, N.H. et al., 2001, {\it PNAS}, 98, 3666–3672
\refs 
Smith, J.V. et al., 1970, {\it Proc. Apollo 11 Lunar Sci. Conf.}, 1, 897–925
\refs 
Snape, J.F. et al., 2016, {\it GCA} 174, 13–29
\refs 
Solomatov, V., 2015, {\it Treatise on Geophysics} (G. Schubert, Elsevier), 81–104
\refs 
Solomatov, V.S. \& Stevenson, D.J., 1993, {\it JGR}, 98, 5391–5406
\refs 
Solomatov, V.S., 2000, {\it Origin of the Earth and Moon}, 1, 323–338
\refs 
Solomatova, N.V, \& Caracas, R., 2021, {\it Sci. Adv.}, 7, eabj0406
\refs 
Solomon, S.C., 1979, {\it PEPI}, 19, 168–182
\refs 
Sonett, C.P.  et al., 1970, {\it Ast. Space Sci.}, 7, 446–488
\refs 
Sossi, P.A. et al., 2019, {\it GCA}, 260, 204–231
\refs 
Sossi, P.A. et al., 2020, {\it Sci. Adv.}, 6, eabd1387
\refs 
Spalding, C. \& Adams, F.C., 2020, {\it PSJ} 1, 7
\refs 
Spencer, D.C. et al., 2020, {\it JGR:P}, 125, e2020JE006604
\refs 
Šrámek, O. et al., 2012, {\it Icarus}, 217, 339–354
\refs 
Stähler, S.C. et al., 2021, {\it Science}, 373, 438–443
\refs 
Stamenković, V. \& Breuer, D.., 2014, {\it Icarus}, 234, 174–193
\refs 
Stephant, A. et al., 2021, {\it GCA}, 297, 203–219
\refs 
Stern, R. J. et al., 2018, {\it Geosci. Front.}, 9, 103–119
\refs 
Stern, R. J., 2018, {\it Philos. Trans. R. Soc. A}, 376, 20170406 
\refs 
Stevenson, D.J., 1981, {\it Science}, 214, 611–619
\refs 
Stevenson, D.J., 1988, {\it Origin of the Earth} (vol. 681, Lunar and Planetary Institute), 87–88
\refs 
Stevenson, D.J., 1989, {\it Mantle Convection Plate Tectonics and Global Dynamics} (W.R. Peltier, ed., Gordon \& Breach) 817 - 873
\refs 
Stevenson, D.J., 2014, {\it ACCRETE Group Meeting}
\refs 
Stewart, A.J. et al., 2007, {\it Science}, 316, 1323
\refs 
Stewart, S.T. et al., {\it AIPCP}, 2272, 080003
\refs 
Stixrude, L. et al., 2009, {\it EPSL} 278, 226–232
\refs 
Stixrude, L. et al., 2020, {\it Nat. Commun.}, 11, 935
\refs 
Stixrude, L., 2014, {\it Philos. Trans. R. Soc. A}, 372, 20130076
\refs 
Stökl, A. et al., 2015, {\it A\&A}, 576, A87
\refs 
Suer, T.-A. et al., 2017, {\it EPSL}, 469, 84–97
\refs 
Sugiura, N. \& Fujiya, W., 2014, {\it M\&PS}, 49, 772–787
\refs 
Tackley, P. J. et al., 2013, {\it Icarus}, 225, 50–61
\refs 
Tagawa, S. et al., 2021, {\it Nat. Comm.}, 12, 1–8
\refs 
Takafuji, N. et al., 2004, {\it EPSL}, 224, 249–257
\refs 
Tang, H. \& Dauphas, N., 2012. EPSL, 359, 248–263
\refs 
Tang, H. \& Young, E.D., 2020, {\it PSJ} 1, 49
\refs 
Tarduno, J.A. et al., 2015, {\it Science}, 349, 521–524
\refs 
Teachey, A. \& Kipping, D.M., 2018, {\it Sci. Adv.} 4, eaav1784
\refs 
Tera, F. et al., 1974, {\it EPSL} 22, 1-21
\refs 
Terasaki, H. et al., 2007, {\it PEPI}, 161, 170–176
\refs 
Terasaki, H. et al., 2012, {\it PEPI}, 202–203, 1–6
\refs 
Thiemens, M.M. et al., 2019, {\it Nat. Geosci.}, 12, 696–700
\refs 
Thompson, C. \& Stevenson, D.J., 1988, {\it ApJ} 333, 452–481
\refs 
Thompson, M.A. et al., 2019, {\it ApJ}, 875, 45
\refs 
Throngren, D.P. \& Fortney, J.F., 2018, {\it ApJ}, 155, 214
\refs 
Timms, N.E. et al., 2017, {\it Earth Sci. Rev.}, 165, 185–202
\refs 
Tkalcec, B.J. et al., 2013, {\it Nat. Geosci.}, 6, 93–97
\refs 
Tonks, W.B. \& Melosh, H.J., 1992, {\it Icarus}, 100, 326–346
\refs 
Tonks, W.B. \& Melosh, H.J., 1993, {\it JGR},  98, 5319–5333
\refs 
Tosi, N. \& Padovan, S., 2021, {\it Mantle Convection and Surface Expressions},  (AGU), Ch. 17
\refs 
Tosi, N. et al., 2017, {\it A\&A}, 605, A71
\refs 
Touboul, M. et al., 2007, {\it Nature}, 450, 1206–1209
\refs 
Touboul, M. et al., 2015, {\it Nature}, 520, 530–533
\refs 
Trail, D. et al., 2011, {\it Nature}, 480, 79–82
\refs 
Trappitsch, R. et al., 2018, {\it ApJL}, 857, L15
\refs 
Trinquier, A. et al., 2007, {\it ApJ}, 655, 1179
\refs 
Trinquier, A. et al., 2009, {\it Science}, 324, 374–376
\refs 
Tsai, S.M. et al., 2021, {\it ApJL}, 922, L27
\refs 
Tsuno, K. et al., 2013, {\it GRL}, 40, 66–71
\refs 
Tucker, J.M. \& Mukhopadhyay, S., 2014, {\it EPSL}, 393, 254–265
\refs 
Turcotte, D.L. \& Schubert, G., 2014, {\it Geodynamics}, Cambridge University Press
\refs 
Turbet, M. et al., 2019, {\it A\&A}, 628, A12
\refs 
Turbet, M. et al., 2021, {\it Nature}, 598, 276-280
\refs 
Turner, S. et al., 2020, {\it Nat. Commun.}, 11, 1241
\refs 
Ulvrova, M. et al., 2012, {\it PEPI}, 206, 51–66
\refs 
Unterborn, C.T. et al., 2018, {\it Nat. Astron.}, 2, 297–302
\refs 
Valencia, D. \& O’Connell, R. J., 2009, {\it EPSL}, 286, 492–502 
\refs 
Valencia, D. et al., 2007, {\it ApJL}, 670, L45
\refs 
Van Eylen, V. et al., 2018, {\it MNRAS}, 479, 4786–4795
\refs 
Van Hoolst, T. et al., 2020, {\it JGR:P}, 125, e2020JE006473
\refs 
Varas-Reus, M.I. et al., 2019, {\it Nat. Geosci.}, 12, 779–782
\refs 
Vazan, A. et al., 2018, {\it ApJ}, 869, 163
\refs 
Vazan, A. et al., 2022, {\it ApJ}, 926, 150
\refs 
Venturini, J. et al., 2020, {\it A\&A}, 643, L1
\refs 
Villeneuve, J. et al., 2009, {\it Science}, 325, 985–988
\refs 
Visser, R.G. \& Ormel, C.W., 2016, {\it A\&A}, 586, A66
\refs 
Wade, J. \& Wood, B.J., 2005, {\it EPSL}, 236, 78–95
\refs 
Wade, J. et al., 2012, {\it GCA}, 85, 58–74     
\refs 
Wahl, S.M. \& Militzer, B., 2015, {\it EPSL} 410, 25–33
\refs 
Wakita, S. \& Sekiya, M., 2011, {\it EPS}, 63, 1193–1206
\refs 
Walker, J.C.G. et al., 1981, {\it JGR:O}, 86, 9776–9782
\refs 
Walker, R.J., 2009, {\it Geochem. J.}, 69, 101–125
\refs 
Wang, J. et al., 2019, {\it BAAS}, 521, 200
\refs 
Wang, Z. \& Becker, H., 2013, {\it Nature}, 499, 328–331
\refs 
Warren, P.H., 2011, {\it EPSL}, 311, 93–100
\refs 
Way, M.J. et al., 2016, {\it GRL}, 43, 8376–838
\refs 
Weiss, B.P. et al., 2010, {\it SSR}, 152, 341–390
\refs 
Weiss, B.P. et al., 2021, {\it Sci. Adv.}, 7, eaba5967
\refs 
Wicks, J.K. et al., 2018, {\it Sci. Adv.}, 4, eaao5864
\refs 
Wieczorek, M.A. et al., 2013, {\it Science}, 339, 671–675
\refs 
Wilde, S.A. et al., 2001, {\it Nature}, 409, 175–178
\refs 
Williams, C.D. \& Mukhopadhyay, S., 2019, {\it Nature}, 565, 78–81
\refs 
Williams, H.M. et al., 2021, {\it Sci. Adv.} 7, eabc7394
\refs 
Williams, Q., 2009, {\it EPSL}, 284, 564–569
\refs 
Williams, Q., 2018, {\it AREPS}, 46, 47–66
\refs 
Wilson, L. \& Keil, K., 1991, {\it EPSL}, 104, 505–512
\refs 
Wilson, L. \& Keil, K., 2017, {\it Planetesimals: Early Differentiation and Consequences for Planets} (L.T. Elkins-Tanton and B.P. Weiss, Cambridge Planetary Science), 159–179
\refs 
Wohlers, A. \& Wood, B.J., 2015, {\it Nature}, 520, 337–340
\refs 
Wolf, E.T. \& Toon, O. B., 2010, {\it Science}, 328, 1266–1268
\refs 
Wood, J.A. et al., 1970, {\it GCA}, Supplement 1., Proc. Apollo 11 Lunar Sci. Conf., 965
\refs 
Wordsworth, R. \& Pierrehumbert, R. T., 2013b, {\it ApJ}, 778, 154
\refs 
Wordsworth, R. \& Pierrehumbert, R., 2013a, {\it Science}, 339, 64–67
\refs 
Wordsworth, R. \& Kreidberg, L., 2022, {\it ARAA}, , {\it arXiv}:2112.04663
\refs 
Wordsworth, R. et al., 2018, {\it AJ},   155, 195
\refs 
Wordsworth, R., 2016, {\it AREPS}, 44, 381–408
\refs 
Wu, J. et al., 2018, {\it JGR:P}, 123, 2691–2712
\refs 
Xu, S. \& Bonsor, A., 2021, {\it Elements}, 17, 241–244
\refs 
Yang, J. et al., 2013, {\it ApJL}, 771, L45
\refs 
Yokochi, R. \& Marty, B., 2004, {\it EPSL}, 225, 77–88
\refs 
Yoshino, T. et al., 2003, {\it Nature}, 422, 154–157
\refs 
Young, E.D. et al., 1999, {\it Science}, 286, 1331–1335
\refs 
Young, E.D. et al., 2019, {\it Icarus}, 323, 1–15
\refs 
Yu, X. et al., 2021, {\it ApJ}, 914, 38
\refs 
Zahnle, K.J., 2006, {\it Elements}, 2, 217–222
\refs 
Zahnle, K. et al., 2010, {\it Cold Spring Harb. Perspect. Biol.}, 2, a004895
\refs 
Zahnle, K.J. \& Carlson, R.W., 2020, {\it Planetary Astrobiology} (V.S. Meadows, G.N. Arney, B.E. Schmidt, and D.J. Des Marais, University of Arizona Press), 3–36
\refs 
Zahnle, K.J. et al., 1988, {\it Icarus}, 74, 62–97
\refs 
Zahnle, K.J. et al., 2015, {\it EPSL}, 427, 74–82
\refs 
Zahnle, K.J. et al., 2020, PSJ, 1, 11
\refs 
Zaranek, S.E. \& Parmentier, E.M., 2004, {\it JGR},  109, B03409
\refs 
Zellner, N.E., 2017, {\it Orig. Life Evol. Biosph.}, 47, 261–280
\refs 
Zeng, L. et al., 2019, {\it PNAS}, 116, 9723–9728
\refs 
Zhang, L. \& Fei, Y., 2008, {\it GRL}, 35, L13302
\refs 
Zhang, Y. et al., 2019, {\it JGR:SE}, 124, 10954
\refs 
Zhang, Z. et al., 2021, {\it JGR:P}, 126, e2020JE006754
\refs 
Zhu, K. (\begin{CJK*}{UTF8}{gbsn}朱柯\end{CJK*}) et al., 2020, {\it ApJ}, 888, 126
\refs 
Zube, N.G. et al., 2019, {\it EPSL}, 522, 210–21
%\bibliographystyle{pp7}
%\bibliographystyle{ppvi}
%\bibliographystyle{apj}
%\bibliographystyle{aasjournal}
%\bibliography{pp7.bib}
}
\end{document}